\documentclass[11pt]{article}

\usepackage{color}

\usepackage{amssymb}   
\usepackage{amsthm}    
\usepackage{amsmath}   
\usepackage{stmaryrd}  
\usepackage{titletoc}  
\usepackage{mathrsfs}  
\usepackage{graphicx}
\usepackage{multicol}
\usepackage{multirow}
\usepackage{subcaption}
\usepackage{caption}
\usepackage{float}
\usepackage{hyperref}
\usepackage{cleveref}
\usepackage{xcolor}
\usepackage{hyperref}
\usepackage[english]{babel}
\addto\extrasenglish{
  
}
\addto\extrasenglish{
  
}
\addto\extrasenglish{
  
}
\usepackage{comment}

\hypersetup{
  colorlinks   = true, 
  urlcolor     = red, 
  linkcolor    = blue, 
  citecolor   = blue 
}
\usepackage[round]{natbib}   
\newlength{\defbaselineskip}
\newcommand{\setlinespacing}[1]%
           {\setlength{\baselineskip}{#1 \defbaselineskip}}

\theoremstyle{plain}

\theoremstyle{definition}

\makeatletter\@addtoreset{equation}{section} \makeatother
 \allowdisplaybreaks
\begin{document}

\title{Penalized Variable Selection with Broken Adaptive Ridge Regression for Semi-competing Risks Data }

\author{Fatemeh Mahmoudi\footnotemark[2]  \and Xuewen Lu\footnotemark[2]
}

\maketitle

\begin{abstract}
Semi-competing risks data arise when both non-terminal and terminal events are considered in a model. Such data with multiple events of interest are frequently encountered in medical research and clinical trials. In this framework, terminal event can censor the
non-terminal event but not vice versa. It is known that variable selection
is practical in identifying significant risk factors in high-dimensional data.
While some recent works on penalized variable selection deal with these
competing risks separately without incorporating possible correlation between
them, we perform variable selection in an illness-death model using shared
frailty where semiparametric hazard regression models are used to model the
effect of covariates. We propose a broken adaptive ridge (BAR) penalty to
encourage sparsity and conduct extensive simulation studies to compare its
performance with other popular methods. We perform variable selection in an event-specific manner so that the potential risk factors can be selected and their effects can be estimated simultaneously corresponding to each event in the study. The grouping effect, as well as the oracle property
of the proposed BAR procedure are investigated using simulation studies.
The proposed method is then applied to real-life data arising from a
Colon Cancer study.
\end{abstract}


\section{Introduction}\label{sec:intro}
Multiple failure types can occur in survival analysis, as is well known. Semi-competing risks can classify data in this category in dealing with more than one event of interest. Two other popular settings that require analyzing multiple events of interest are competing risks and multivariate failure time data. The former is when an event precludes the other events from happening \citep{austin2017practical}, while in the latter, multiple types of events may happen to an individual \citep{thall2012recent}. However, semi-competing risks data arise when the so-called non-terminal event (e.g., disease progression) can censor the so-called terminal event (e.g., death), but not vice versa \citep{fine2001semi}. This type of data is commonly encountered in cancer clinical trials.
An example is a colon cancer study. The goal is to determine the effectiveness of two adjuvant therapy regimens in improving surgical cure rates in stage III colon cancer \citep{moertel1995fluorouracil}. In this example, three stochastic processes are of interest: time to cancer recurrence, time to death while being free of the recurrence, and time to death after cancer recurrence. Since this setting has a potential path from the non-terminal event to the terminal event, analyzing it under multivariate failure time or competing risks setting is oversimplifying the model and ignoring that possible transition. However, semi-competing risks data is different. In this type of data, movements between the states are modeled simultaneously without excluding the other possible transitions. For instance, in cancer studies, it is known that patients may be at risk of disease recurrence followed by death. Specifically, some examples of ignoring the non-terminal to terminal event transition in the colon cancer data include the works by \cite{cai2022adaptive} and \cite{lin1994cox} who considered it under multivariate failure times and \citet{bouvier2015incidence} under competing risks settings. In this work, we incorporate the natural format of semi-competing risks data to engage the cancer recurrence information (before death) in the model. 

There are various research works on semi-competing risks data analysis in the literature. An overlapping principal consideration in works related to this field is to figure out how to deal with the dependence between the events of interest. A primitive method is to model time to the events of interest through two marginal distributions where there is no constraint on their dependence structure \citep{ghosh2000nonparametric}. The other popular method is to step forward and utilize copula to account for the dependence \citep{ghosh2006semiparametric,fu2013joint}. Another technique uses the conditional modeling approach to model the transition-specific hazard functions for the terminal and non-terminal events. The innovative work by \cite{xu2010statistical} is an example of this approach in which a frailty model for semi-competing risks data has been proposed. Furthermore, the multiplicative Cox model was employed for the corresponding three transitions. Also, the Gamma frailty and the non-parametric maximum likelihood estimation were utilized to manipulate the estimation task. Recently, \cite{lee2021fitting} extended the shared-frailty illness death model to right-censored and left-truncated semi-competing risks data under the multi-state modeling approach. The illness-death model is a simple non-trivial example of multi-state models in which individuals may undergo a transient (diseased) state before reaching a terminal (dead) state \citep{vakulenko2016comparing}. 

Another complication that has been increasingly arising in the era of big data is to deal with a large number of covariates in high-dimensional data sets. The importance of selecting relevant covariates has led to ongoing progress in developing variable selection methods. However, most existing works only apply to a unique event of interest. Among others that include multiple events of interest, \cite{cai2022adaptive} considered an adaptive bi-level variable selection method to analyze multivariate failure time data. In addition, \cite{ha2014variable} worked on a variable selection problem for clustered competing risks data under proportional sub-distribution hazards (PSH) frailty models, and \cite{fu2017penalized} proposed a generalized variable selection under the PSH model and investigated its theoretical oracle properties. 

Penalized variable selection methods and various penalty functions have been widely investigated under different models. Some of the popular penalties proposed in the literature include least absolute shrinkage and selection operator (LASSO) proposed by \cite{tibshirani1996regression} for linear models. \cite{zou2006adaptive} proposed Adaptive LASSO to improve the performance of LASSO by incorporating some adaptive weights to achieve oracle properties. A non-convex penalty, smoothly-clipped absolute deviation (SCAD), was proposed by \cite{fan2001variable} for linear models. \cite{fan2002variable} used SCAD under the Cox regression model. Among the existing penalties, it is well-known that $L_0$ penalization enjoys the most excellent optimal properties for estimation and variable selection as it directly penalizes the cardinality of the model \citep{shen2012likelihood}. However, working with this penalty in high-dimensional data is not feasible. Variable selection, in that case, would be an NP-hard problem, and searching for the best subset with a non-convex penalty function makes it impractical to select essential variables.

Recently, an innovative method, namely broken adaptive ridge (BAR) regression, has been proposed for variable selection. It can be defined as an iteratively reweighted squared $L_2$-penalized regression that approximates the $L_0$-penalized regression. \cite{liu2016efficient} initiated the first work on BAR under the context of generalized linear models with uncensored data. It was then investigated in  \cite{kawaguchi2017scalable} under the Cox model with right-censored data, in \cite{dai2018broken} for linear models, and in \cite{zhao2019simultaneous} for the Cox model with interval-censored data, respectively. BAR has been shown to have great computational feasibility in these studies as it converges fast and can significantly accelerate the process. In addition, it has an excellent property of group effects. A complication frequently occurs in high-dimensional data is to deal with a high correlation among the covariates. It can make variable selection more complicated in such cases as it is natural for the variables clustered in a group to share similar properties and they should be selected together. Explicitly, this happens in many gene expression data where gene pathways can be grouped. It can be troublesome to solve variable selection problems as almost all the existing penalty functions only possess grouping effects if one incorporates the group structure into the regularization procedure. In contrast with other existing methods, BAR is specifically functional in recognizing and estimating significant grouped covariate effects simultaneously and automatically.  
In addition to the works mentioned earlier, BAR has been discussed by \cite{zhao2019simultaneous} for right-censored recurrent event data. Furthermore, \cite{li2020penalized} studied BAR under the semiparametric transformation models with interval-censored data and \cite{sun2022broken} extended it to the semiparametric accelerated failure time model with right-censored data.

In this paper, we employ BAR for variable selection problems in the illness-death model with semi-competing risks data. Furthermore, we assume data to be right-censored and potentially left-truncated data. The model we considered is the illness-death model studied by \cite{xu2010statistical}, \cite{lee2021fitting}, and \cite{vakulenko2016comparing} under parametric and semiparametric model assumptions, respectively. 
In the illness-death model, a shared frailty term is exploited to model the dependence among different events, and the Cox proportional hazards model is utilized to model three state transitions. 
Depending on the knowledge of the baseline hazard functions in these transitions, we take two different approaches: parametric and semiparametric. We assume the Weibull baseline hazard functions in the parametric approach, where a standard parametric likelihood-based variable selection method can be formulated. In the semiparametric approach, we assume unknown baseline hazard functions and adopt the sieve method considered in \cite{zhao2019simultaneous} to construct a penalized sieve likelihood for variable selection. Therefore, we approximate the baseline hazard functions by Bernstein polynomials that possess some significant advantages over similar methods. More discussion can be found in \autoref{sec:discussion}.
We propose an optimization algorithm that takes advantage of an iteratively reweighted least square method to implement the proposed method. This strategy approximates the likelihood function for a complicated model with a simple least squares function. An iterative optimization leads to a straightforward application of BAR in a simple linear model. 
We evaluate the performance of the proposed variable selection method using BAR in extensive simulation studies and show its superior performance to other existing methods. 
A generalized cross-validation (GCV) method is established for the tuning parameter selection. Finally, we apply our method to a colon cancer study for illustration.  

To the best of our knowledge, in the literature, there are very few papers for 
variable selection for the frailty-based illness-death model.
A recent work by \cite{reeder2022penalized} presented 
an approach that combines non-convex and structured fusion penalization, inducing global sparsity and parsimony across submodels, and proved the statistical error bound results. Their method can handle high-dimensional data where the number of regression parameters exceeds the number of observations. From a different motivation, instead of fusing regression parameters to force the selection parameters to be the same across submodels, our work focuses on group effects. That is, the covariates in each submodel may be highly correlated, and the parameters in each group can be estimated to be the same. Another motivation is that the likelihood-based loss function is non-convex in this setting; locally, we can approximate the loss function by a least squares loss function, which is a convex function. BAR penalty is also a convex function; therefore, the penalized loss function is convex. This convexity greatly impacts computation. 
Based on the above discussion, our contributions of this paper are fourfold. First, we propose a framework for selecting sparse covariate sets for each submodel via BAR penalties which are convex and possess the so-called group effects. Secondly, we develop an efficient optimization algorithm by approximating the non-convex loss function by a quadratic least squares type of loss function so that the problem boils down to a pure convex optimization problem. Further, we use Bernstein polynomials to approximate unknown baseline hazard functions for semiparametric models to facilitate computation. Bernstein polynomial is an approximation tool for modeling non-parametric components in statistical models, and it boasts some specific advantages over some of its competing methods. Thirdly, our method works for a diverging number of covariates, i.e., the dimension of covariates or regression parameters $p$ is less than the sample size $n$, but grows with $n$, i.e., $p$ tends to infinity when $n$ tends to infinity. When the dimension is high or $p > n$, we suggest reducing the dimension using some screening methods such as the sure independence screening \citep{desboulets2018review}, then applying our method. Finally, our method can handle both right-censored and left-truncated data. 
A noteworthy point about the variable selection in this work is not just about dealing with more than one event of interest in semi-competing risks setting but also treating the covariates corresponding to the non-terminal and terminal events separately. Hence, We can assess the significance of the effects of covariates differently, corresponding to the states (transient or absorbing) in the model. 

The remainder of this article is organized as follows. First, \autoref{sec:modelsandmethods} presents an overview of semi-competing risks data as well as the shared frailty multi-state modeling approach and the BAR estimation methodology. Next, \autoref{sec:bar} clarifies the proposed variable selection procedure along with its computation algorithm. \autoref{sec:sim} reports an extensive simulation study on assessing the performance of the proposed method from both individual and clustered variables perspectives. Finally, \autoref{sec:realdata} analyzes a real-life data set to illustrate the method, and \autoref{sec:discussion} concludes the paper. In addition, \autoref{sec:additionalresults} covers the supplementary materials, including the analysis based on the parametric method using the Weibull distribution.

\section{Estimation Structure}
\label{sec:modelsandmethods}

\subsection{Notation, Data, and Likelihood Construction}\label{subsec:notdatlikcons}

Among the different techniques introduced for analyzing semi-competing risks data, we adopt an illness-death model to jointly exploit the information on terminal and non-terminal events of interest. More details on illustration of this setting can be found in \cite{putter2007tutorial} and \cite{xu2010statistical}. Illness-death model is a particular form of multi-state modeling approach in which patients' starting state is an initial condition where they are at risk of either moving to the state of a non-terminal event prior to moving to the absorbing state or directly transitioning to the terminal state. They can also move from the non-terminal state to the terminal one.

We consider a multiple failure time study containing $n$ independent subjects. For the $i\textsuperscript{th}$ subject, there exist three sets of $d_k$-dimensional vectors of covariates $\boldsymbol{Z}_{ik}=(Z_{i1},\ldots,Z_{id_k})^\top$ for $k=1,2,3$ and $i=1,\ldots,n$. $T_{i1}$ and $T_{i2}$ denote the time to the non-terminal and terminal events, respectively, for the $i\textsuperscript{th}$ subject. 
One key concept that needs to be characterized formally in illness-death models is realizing the form of dependence between $T_1$ and $T_2$. A popular procedure for structuring this dependence is introducing a frailty term \citep{gorfine2021marginalized, jiang2017semi, xu2010statistical} to absorb the information on the dependence of non-terminal and terminal events.

Let $\omega$ denote the subject-specific frailty term and $\lambda_{k}(\cdot)$ represent the hazard functions or intensities of moving between the states for $k=1,2,3$. While $\lambda_{1}(\cdot)$ and $\lambda_{2}(\cdot)$ correspond to the cause-specific hazards in the competing risks setting, 
$\lambda_{3}(\cdot)$ is responsible for carrying the hazard information of moving from the non-terminal status to the terminal status, which is exclusive to the illness-death model setting. 

In this work, we follow the semi-Markov approach with frailty $\omega$ to account for the dependence of $T_{1}$ and $T_{2}$. Semi-Markov approach encompasses setting the time back to 0 at each state entry time. An example of making this assumption in the recent literature is the work by \cite{jazic2020estimation} on analyzing a nested case-control study. Considering the proportional hazards Cox model for intensities of moving between states and assuming that the frailty term is independent of covariates, the hazard functions of the model are given by 
\begin{eqnarray}
\label{haztransition1}
\lambda_1(t_{1}|\boldsymbol{Z}_{1}, \omega)&=&\omega\lambda_{01}(t_{1})\exp(\boldsymbol{\beta}_{1}^\top\boldsymbol{Z}_{1}),\\
\label{haztransition2}
\lambda_2(t_{2}|\boldsymbol{Z}_{2}, \omega)&=&\omega_i\lambda_{02}(t_{2})\exp(\boldsymbol{\beta}_2^\top\boldsymbol{Z}_{2}),\\
\label{haztransition3}
\lambda_{3}(t_{2}|T_{1}=t_{1},\boldsymbol{Z}_{3}, \omega)&=&\omega_i\lambda_{03}(t_{2}|T_{1}=t_{1})\exp(\boldsymbol{\beta}_3^\top\boldsymbol{Z}_{3}),~0<t_{1}<t_{2},
\end{eqnarray}
where $\lambda_{0k}$ and $\boldsymbol{\beta}_k$, 
$k=1,2,3$, denote the true baseline hazards functions and the vector of regression coefficients parameters in the Cox model for time to the non-terminal, terminal, and from non-terminal to the terminal state, respectively.
The semi-Markov approach means 
$\lambda_{03}(t_2 \mid T_1=t_1)=\lambda_{03}(t_2-t_1)$. The frailty term $\omega$ is not observable. Therefore, to derive the conditional likelihood function of the model and construct the foundation for performing variable selection, a conventional choice is to consider the Gamma distribution function for the frailty and assume that $\omega \sim \Gamma(1/\gamma,1/\gamma)$. Furthermore, to generalize the model to encompass more complex data types, as in \cite{lee2021fitting}, we assume that the observed data are subject to right-censoring and left-truncation. For the $i\textsuperscript{th}$ subject, assume
$C_i$ and $L_i$ are censoring and truncation times, respectively, and $T_{i1}$ and $T_{i2}$ are independent of $C_i$ and $L_i$ conditional on $\boldsymbol{Z}_i$.

Suppose that one  observes a possibly  right-censored and left-truncated random sample, 
\begin{eqnarray*}
\mathcal{D}=\{L_i,\delta_{i1}, Y_{i1},\delta_{i2},Y_{i2}, (\boldsymbol{Z}_{i1}, \boldsymbol{Z}_{i2}, \boldsymbol{Z}_{i3}) \}_{i=1}^n,
\end{eqnarray*}
for $k=1,2,3$, where
\begin{eqnarray*}
  Y_{i1}&=&\text{min}(T_{i1},T_{i2},C_i)\\
  &=&\begin{cases}
    T_{i1}, & \text{non-terminal event is the first state to transition},\\
    T_{i2}, & \text{terminal event is the first state to transition},\\
    C_i, & \text{censored before any transitioning},
    \end{cases}\\
    \\
    \delta_{i1}&=&I\left\{T_{i1}\leq\text{min}\left(T_{i2},C_i\right)\right\}\\
    &=&\begin{cases}
    1, & \text{non-terminal event is observed},\\
    0, & \text{censored or transitioning to the terminal event before non-terminal event},
    \end{cases}\\
    \\
    Y_{i2}&=&\text{min}(T_{i2},C_i)\\
  &=&\begin{cases}
    T_{i2}, & \text{terminal event is the second state to transition},\\
    C_i, & \text{censored before transitioning to the terminal event from non-terminal state},
    \end{cases}\\
    \\
    \delta_{i2}&=&I\left\{T_{i2}\leq C_i\right\}\\
  &=&\begin{cases}
    1, & \text{terminal event is observed},\\
    0, & \text{censored prior to transitioning to terminal event},
    \end{cases} \\
  L_i &<& Y_{i1}.
\end{eqnarray*}
Hence, there are four possible scenarios for each subject in the study:
\begin{itemize}
\item [1.] Non-terminal and terminal events are both observed:\\
$\left\{L_i,\delta_{i1}=1,T_{i1},\delta_{i2}=1,T_{i2}, (\boldsymbol{Z}_{i1}, \boldsymbol{Z}_{i2}, \boldsymbol{Z}_{i3}) \right\}_{i=1}^n$.
\item [2.] Non-terminal event is observed followed with censoring:\\
$\left\{L_i,\delta_{i1}=1,T_{i1},\delta_{i2}=0,C_i, (\boldsymbol{Z}_{i1}, \boldsymbol{Z}_{i2}, \boldsymbol{Z}_{i3}) \right\}_{i=1}^n$.
\item [3.] Only terminal event is observed:\\ $\left\{L_i,\delta_{i1}=0,T_{i1},\delta_{i2}=1,T_{i2}, (\boldsymbol{Z}_{i1}, \boldsymbol{Z}_{i2}, \boldsymbol{Z}_{i3})\right\}_{i=1}^n$.
\item [4.] No event is observed:\\
$\left\{L_i,\delta_{i1}=0,C_i,\delta_{i2}=0,C_i, (\boldsymbol{Z}_{i1}, \boldsymbol{Z}_{i2}, \boldsymbol{Z}_{i3})\right\}_{i=1}^n$.
\end{itemize}
Denoting the contribution of each of the above-mentioned scenarios by  $f_r(\cdot);r=1,2,3,4$, the likelihood function can be defined as
\begin{eqnarray}
\label{likelihood-semip}
\mathcal{L}_n(\boldsymbol{\beta},\boldsymbol{\Phi})=\prod_{i=1}^n\Big[\{f_1(y_{i1},y_{i2}|\boldsymbol{Z}_i)\}^{\delta_{i1}\delta_{i2}} \{f_2(y_{i1},y_{i2}|\boldsymbol{Z}_i)\}^{\delta_{i1}(1-\delta_{i2})}\nonumber\\
\{f_3(y_{i1},y_{i2}|\boldsymbol{Z}_i)\}^{(1-\delta_{i1})\delta_{i2}} \{f_4(y_{i1},y_{i2}|\boldsymbol{Z}_i)\}^{(1-\delta_{i1})(1-\delta_{i2})}\Big],
\end{eqnarray}
where $\boldsymbol{\Phi}=(\boldsymbol{\xi}_1^\top,\boldsymbol{\xi}_2^\top,\boldsymbol{\xi}_3^\top,\gamma)^\top$ denotes the parameter vector for three baseline hazard functions and the frailty distribution. Each of the contributions can be derived by integrating the frailty term out as
\begin{eqnarray}
\label{each-likelihood}
f_r(y_{i1},y_{i2}|\boldsymbol{Z}_i)=\int_{0}^\infty \mathcal{L}_r(y_{i1},y_{i2}|\boldsymbol{Z}_i,\omega_i)f(\omega_i|\gamma)d\omega_i,
\end{eqnarray}
where $f(\omega_i|\gamma)$ represents the Gamma density function of the subject-specific frailty $\omega_i$ given by
\begin{eqnarray*}
f(\omega_i|\gamma)=\frac{\gamma^{-\frac{1}{\gamma}}}{\Gamma(\frac{1}{\gamma})}\omega_i^{\frac{1}{\gamma} -1}\exp\{ -\frac{\omega_i}{\gamma}\}.
\end{eqnarray*}
Integrating out the gamma frailty term (based on \eqref{each-likelihood}) is a straightforward task due to the closed form of Gamma distribution. Eventually, the logarithm of \eqref{likelihood-semip} has the form
\begin{eqnarray}
\label{loglikelihood-semip}
    \ell_n(\boldsymbol{\beta},\boldsymbol{\Phi})&=&\sum_{i=1}^n\Bigg[\delta_{i1}\delta_{i2}\Bigg\{\log(\lambda_{01}(y_{i1}))+\boldsymbol{\beta}_1^\top\boldsymbol{Z}_{i1}+\log(\lambda_{03}(y_{i2}))+\boldsymbol{\beta}_3^\top\boldsymbol{Z}_{i3}\nonumber\\
    &&-(\frac{1}{\gamma}+2)\log\bigg(\gamma[g_{i1}+g_{i2}]+1 \bigg)\Bigg\}\nonumber\\
    &&+(\delta_{i1})(1-\delta_{i2})\Bigg\{\log(\lambda_{02}(y_{i1}))+\boldsymbol{\beta}_2^\top\boldsymbol{Z}_{i2}-(\frac{1}{\gamma}+1)\log[\gamma g_{i2}+1]\Bigg\}\nonumber\\
    &&+(1-\delta_{i1})(\delta_{i2})\Bigg\{\log\lambda_{01}(y_{i1})+\boldsymbol{\beta}_1^\top\boldsymbol{Z}_{i1}-(\frac{1}{\gamma}+1)\log\bigg(\gamma[g_{i1}+g_{i2}]+1 \bigg)\Bigg\}\nonumber\\
    &&+(1-\delta_{i1})(1-\delta_{i2})\Bigg\{\frac{1}{\gamma}\log[\gamma g_{i2}+1]\Bigg\} \Bigg].
\end{eqnarray}
Denote by $\Lambda_{0k}(t)= \int_0^t\lambda_{0k}(u)du$ the cumulative baseline hazard functions for $k=1,2,3$, then
\begin{eqnarray*}g_{i1}=\int_0^{y_{i2}-y_{i1}}\lambda_3(u_i|y_{i1},\boldsymbol{Z}_{i3})du_i=\Lambda_{03}(y_{i2}-y_{i1})e^{\boldsymbol{\beta}_3\boldsymbol{Z}_{i3}},
\end{eqnarray*}
and 
\begin{eqnarray*}g_{i2}=\int_{l_i}^{y_{i2}}\left(\lambda_1(u_i|\boldsymbol{Z}_{i1})+\lambda_2(u_i|\boldsymbol{Z}_{i2})\right)du_i=\Lambda_{01}(y_{i2}-l_i)e^{\boldsymbol{\beta}_1\boldsymbol{Z}_{i1}}+\Lambda_{02}(y_{i2}-l_i)e^{\boldsymbol{\beta}_2\boldsymbol{Z}_{i2}}.
\end{eqnarray*}
For more details on the likelihood function, we refer to \cite{lee2021fitting}, \cite{vakulenko2016comparing}, and \cite{xu2010statistical}.

\subsection{Baseline Hazard Function Specification}
\label{subsec:baseline}

In this section, we present the specification of the baseline hazard functions in the Cox model. The general approach to estimate the set of parameters $\boldsymbol{\nu}
=(\boldsymbol{\beta}^\top,\boldsymbol{\Phi}^\top)^\top$ is to maximize \eqref{loglikelihood-semip}. However, prior to going further into the maximization procedure, the form of the unknown parameters of baseline hazard functions must be specified, and the estimation strategy of these functions should be established. 
In the literature, characterizing the baseline hazard function in the Cox model falls into two categories of approaches: parametric and non-parametric. 
For the parametric approach, a popular assumption in survival analysis is to fit Weibull distribution to the baseline hazard function \citep{lee2021fitting} and assume that $\lambda_{0k}(\cdot)\sim \text{Weibull}(\alpha_k,\tau_k)$. Thus, under a semi-Markovian framework, baseline hazard components of the log-likelihood function in \eqref{loglikelihood-semip} can be rewritten as 
\begin{eqnarray*}
\lambda_{0k}(t)=\alpha_k\tau_kt^{\alpha_k-1},
\end{eqnarray*}
for $k=1,2,3$. The parametric approach is straightforward in terms of computational feasibility. However, its main disadvantage is the strict assumptions that may be unrealistic in some applications.
For the non-parametric approach, the functional forms of the baseline hazard functions are unknown and infinite dimensional. We approximate them by finite-dimensional functions using the so-called sieve method. The main idea behind the sieve method is to approximate $\boldsymbol{\lambda}_{0k},~k=1, 2, 3$, in the infinite-dimensional parameter space using a sequence of finite-dimensional parameters. Let $\Omega$ denote the parameter space
\begin{eqnarray*}
\Omega=\left\{\boldsymbol{\nu}^\top
=(\boldsymbol{\beta}^\top,\boldsymbol{\Phi}^\top)
=(\boldsymbol{\beta}^\top,\gamma,\lambda_{01},\lambda_{02},\lambda_{03}) \in\mathcal{B}\otimes\mathcal{M}^1\otimes\mathcal{M}^2\otimes\mathcal{M}^3\right\},
\end{eqnarray*}
where 
\begin{eqnarray*}
\mathcal{B}=\left\{(\boldsymbol{\beta}^\top,\gamma)\in \mathbb{R}^{d_1}  \times  \mathbb{R}^{d_2}  \times  \mathbb{R}^{d_3}  \times \mathbb{R}^+,||\boldsymbol{\beta}||+|\gamma|\leq M\right\},
\end{eqnarray*}
with $M$ being a positive constant, and $\mathcal{M}^j$ denotes the collection of all bounded non-decreasing non-negative functions over the range of observed data for $j=1,2,3$. In order to apply the sieve method, one needs to choose a function that is known up to finite-dimensional parameters \citep{sun2006statistical}. Here, we employ Bernstein polynomials to approximate the transition-specific baseline hazard functions. Hence, the sieve space $\Omega_n$ can be defined as
\begin{eqnarray*}
\Omega_n=\left\{\boldsymbol{\nu}_n^\top
=(\boldsymbol{\beta}^\top,\gamma,\lambda_{01n},\lambda_{02n},\lambda_{03n})\in\mathcal{B} \otimes\mathcal{M}_n^1 \otimes\mathcal{M}^2_n \otimes\mathcal{M}^3_n\right\},
\end{eqnarray*}
with $\mathcal{M}_n^j$ defined as
\begin{eqnarray}
\label{sievespace}
\mathcal{M}_n^j=\left\{\lambda_{0jn}(t)=\exp\left(\sum_{k=0}^m \phi_{jk}^* B_k(t,m,c_j,u_j)\right):\sum_{0\leq k\leq m}|\phi_{jk}^*|\leq M_n\right\}.
\end{eqnarray}
In \eqref{sievespace}, $B_k(t,m,c_j,u_j)$ denotes the Bernstein basis polynomials with degree of freedom $m$ defined as
\begin{eqnarray*}
B_k(t,m,c_j,u_j)=\dbinom{m}{k}\left(\frac{t-c_j}{u_j-c_j} \right)^k\left( 1-\frac{t-c_j}{u_j-c_j}\right)^{m-k},
\end{eqnarray*}
for $j=1,2,3$ and $k=0,1,\ldots,m$. We define $c_j$ and $u_j$ as the first and last time to follow up in the survival analysis study for the $j\textsuperscript{th}$ hazard function with $j=1,2,3$. The Bernstein polynomial coefficients to be estimated are $\phi_{jk}^*$.

Finally, the set of parameters to be estimated is $\boldsymbol{\nu}=(\boldsymbol{\beta}^\top,
\boldsymbol{\Phi}^\top)^\top$.  
We denote $\boldsymbol{\Phi}=\boldsymbol{\Phi}^W$ for the parametric approach and $\boldsymbol{\Phi}=\boldsymbol{\Phi}^{BP}$ for the non-parametric approach, where
$
\boldsymbol{\Phi}^{BP}
=(\phi_{11}^*,\ldots,\phi_{1m_1}^*,\phi_{21}^*,\ldots,\phi_{2m_2}^*,\phi_{31}^*,\ldots,\phi_{3m_3}^*,\gamma)^\top
$
with $m_j$ representing the Bernstein polynomial degree corresponding to three transitions, $j=1,2,3$, and 
$\boldsymbol{\Phi}^{W}=(\alpha_1,\tau_1,\alpha_2,\tau_2,\alpha_3,\tau_3,\gamma)^\top$.

\section{BAR Penalized Estimation}\label{sec:bar}

For the penalized variable selection, it is natural to construct an objective function and then optimize it. In order to construct the objective function, we denote the estimate of $\boldsymbol{\beta}$ and $\boldsymbol{\Phi}$ without penalty by $\tilde{\boldsymbol{\beta}}$ and $\tilde{\boldsymbol{\Phi}}$, respectively. We construct 
$\ell_n(\boldsymbol{\beta})=\ell_n(\boldsymbol{\beta},\tilde{\boldsymbol{\Phi}})$ for variable selection. To do this, we propose to adopt the penalized likelihood method:
\begin{eqnarray}
\label{objective}
\ell_p(\boldsymbol{\beta})
=-\ell_n(\boldsymbol{\beta})+\sum_{k=1}^{K}\sum_{j=1}^{d_k} P_{\lambda_n}(\boldsymbol{\beta}_{j,k})
=-\ell_n(\boldsymbol{\beta})
 +\lambda_n\sum_{k=1}^{K}\sum_{j=1}^{d_k} \frac{ \beta_{j,k}^2}{\check{\beta}_{j,k}^2},
\end{eqnarray}
where $\lambda_n$ is a tuning parameter controlling a trade-off between the intensity of the sparsity of the model and the bias of the resulting estimates, 
$\check{\boldsymbol{\beta}}$ represents a consistent estimator of $\boldsymbol{\beta}$ with all the components being non-zero. The strength of BAR consists of two layers. First, the term 
$\beta_{j,k}^2/\check{\beta}_{j,k}$ converges to $I(|\beta_{j,k}|\neq 0)$ in probability as $n$ goes to infinity. This is why the method of BAR can be regarded as a surrogate of the $L_0$ penalization approach in an asymptotic sense. At the same time, it enjoys a simple closed form and computational efficiency. Second, it works with an adaptively reweighting and updating method. A procedure that can intelligently grow the weighted penalty for the zero components to shrink the non-relevant components to zero with great accuracy and outperforms other penalty functions such as LASSO and ALASSO (adaptive LASSO) \citep{zhao2019simultaneous,kawaguchi2017scalable}. 

To complete the task of selecting important variables, we need to minimize the objective function \eqref{objective}. We propose using an iteratively reweighted least square algorithm that involves a Newton-Raphson update. This algorithm approximates the nonlinear log-likelihood function with linear regression and hence, would be a considerable improvement in terms of diminishing the complexity of the optimization procedure. 

The iterative algorithm is provided below.
\begin{itemize}
    \item [\textit{Step 1.}] Use the parametric or semiparametric approach described in \eqref{sec:modelsandmethods} to get the estimate, $\tilde{\boldsymbol{\nu}}^\top=(\tilde{\boldsymbol{\beta}}^\top,\tilde{\boldsymbol{\Phi}}^\top)^\top$.
    \item[\textit{Step 2.}] Fix $\tilde{\boldsymbol{\Phi}}$ and set the initial estimator $\hat{\boldsymbol{\beta}}^{(0)}=\tilde{\boldsymbol{\beta}}=(\tilde{\boldsymbol{\beta}}_1^\top,\tilde{\boldsymbol{\beta}}_2^\top,\tilde{\boldsymbol{\beta}}_3^\top)^\top$ when $m=0$.
    \item[\textit{Step 3.}]At step $m+1$, compute $\boldsymbol{u}$, $\boldsymbol{H}$, $\boldsymbol{X}$, and $\boldsymbol{W}$ based on the current values of $\boldsymbol{\beta}=\hat{\boldsymbol{\beta}}^{(m)}$, where:
    \begin{eqnarray*}
    \boldsymbol{u}=(\boldsymbol{u}_1^\top,\boldsymbol{u}_2^\top,
    \boldsymbol{u}_3^\top)=(\partial \ell_n(\boldsymbol{\beta},\tilde{\Phi})/{\partial \boldsymbol{\beta}_1^\top},\partial \ell_n(\boldsymbol{\beta},\tilde{\Phi})/{\partial\boldsymbol{\beta}_2^\top},\partial \ell_n(\boldsymbol{\beta},\tilde{\Phi})/{\partial\boldsymbol{\beta}_3^\top})_{(\sum_{k=1}^3 d_k) \times 1}^\top
    \end{eqnarray*}
    is the gradient vector, and $d_k$ denotes the number of covariates corresponding to each transition for $k=1,2,3$. 
    $\boldsymbol{H}$ represents the Hessian matrix:
    \begin{eqnarray*}
    \boldsymbol{H} =\begin{bmatrix}
    {\boldsymbol{H}^{11}_{(d_1\times d_1)}} & {\boldsymbol{H}^{12}_{(d_1\times d_2)}}& \boldsymbol{H}^{13}_{(d_1\times d_3)}\\
    \boldsymbol{H}^{21}_{(d_2\times d_1)}& \boldsymbol{H}^{22}_{(d_2\times d_2)}& \boldsymbol{H}^{23}_{(d_2\times d_3)}\\
    \boldsymbol{H}^{31}_{(d_3\times d_1)}& \boldsymbol{H}^{32}_{(d_3\times d_2)} & \boldsymbol{H}^{33}_{(d_3\times d_3)}
    \end{bmatrix}_{(\sum_{k=1}^3 d_k\times \sum_{k=1}^3 d_k)},
    \end{eqnarray*}
    where $\boldsymbol{H}^{kk^\prime}=\frac{\partial^2\ell_n(\boldsymbol{\beta},\tilde{\boldsymbol{\Phi}})}{\partial\boldsymbol{\beta}_k \partial\boldsymbol{\beta}_{k^\prime}^\top}$ for $k=1,2,3,k^\prime=1,2,3$.
    The pseudo response vector is:
    $$\boldsymbol{W} =(\boldsymbol{X^\top})^{-1}\left\{ \boldsymbol{H} \boldsymbol{\beta} -\boldsymbol{u}\right\},
    $$
    where $-\boldsymbol{H} =\boldsymbol{X}^\top \boldsymbol{X}$ and $\boldsymbol{X}$ is an upper triangular matrix that is computed using the Cholesky decomposition of $\boldsymbol{H}$.
    \item[\textit{Step 4.}] Use the second-order Taylor expansion to approximate the objective function \eqref{objective} and rewrite the log-likelihood function as
    \begin{eqnarray*}
    -\ell_n(\boldsymbol{\beta})
    =-\ell_n(\boldsymbol{\beta},\tilde{\boldsymbol{\Phi}})=\frac{1}{2} (\boldsymbol{W} -\boldsymbol{X}\boldsymbol{\beta})^\top (\boldsymbol{W} -\boldsymbol{X}\boldsymbol{\beta}).
    \end{eqnarray*}
    \item[\textit{Step 5.}]Minimize the approximated objective function and obtain
    \begin{eqnarray*}
    \hat{\boldsymbol{\beta}}^{(m+1)}=\arg\min_{\boldsymbol{\beta}}\left\{\frac{1}{2} (\boldsymbol{W} -\boldsymbol{X}\boldsymbol{\beta})^\top (\boldsymbol{W} -\boldsymbol{X}\boldsymbol{\beta})\right\}+\sum_{k=1}^{K}\sum_{j=1}^{d_k}\lambda_n\frac{\beta_{j,k}^2}{\{\hat{\beta}_{j,k}^{(m)}\}^2} 
    \end{eqnarray*}
    for $K=3$, the closed-form solution for finding the BAR penalized estimator
    \begin{eqnarray*}
     \hat{\boldsymbol{\beta}}^{(m+1)}&=&
    \left\{\boldsymbol{X}^\top\boldsymbol{X}+\lambda_n D \right\}^{-1}\boldsymbol{X}^\top \boldsymbol{W},
    \end{eqnarray*}
    where 
    \begin{eqnarray*}
    D=\text{diag}
    \left(\frac{1}{
    {(\hat{\beta}^{(m)}_{1,1})}^2},\ldots,
    \frac{1}{{(\hat{\beta}^{(m)}_{d_1,1})}^2},
    \frac{1}{
    {(\hat{\beta}^{(m)}_{1,2})}^2},\ldots,
    \frac{1}{
    {(\hat{\beta}^{(m)}_{d_2,2})}^2},
    \frac{1}{
    {(\hat{\beta}^{(m)}_{1,3})}^2},\ldots,
    \frac{1}{
    {(\hat{\beta}^{(m)}_{d_3,3})}^2}\right)
    \end{eqnarray*}
    is a square matrix with $\sum_{k=1}^3 d_k$ rows and columns. 
    \item[\textit{Step 6.}] 
    Go back to \textit{Step 3} and reiterate until the convergence criterion is met. 
Then, the penalized BAR estimator can be found by iterating the above procedure until it converges, i.e.,  $\hat{\boldsymbol{\beta}}^{\text{BAR}}=\lim_{m \to \infty}\hat{\boldsymbol{\beta}}^{(m)}$.
\end{itemize}
The value of a tuning parameter, $\lambda_n$ can affect the performance of a penalized variable selection method in a large scale \citep{fan2013tuning}. Therefore, the remaining task is to select the optimal tuning parameter. There are different methods proposed in the literature to find the optimal tuning parameter, such as the Akaike information criterion (AIC) \citep{akaike1974new}, and the Bayesian information criterion (BIC) \citep{schwarz1978estimating}. Another popular method is the generalized cross-validation (GCV) \citep{craven1978smoothing}, which was first proposed to overcome the computational burden of the cross-validation (CV) method and was employed as a tuning parameter selection method under different models \citep{zhang2007adaptive,liu2013variable, huang2009group} afterwards. While CV requires dividing data into multiple subsets that impose heavy computation, GCV can handle the problem in only one iteration, increasing computation speed. This is specifically desirable for high-dimensional data. Working with semi-competing risks data for variable selection in the illness-death model means that the number of covariates is tripled. Therefore, it is desirable to use a computationally less expensive method. On the other hand, it has been advised in the literature that AIC, BIC, and GCV would produce similar results \citep{cai2020group}. Here, we denote the penalty function by $p_{\lambda_n}(\boldsymbol{\beta})$ where $\boldsymbol{\beta}=(\boldsymbol{\beta}^\top_1,\boldsymbol{\beta}^\top_2,\boldsymbol{\beta}_3^\top)^\top$ and $\boldsymbol{\beta}_k=(\beta_{1,k},\beta_{2,k},\ldots,\beta_{d_k,k})^\top$ for $k=1,2,3$. Then, the number of effective parameters can be computed by
\begin{eqnarray*}
s(\lambda_n)=\text{tr}\left[\{\boldsymbol{H}(\hat{\boldsymbol{\beta}})-v(
\lambda_n)\}^{-1}\boldsymbol{H}(\hat{\boldsymbol{\beta}})\right],
\end{eqnarray*}
where $\hat{\boldsymbol{\beta}}$ represents the penalized estimate of the vector of regression coefficients parameters, and $\boldsymbol{H}(\hat{\boldsymbol{\beta}})=\boldsymbol{H}$, the second derivative of the log-likelihood function at $\hat{\boldsymbol{\beta}}$,
\begin{eqnarray*}
v(\lambda_n)=\lambda_n r(\hat{\boldsymbol{\beta}}),
\end{eqnarray*}
and 
\begin{eqnarray*}
r(\hat{\boldsymbol{\beta}})=\text{diag}\left(\frac{\nabla p_{\lambda_n}(\hat{\boldsymbol{\beta}})}{|\hat{\boldsymbol{\beta}}_{1,1}\mid},\ldots,\frac{\nabla p_{\lambda_n}(\hat{\boldsymbol{\beta}})}{\mid\hat{\boldsymbol{\beta}}_{d_1,1}\mid},\frac{\nabla p_{\lambda_n}(\hat{\boldsymbol{\beta}})}{\mid\hat{\boldsymbol{\beta}}_{1,2}\mid},\ldots,\frac{\nabla p_{\lambda_n}(\hat{\boldsymbol{\beta}})}{\mid\hat{\boldsymbol{\beta}}_{d_2,2}\mid},\frac{\nabla p_{\lambda_n}(\hat{\boldsymbol{\beta}})}{\mid\hat{\boldsymbol{\beta}}_{1,3}\mid},\ldots,\frac{\nabla p_{\lambda_n}(\hat{\boldsymbol{\beta}})}{\mid\hat{\boldsymbol{\beta}}_{d_3,3}\mid}\right),
\end{eqnarray*}
where $\nabla$ denotes the first derivative.
Finally, the selected optimal tuning parameter is the one that minimizes the GCV criterion:
\begin{eqnarray*}
\text{GCV}(\lambda_n)&=&\frac{-\ell_n(\hat{\boldsymbol{\beta}})}{n\left[1-s(\lambda_n)/n\right]^2}.
\end{eqnarray*}

\section{Simulation Study}\label{sec:sim}
In this section, we conduct simulation studies to investigate the finite-sample performance of the proposed variable selection method. Our primary focus is to explore two aspects of this proposed method: first, its ability to handle covariate selection and estimation of covariate effects simultaneously, and second, its performance under a grouping effect structure where variables are clustered in groups with very high correlation values.  
We do experiments under known and unknown baseline hazard specification settings presented in \autoref{subsec:baseline} for which we consider 100 replications for each experiment. We generate data with the baseline hazard functions for both settings following the Weibull distribution. Afterwards, we employ both parametric and semiparametric methods using Weibull distribution and Bernstein polynomials to model the baseline hazard functions. Following the transition hazard functions in \eqref{haztransition1}, \eqref{haztransition2}, and \eqref{haztransition3}, we generate $T_{i1}$ with probability $1-Pr(T_{i1}=\infty)$ (time to the non-terminal event) from Weibull distribution. We utilize the inverse probability method to generate $T_{i1}$ as follows,
\begin{eqnarray}
T_{i1}=\Lambda_{01}^{-1}\left(\frac{-\log (1-U_{i1})}{\omega \exp (\boldsymbol{\beta}_1^\top \boldsymbol{Z}_{i1})} \right).
\end{eqnarray}
With $Pr(T_{i1}=\infty)$, we generate $T_{i2}$ (time to the terminal event directly moving from the initial state without observing the non-terminal event) using the hazard function  $ \lambda_{02}(t)=\alpha_2 \tau_2 t^{\alpha_2-1}$, and
\begin{eqnarray}
T_{i2}=\Lambda_{02}^{-1}\left(\frac{-\log (1-U_{i2})}{\omega \exp (\boldsymbol{\beta}_2^\top \boldsymbol{Z}_{i2})} \right).
\end{eqnarray}
Similarly, time to the terminal event following a non-terminal event $T_{i2}$ when $t_{i1}<\infty$ can be generated using the hazard function $\lambda_{03}(t)=\alpha_3\tau_3 t^{\alpha_3-1}$ conditioning on the observed value of $T_{i1}=t_{i1}$. In the above definitions, $U_k\sim\text{Uniform}(0,1)$ and $\Lambda_{0k}^{-1}(\cdot)$ is the inverse of the cumulative baseline hazard function defined by
\begin{eqnarray}
\Lambda_{0k} (x)=(\frac{x}{\tau_k})^{1/{\alpha_k}},
\end{eqnarray}
for $k=1,2,3$. It is worth noting that under a semi-Markovian setting, those observations of 
$T_2$ satisfying $T_1 \ge  T_2$, $T_2$ is taken as terminal event without experiencing
non-terminal event, while for those $T_2$ satisfying $T_1 <  T_2$, $T_2$ is replaced
by $T_2=T_1+T_3$. $T_3$ is the third transition's time (time to the absorbing state moving from the non-terminal state). Then, $T_2$ is adjusted by adding $T_1$ (time to the non-terminal event) to $T_3$ for those subjects with the observed intermediate event. Weibull parameters are set as $\text{log}(\alpha_1)=0.18$, $\text{log}(\tau_1)=-4$, $\text{log}(\alpha_2)=0.2$, $\text{log}(\tau_2)=-4$, $\text{log}(\alpha_3)=1.7$, $\text{log}(\tau_3)=-11$, and the frailty variable is generated from gamma distribution with $\gamma=0.25$. The left truncation time is generated independently from a uniform distribution. Our approach to handling the prevalent cases is to exclude them from the study. One can choose to let them stay in the model either by updating the likelihood function and adding two more terms to it as mentioned in \cite{lee2021fitting}, or using the approach that works via conditioning on the left truncation time as discussed in \cite{saarela2009joint}. However, either approach imposes more computational complexity on the model.

We study two levels of right censoring rate approximately at $50\%$ and $70\%$. As a specific case, all the $K=3$ sets of covariates are assumed to be the same and generated from the marginal standard normal distribution with pairwise correlation $\text{corr}(Z_{jk},Z_{j'k})=\rho^{\mid j-j'\mid}$ with $\rho=0.5$. This implies $d_1=d_2=d_3$. However, our method can be readily applied to cases where the sets of covariates may differ.

To test the proposed method in the case of a diverging number of covariates, we set the number of covariates $d_k=\lfloor 6n^{1/6}\rfloor$ and $p_n=p=3d_k$, for $k=1,2,3$, where the output of the floor function $f(x)=\lfloor x \rfloor$ is the largest integer less than or equal to $x$. The sample sizes are $n=100$, 300, and 500. Sample sizes and the number of covariates to select from in the scenario of diverging number of covariates are set below
\begin{eqnarray*}
\centering
\text{Scenario of diverging number of covariates:}
\begin{cases}
n=100,~p_n=12\times 3=36,\\
n=300,~p_n=15\times 3=45,\\
n=500,~p_n=16\times 3=48.
\end{cases}
\end{eqnarray*} 
True values of $\boldsymbol{\beta}_k$, $k=1,2,3$, are set as \begin{eqnarray*}
\boldsymbol{\beta}_{01}&=&(-0.8,1,1,0.9,\boldsymbol{0}_{d_1-q_n})^\top,\\
\boldsymbol{\beta}_{02}&=&(1,1,1,0.9,\boldsymbol{0}_{d_2-q_n})^\top,\\
\boldsymbol{\beta}_{03}&=&(-1,1,0.9,1,\boldsymbol{0}_{d_3-q_n})^\top,\end{eqnarray*}
where $q_n$ denotes the number of non-zero covariates in each sub-model which is 4 in this specific experiment. We set degrees of Bernstein polynomials to be $\boldsymbol{m}=(2,2,3)$.
In order to assess the performance of BAR in different scenarios and to summarize the simulation results for simultaneous covariate selection and estimation of covariate effects, five measures are reported. True positive (TP) - the averaged number of non-zero estimates whose true values are non-zero, false positive (FP) - the averaged number of non-zero estimates whose true values are zero, mean of misclassified variables (MCV), median of mean squared errors (MMSE) and standard deviation of MSE (SD).
MSE is defined as $(\boldsymbol{\hat{\beta}}_k-\boldsymbol{\beta}_{0k})^\top\boldsymbol{\Sigma}_k(\boldsymbol{\hat{\beta}}_k-\boldsymbol{\beta}_{0k})$, where $\boldsymbol{\hat{\beta}}_k$ and $\boldsymbol{\Sigma}_k$ represent the estimates of $\boldsymbol{\beta}_{0k}$ and the estimated population covariance matrix corresponding to the $k$th risk, respectively.
\begin{table}[H]
\caption{Summary of variable selection results with $n$=100, 300, and 500 and baseline hazard functions approximated using Bernstein Polynomials.}
\label{summary-BPBH}
\resizebox{\textwidth}{!}{%
\begin{tabular}{ccccccccccc}
\hline
       &  & \multicolumn{4}{c}{$50\%$ censoring rate} &  & \multicolumn{4}{c}{$70\%$ censoring rate} \\ \cline{3-6} \cline{8-11} 
       &  & \multicolumn{9}{c}{$n=100, p=12 \times 3$}                                                          \\ \cline{3-11} 
method &  & TP      & FP     & MCV   & MMSE (SD)      &  & TP      & FP     & MCV   & MMSE (SD)      \\ \cline{1-1} \cline{3-6} \cline{8-11} 
BAR    &  & 11.32   & 0.65 &   1.32 &  1.007 (1.062) &  &   10.37    &  0.95 & 2.57 & 2.478 (2.301)  \\
Lasso  &  &  11.59  & 4.23  & 4.63  & 2.819 (1.640)  &  & 10.51  &  3.15  & 4.64  & 5.728 (2.769)  \\
ALasso &  &  11.58  &  1.95  & 2.36  & 1.645 (1.322)  &  &  11.11  &  2.32  & 2.56  & 2.793 (2.162)  \\
Oracle &  & 12.00      & 0.00      & 0.00     & 0.722 (0.717)  &  & 12.00      & 0.00      & 0.00     & 1.025 (1.311)  \\ \cline{3-11} 
       &  & \multicolumn{9}{c}{$n=300, p=15 \times 3$}                                                          \\ \cline{3-11} 
BAR    &  & 12.00    & 0.43   & 0.43 & 0.482 (0.385)  &  & 11.99  & 0.65  & 0.66  & 0.502 (0.381)  \\
Lasso  &  & 12.00  & 8.69 & 8.69  & 1.553 (0.607)  &  & 12.00 & 7.12 & 7.14 & 1.427 (0.916)  \\
ALasso &  & 12.00  & 1.80  & 1.80  & 0.832 (0.452)  &  & 12.00   & 1.18   & 1.18  & 0.691 (0.646)  \\
Oracle &  & 12.00 & 0.00  & 0.00     & 0.445 (0.310)  &  & 12.00 & 0.00      & 0.00     & 0.420 (0.341)  \\ \cline{3-11} 
       &  & \multicolumn{9}{c}{$n=500, p=16 \times 3$}                                                          \\ \cline{3-11} 
BAR    &  & 12.00    & 0.34   & 0.34  &  0.493 (0.247)  &  & 12.00 &  0.34  & 0.34 & 0.284 (0.227)  \\
Lasso  &  & 12.00    & 13.51  & 13.51  & 0.745 (0.330)  &  &  12.00  & 12.14  &  12.14& 0.738 (0.395) \\
ALasso &  & 12.00   & 1.21   & 0.80  & 0.538 (0.288)  &  & 12.00 & 1.30  & 1.30 & 0.454 (0.328)  \\
Oracle &  & 12.00      & 0.00      & 0.00    & 0.402 (0.242)  &  & 12.00    & 0.00      & 0.00     & 0.288 (0.299)  \\ \hline
\end{tabular}%
}
\end{table}
In addition to assessing the performance of BAR using the abovementioned measures, we also present the results for two popular $L_1$ based penalty functions, LASSO and Adaptive LASSO (ALASSO), for comparison. In addition, another row is presented in the tables, along with the penalty functions. Oracle refers to the ideal case where the true model is assumed to be known, and therefore, the estimation is performed with data that contain the non-zero covariates only. Hence, no variable selection is performed.
The results shown in \autoref{summary-BPBH} present a summary of the variable selection and estimation performance of BAR, LASSO, and ALASSO in the diverging number of covariates under the semiparametric model, where the Bernstein polynomials approximate the baseline hazard functions. Similar results under the parametric model with Weibull distribution are represented in \autoref{summary-WBH} in \autoref{sec:additionalresults}. The baseline hazard functions are generated from Weibull distribution; hence, using the parametric approach is expected to have superior performance compared to the semiparametric method. Comparing \autoref{summary-BPBH} and \autoref{summary-WBH} shows that these two methods have a similar performance. Hence, the proposed semiparametric approach is robust to the model assumption and is preferred in practice when little information is available for the underlying data distribution. 

As expected, when the sample size grows, the performance of BAR improves. This is aligned with the established oracle property of BAR in the literature \citep{zhao2019simultaneous, dai2018broken, zhao2018variable, kawaguchi2017scalable, sun2022broken}. The results of these two tables also suggest that BAR maintains excellent performance based on selecting the correct variables with much lower MCV in the case of a high censoring rate like $70\%$. In terms of MMSE or estimation accuracy, although ALASSO has a competing strength in some cases, BAR is still performing better in general. In addition, as it is expected, BAR is more conservative in false positive rate. It means it performs well in excluding unimportant variables from the model. This is consistent with BAR's tendency to produce a more sparse model.

In order to have a more intense assessment of BAR, in addition to the summary of variable selection results, we report the selection frequencies and estimates in the scenario of diverging number of covariates under the semiparametric framework in \autoref{selecfreqB1}, \autoref{selecfreqB2}, and \autoref{selecfreqB3} for the case of 
$(n=100, p_n=36)$, $(n=300,p_n=45)$, and $(n=500, p_n=48)$, respectively. Similar results under the parametric method are reported in \autoref{selecfreqW1}, \autoref{selecfreqW2}, and \autoref{selecfreqW3} in \autoref{sec:additionalresults}, further confirming the similarity of the performance under the parametric and semiparametric approaches. 
\begin{table}[]
\centering
\caption{Performance of BAR, LASSO, and ALASSO in terms of estimation accuracy and variable selection frequencies among 100 replications when $n=100$ and $p=36$ under the semiparametric approach.   $\boldsymbol{\beta}_{01}$, $\boldsymbol{\beta}_{02}$, and $\boldsymbol{\beta}_{03}$ denote the true parameter values corresponding to the first, second, and the third transitions.}
\label{selecfreqB1}
\resizebox{\textwidth}{!}{%
\begin{tabular}{cccccccccccccccc}
\hline
\multicolumn{16}{c}{$n=100, p=12 \times 3$} \\ \hline
Method &
  \begin{tabular}[c]{@{}c@{}}Type of \\ Assessment\end{tabular} &
  Transition &
  $\boldsymbol{\beta}$ &
  $X_1$ &
  $X_2$ &
  $X_3$ &
  $X_4$ &
  $X_5$ &
  $X_6$ &
  $X_7$ &
  $X_8$ &
  $X_9$ &
  $X_{10}$ &
  $X_{11}$ &
  $X_{12}$ \\ \hline
- &
  True Values &
  $1\rightarrow 2$ &
  $\boldsymbol{\beta}_{01}$ &
  -0.80 &
  1.00 &
  1.00 &
  0.90 &
  0.00 &
  0.00 &
  0.00 &
  0.00 &
  0.00 &
  0.00 &
  0.00 &
  0.00 \\
 &
   &
  $1\rightarrow 3$ &
  $\boldsymbol{\beta}_{02}$ &
  1.00 &
  1.00 &
  1.00 &
  0.90 &
  0.00 &
  0.00 &
  0.00 &
  0.00 &
  0.00 &
  0.00 &
  0.00 &
  0.00 \\
 &
   &
  $2\rightarrow 3$ &
  $\boldsymbol{\beta}_{03}$ &
  -1.00 &
  1.00 &
  1.00 &
  0.90 &
  0.00 &
  0.00 &
  0.00 &
  0.00 &
  0.00 &
  0.00 &
  0.00 &
  0.00 \\
 &
   &
   &
   &
   &
   &
   &
   &
   &
   &
   &
   &
   &
   &
   &
   \\
BAR &
  Estimate &
  $1\rightarrow 2$ &
  $\boldsymbol{\beta}_1$ &
  -0.71 &
  0.83 &
  1.04 &
  0.84 &
  0.03 &
  0.00 &
  0.01 &
  0.01 &
  0.00 &
  0.00 &
  -0.01 &
  0.00 \\
 &
   &
  $1\rightarrow 3$ &
  $\boldsymbol{\beta}_2$ &
  1.02 &
  0.87 &
  1.00 &
  0.73 &
  0.01 &
  0.01 &
  0.01 &
  0.00 &
  0.00 &
  0.00 &
  0.00 &
  0.00 \\
 &
   &
  $2\rightarrow 3$ &
  $\boldsymbol{\beta}_3$ &
  -0.77 &
  0.86 &
  0.93 &
  1.00 &
  0.03 &
  0.02 &
  0.00 &
  0.00 &
  0.00 &
  -0.01 &
  -0.02 &
  0.02 \\
 &
   &
   &
   &
   &
   &
   &
   &
   &
   &
   &
   &
   &
   &
   &
   \\
 & Selection &
  $1\rightarrow 2$ &
  $\boldsymbol{\beta}_1$ &
  0.75 &
  0.89 &
  0.92 &
  0.89 &
  0.08 &
  0.05 &
  0.04 &
  0.04 &
  0.01 &
  0.03 &
  0.03 &
  0.05 \\
 & Frequency
   &
  $1\rightarrow 3$ &
  $\boldsymbol{\beta}_2$ &
  0.99 &
  0.94 &
  0.96 &
  0.88 &
  0.04 &
  0.02 &
  0.05 &
  0.01 &
  0.01 &
  0.01 &
  0.00 &
  0.00 \\
 &
   &
  $2\rightarrow 3$ &
  $\boldsymbol{\beta}_3$ &
  0.76 &
  0.72 &
  0.81 &
  0.88 &
  0.04 &
  0.06 &
  0.06 &
  0.05 &
  0.05 &
  0.06 &
  0.07 &
  0.07 \\
 &
   &
   &
   &
   &
   &
   &
   &
   &
   &
   &
   &
   &
   &
   &
   \\
LASSO &
  Estimate &
  $1\rightarrow 2$ &
  $\boldsymbol{\beta}_1$ &
  -0.18 &
  0.34 &
  0.83 &
  0.56 &
  0.04 &
  0.01 &
  0.01 &
  0.01 &
  0.00 &
  0.00 &
  0.00 &
  0.00 \\
 &
   &
  $1\rightarrow 3$ &
  $\boldsymbol{\beta}_2$ &
  0.89 &
  0.70 &
  0.81 &
  0.52 &
  0.02 &
  0.01 &
  0.01 &
  0.00 &
  0.01 &
  0.00 &
  0.00 &
  0.00 \\
 &
   &
  $2\rightarrow 3$ &
  $\boldsymbol{\beta}_3$ &
  -0.28 &
  0.30 &
  0.57 &
  0.61 &
  0.05 &
  0.01 &
  0.00 &
  0.00 &
  0.01 &
  0.00 &
  0.00 &
  0.00 \\
 &
   &
   &
   &
   &
   &
   &
   &
   &
   &
   &
   &
   &
   &
   &
   \\
 & Selection &
  $1\rightarrow 2$ &
  $\boldsymbol{\beta}_1$ &
  0.64 &
  0.91 &
  1.00 &
  0.97 &
  0.18 &
  0.15 &
  0.21 &
  0.12 &
  0.08 &
  0.08 &
  0.11 &
  0.18 \\
 & Frequency
   &
  $1\rightarrow 3$ &
  $\boldsymbol{\beta}_2$ &
  1.00 &
  0.99 &
  1.00 &
  0.97 &
  0.24 &
  0.08 &
  0.11 &
  0.09 &
  0.16 &
  0.06 &
  0.05 &
  0.05 \\
 &
   &
  $2\rightarrow 3$ &
  $\boldsymbol{\beta}_3$ &
  0.65 &
  0.67 &
  0.85 &
  0.87 &
  0.22 &
  0.16 &
  0.17 &
  0.10 &
  0.15 &
  0.11 &
  0.12 &
  0.14 \\
 &
   &
   &
   &
   &
   &
   &
   &
   &
   &
   &
   &
   &
   &
   &
   \\
ALASSO &
  Estimate &
  $1\rightarrow 2$ &
  $\boldsymbol{\beta}_1$ &
  -0.43 &
  0.65 &
  0.97 &
  0.74 &
  0.03 &
  0.01 &
  0.00 &
  0.01 &
  -0.01 &
  0.01 &
  0.01 &
  0.00 \\
 &
   &
  $1\rightarrow 3$ &
  $\boldsymbol{\beta}_2$ &
  0.95 &
  0.81 &
  0.93 &
  0.64 &
  0.01 &
  0.01 &
  0.01 &
  0.00 &
  0.00 &
  0.00 &
  0.00 &
  0.00 \\
 &
   &
  $2\rightarrow 3$ &
  $\boldsymbol{\beta}_3$ &
  -0.63 &
  0.77 &
  0.78 &
  0.95 &
  0.02 &
  0.03 &
  0.01 &
  0.00 &
  -0.01 &
  -0.01 &
  0.02 &
  0.00 \\
 &
   &
   &
   &
   &
   &
   &
   &
   &
   &
   &
   &
   &
   &
   &
   \\
 & Selection &
  $1\rightarrow 2$ &
  $\boldsymbol{\beta}_1$ &
  0.84 &
  0.95 &
  0.97 &
  0.92 &
  0.15 &
  0.10 &
  0.14 &
  0.08 &
  0.20 &
  0.09 &
  0.09 &
  0.10 \\
 & Frequency
   &
  $1\rightarrow 3$ &
  $\boldsymbol{\beta}_2$ &
  0.97 &
  0.99 &
  0.98 &
  0.95 &
  0.06 &
  0.04 &
  0.06 &
  0.04 &
  0.08 &
  0.05 &
  0.03 &
  0.04 \\
 &
   &
  $2\rightarrow 3$ &
  $\boldsymbol{\beta}_3$ &
  0.82 &
  0.89 &
  0.86 &
  0.93 &
  0.10 &
  0.15 &
  0.18 &
  0.13 &
  0.12 &
  0.18 &
  0.11 &
  0.19 \\ \hline
\end{tabular}%
}
\end{table}
Selection frequency refers to the number of times a variable is selected in 100 replications. Reporting both selection frequencies and estimates gives a more precise insight into comparing the methods regarding both variable selection and estimation accuracy. As shown in the tables mentioned earlier, BAR has removed almost all of the variables that are unimportant to the three events of interest, and LASSO has the worst performance from this point of view. While ALASSO seems to report better results for the selection frequency of the non-zero covariates, its selection frequency for the zero covariates is detrimental to the model compared to BAR. This trade-off between TP and FP can be validated using MCV, compromising the overall classification accuracy.
\begin{table}[]
\caption{Performance of BAR, LASSO, and ALASSO in terms of estimation accuracy and variable selection frequencies among 100 replications when $n=300$ and $p=45$ under the semiparametric approach.   $\boldsymbol{\beta}_{01}$, $\boldsymbol{\beta}_{02}$, and $\boldsymbol{\beta}_{03}$ denote the true parameter values corresponding to the first, second, and the third transitions.}
\label{selecfreqB2}
\resizebox{\textwidth}{!}{%
\begin{tabular}{ccccccccccccccccccc}
\hline
\multicolumn{19}{c}{$n=300, p=15 \times 3$} \\ \hline
Method &
  \begin{tabular}[c]{@{}c@{}}Type of \\ Assessment\end{tabular} &
  Transition &
  $\boldsymbol{\beta}$ &
  $X_1$ &
  $X_2$ &
  $X_3$ &
  $X_4$ &
  $X_5$ &
  $X_6$ &
  $X_7$ &
  $X_8$ &
  $X_9$ &
  $X_{10}$ &
  $X_{11}$ &
  $X_{12}$ &
  $X_{13}$ &
  $X_{14}$ &
  $X_{15}$ \\ \hline
- &
  True Values &
  $1\rightarrow 2$ &
  $\boldsymbol{\beta}_{01}$ &
  -0.80 &
  1.00 &
  1.00 &
  0.90 &
  0.00 &
  0.00 &
  0.00 &
  0.00 &
  0.00 &
  0.00 &
  0.00 &
  0.00 &
  0.00 &
  0.00 &
  0.00 \\
 &
   &
  $1\rightarrow 3$ &
  $\boldsymbol{\beta}_{02}$ &
  1.00 &
  1.00 &
  1.00 &
  0.90 &
  0.00 &
  0.00 &
  0.00 &
  0.00 &
  0.00 &
  0.00 &
  0.00 &
  0.00 &
  0.00 &
  0.00 &
  0.00 \\
 &
   &
  $2\rightarrow 3$ &
  $\boldsymbol{\beta}_{03}$ &
  -1.00 &
  1.00 &
  1.00 &
  0.90 &
  0.00 &
  0.00 &
  0.00 &
  0.00 &
  0.00 &
  0.00 &
  0.00 &
  0.00 &
  0.00 &
  0.00 &
  0.00 \\
 &
   &
   &
   &
   &
   &
   &
   &
   &
   &
   &
   &
   &
   &
   &
   &
   &
   &
   \\
BAR &
  Estimates &
  $1\rightarrow 2$ &
  $\boldsymbol{\beta}_1$ &
  -0.71 &
  0.85 &
  0.90 &
  0.78 &
  0.00 &
  0.00 &
  0.00 &
  0.00 &
  0.00 &
  0.00 &
  0.00 &
  0.00 &
  0.00 &
  0.00 &
  0.00 \\
 &
   &
  $1\rightarrow 3$ &
  $\boldsymbol{\beta}_2$ &
  0.87 &
  0.86 &
  0.89 &
  0.76 &
  0.00 &
  0.00 &
  0.00 &
  0.00 &
  0.00 &
  0.00 &
  0.00 &
  0.00 &
  0.00 &
  0.00 &
  0.00 \\
 &
   &
  $2\rightarrow 3$ &
  $\boldsymbol{\beta}_3$ &
  -0.86 &
  0.94 &
  0.88 &
  0.99 &
  0.00 &
  0.00 &
  0.00 &
  0.00 &
  0.00 &
  0.01 &
  0.00 &
  0.00 &
  0.01 &
  0.00 &
  0.00 \\
 &
   &
   &
   &
   &
   &
   &
   &
   &
   &
   &
   &
   &
   &
   &
   &
   &
   &
   \\
            &
 Selection  &
  $1\rightarrow 2$ &
  $\boldsymbol{\beta}_1$ &
  0.99 &
  0.99 &
  0.99 &
  0.99 &
  0.00 &
  0.00 &
  0.01 &
  0.01 &
  0.01 &
  0.00 &
  0.00 &
  0.01 &
  0.00 &
  0.01 &
  0.02 \\
 &
  Frequency &
  $1\rightarrow 3$ &
  $\boldsymbol{\beta}_2$ &
  0.99 &
  0.99 &
  0.99 &
  0.99 &
  0.01 &
  0.01 &
  0.00 &
  0.01 &
  0.00 &
  0.00 &
  0.00 &
  0.00 &
  0.00 &
  0.01 &
  0.00 \\
 &
   &
  $2\rightarrow 3$ &
  $\boldsymbol{\beta}_3$ &
  0.99 &
  0.99 &
  0.99 &
  0.99 &
  0.01 &
  0.05 &
  0.02 &
  0.02 &
  0.01 &
  0.02 &
  0.01 &
  0.05 &
  0.03 &
  0.01 &
  0.02 \\
 &
   &
   &
   &
   &
   &
   &
   &
   &
   &
   &
   &
   &
   &
   &
   &
   &
   &
   \\
LASSO &
  Estimates &
  $1\rightarrow 2$ &
  $\boldsymbol{\beta}_1$ &
  -0.56 &
  0.72 &
  0.86 &
  0.72 &
  0.01 &
  0.01 &
  0.00 &
  0.00 &
  0.00 &
  0.00 &
  0.00 &
  0.00 &
  0.00 &
  0.01 &
  0.00 \\
 &
   &
  $1\rightarrow 3$ &
  $\boldsymbol{\beta}_2$ &
  0.85 &
  0.81 &
  0.84 &
  0.70 &
  0.02 &
  0.00 &
  0.00 &
  0.00 &
  0.00 &
  0.00 &
  0.00 &
  0.00 &
  0.00 &
  0.02 &
  0.00 \\
 &
   &
  $2\rightarrow 3$ &
  $\boldsymbol{\beta}_3$ &
  -0.69 &
  0.78 &
  0.85 &
  0.91 &
  0.03 &
  0.01 &
  0.00 &
  0.00 &
  0.00 &
  0.01 &
  0.00 &
  0.00 &
  0.01 &
  0.01 &
  0.00 \\
 &
   &
   &
   &
   &
   &
   &
   &
   &
   &
   &
   &
   &
   &
   &
   &
   &
   &
   \\
 & Selection &
  $1\rightarrow 2$ &
  $\boldsymbol{\beta}_1$ &
  0.99 &
  0.99 &
  0.99 &
  0.99 &
  0.37 &
  0.25 &
  0.27 &
  0.30 &
  0.35 &
  0.29 &
  0.30 &
  0.34 &
  0.30 &
  0.29 &
  0.32 \\
 &
   Frequency &
  $1\rightarrow 3$ &
  $\boldsymbol{\beta}_2$ &
  0.99 &
  0.99 &
  0.99 &
  0.99 &
  0.31 &
  0.30 &
  0.30 &
  0.35 &
  0.14 &
  0.26 &
  0.27 &
  0.34 &
  0.29 &
  0.31 &
  0.30 \\
 &
   &
  $2\rightarrow 3$ &
  $\boldsymbol{\beta}_3$ &
  0.99 &
  0.99 &
  0.99 &
  0.99 &
  0.42 &
  0.35 &
  0.29 &
  0.32 &
  0.32 &
  0.22 &
  0.32 &
  0.35 &
  0.26 &
  0.34 &
  0.34 \\
 &
   &
   &
   &
   &
   &
   &
   &
   &
   &
   &
   &
   &
   &
   &
   &
   &
   &
   \\
ALASSO &
  Estimates &
  $1\rightarrow 2$ &
  $\boldsymbol{\beta}_1$ &
  -0.62 &
  0.79 &
  0.86 &
  0.75 &
  0.00 &
  0.00 &
  0.00 &
  0.00 &
  0.00 &
  0.00 &
  0.00 &
  0.00 &
  0.00 &
  0.00 &
  0.00 \\
 &
   &
  $1\rightarrow 3$ &
  $\boldsymbol{\beta}_2$ &
  0.86 &
  0.84 &
  0.85 &
  0.80 &
  0.00 &
  0.00 &
  0.00 &
  0.00 &
  0.00 &
  0.00 &
  0.00 &
  0.00 &
  0.00 &
  0.00 &
  0.00 \\
 &
   &
  $2\rightarrow 3$ &
  $\boldsymbol{\beta}_3$ &
  -0.78 &
  0.87 &
  0.87 &
  0.96 &
  0.01 &
  0.00 &
  0.00 &
  0.00 &
  0.00 &
  0.00 &
  0.00 &
  0.00 &
  0.01 &
  0.00 &
  0.00 \\
 &
   &
   &
   &
   &
   &
   &
   &
   &
   &
   &
   &
   &
   &
   &
   &
   &
   &
   \\
 & Selection &
  $1\rightarrow 2$ &
  $\boldsymbol{\beta}_1$ &
  0.99 &
  0.99 &
  0.99 &
  0.99 &
  0.02 &
  0.04 &
  0.06 &
  0.04 &
  0.06 &
  0.07 &
  0.02 &
  0.05 &
  0.02 &
  0.10 &
  0.06 \\
  & Frequency &
  $1\rightarrow 3$ &
  $\boldsymbol{\beta}_2$ &
  0.99 &
  0.99 &
  0.99 &
  0.99 &
  0.01 &
  0.04 &
  0.01 &
  0.02 &
  0.00 &
  0.01 &
  0.01 &
  0.02 &
  0.00 &
  0.05 &
  0.02 \\
  & & 
  $2\rightarrow 3$ &
  $\boldsymbol{\beta}_3$ &
  0.99 &
  0.99 &
  0.99 &
  0.99 &
  0.06 &
  0.12 &
  0.07 &
  0.07 &
  0.05 &
  0.06 &
  0.07 &
  0.01 &
  0.06 &
  0.06 &
  0.07 \\ \hline
\end{tabular}%
}
\end{table}
Analyzing MCV of BAR in comparison to the other methods in \autoref{summary-BPBH} and \autoref{summary-WBH} can be complementary to the selection frequencies for differentiating between different methods. In addition to the selection frequency, it can be observed that BAR performs better in estimation accuracy, which is confirmed by MMSE values.

Following a similar design to \cite{zhao2019simultaneous}, we reveal the strength of BAR in performing variable selection with highly correlated variables. We categorize the covariates into 4 clusters/groups. In this case, the basic setting of the simulation study is to set $p=30$ and $n=300$ and $500$ with a censoring rate around $50\%$. However, the covariates are generated differently to intensify the correlation within groups. For the correlation values, $\rho$ takes $0.8$, $0.9$ or $0.95$, respectively. The structure of covariates in 4 groups for $k=1,2,3$ is designed as below:
\begin{eqnarray*}
\overbrace{(Z_{1k},Z_{2k})}^\text{group 1},\overbrace{(Z_{3k},Z_{4k})}^\text{group 2},\overbrace{(Z_{5k},Z_{6k},Z_{7k})}^\text{group 3},\overbrace{(Z_{8k},Z_{9k}, Z_{10k})}^\text{group 4},
\end{eqnarray*}
where the first two groups consist of non-zero coefficients and the other two groups contain zero coefficients.
\begin{table}[H]
\caption{Performance of BAR, LASSO, and ALASSO in terms of estimation accuracy and variable selection frequencies among 100 replications when $n=500$ and $p=48$ under the semiparametric approach.   $\boldsymbol{\beta}_{01}$, $\boldsymbol{\beta}_{02}$, and $\boldsymbol{\beta}_{03}$ denote the true parameter values corresponding to the first, second, and the third transitions. }
\label{selecfreqB3}
\resizebox{\textwidth}{!}{%
\begin{tabular}{cccccccccccccccccccc}
\hline
\multicolumn{20}{c}{$n=500, p=16 \times 3$} \\ \hline
Method &
  \begin{tabular}[c]{@{}c@{}}Type of \\ Assessment\end{tabular} &
  Transition &
  $\boldsymbol{\beta}$ &
  $X_1$ &
  $X_2$ &
  $X_3$ &
  $X_4$ &
  $X_5$ &
  $X_6$ &
  $X_7$ &
  $X_8$ &
  $X_9$ &
  $X_{10}$ &
  $X_{11}$ &
  $X_{12}$ &
  $X_{13}$ &
  $X_{14}$ &
  $X_{15}$ &
  $X_{16}$ \\ \hline
- &
  True Values &
  $1\rightarrow 2$ &
  $\boldsymbol{\beta}_{01}$ &
  -0.80 &
  1.00 &
  1.00 &
  0.90 &
  0.00 &
  0.00 &
  0.00 &
  0.00 &
  0.00 &
  0.00 &
  0.00 &
  0.00 &
  0.00 &
  0.00 &
  0.00 &
  0.00 \\
 &
   &
  $1\rightarrow 3$ &
  $\boldsymbol{\beta}_{02}$ &
  1.00 &
  1.00 &
  1.00 &
  0.90 &
  0.00 &
  0.00 &
  0.00 &
  0.00 &
  0.00 &
  0.00 &
  0.00 &
  0.00 &
  0.00 &
  0.00 &
  0.00 &
  0.00 \\
 &
   &
  $2\rightarrow 3$ &
  $\boldsymbol{\beta}_{03}$ &
  -1.00 &
  1.00 &
  1.00 &
  0.90 &
  0.00 &
  0.00 &
  0.00 &
  0.00 &
  0.00 &
  0.00 &
  0.00 &
  0.00 &
  0.00 &
  0.00 &
  0.00 &
  0.00 \\
 &
   &
   &
   &
   &
   &
   &
   &
   &
   &
   &
   &
   &
   &
   &
   &
   &
   &
   &
   \\
BAR &
  Estimates &
  $1\rightarrow 2$ &
  $\boldsymbol{\beta}_1$ &
  -0.72 &
  0.86 &
  0.90 &
  0.82 &
  0.00 &
  0.00 &
  0.00 &
  0.00 &
  0.00 &
  0.00 &
  0.00 &
  0.00 &
  0.00 &
  0.00 &
  0.00 &
  0.00 \\
 &
   &
  $1\rightarrow 3$ &
  $\boldsymbol{\beta}_2$ &
  0.88 &
  0.87 &
  0.89 &
  0.81 &
  0.00 &
  0.00 &
  0.00 &
  0.00 &
  0.00 &
  0.00 &
  0.00 &
  0.00 &
  0.00 &
  0.00 &
  0.00 &
  0.00 \\
 &
   &
  $2\rightarrow 3$ &
  $\boldsymbol{\beta}_3$ &
  -0.88 &
  0.94 &
  0.88 &
  0.91 &
  0.00 &
  0.00 &
  0.00 &
  0.00 &
  0.00 &
  0.01 &
  0.00 &
  0.00 &
  0.00 &
  0.00 &
  0.00 &
  0.00 \\
 &
   &
   &
   &
   &
   &
   &
   &
   &
   &
   &
   &
   &
   &
   &
   &
   &
   &
   &
   \\
 & Selection &
  $1\rightarrow 2$ &
  $\boldsymbol{\beta}_1$ &
  1.00 &
  1.00 &
  1.00 &
  1.00 &
  0.01 &
  0.00 &
  0.01 &
  0.00 &
  0.02 &
  0.01 &
  0.00 &
  0.01 &
  0.00 &
  0.01 &
  0.02 &
  0.00 \\
 &
  Frequency &
  $1\rightarrow 3$ &
  $\boldsymbol{\beta}_2$ &
  1.00 &
  1.00 &
  1.00 &
  1.00 &
  0.01 &
  0.01 &
  0.00 &
  0.01 &
  0.00 &
  0.00 &
  0.00 &
  0.00 &
  0.00 &
  0.00 &
  0.01 &
  0.00 \\
 &
   &
  $2\rightarrow 3$ &
  $\boldsymbol{\beta}_3$ &
  1.00 &
  1.00 &
  1.00 &
  1.00 &
  1.00 &
  0.02 &
  0.02 &
  0.01 &
  0.01 &
  0.00 &
  0.02 &
  0.00 &
  0.01 &
  0.01 &
  0.01 &
  0.00 \\
 &
   &
   &
   &
   &
   &
   &
   &
   &
   &
   &
   &
   &
   &
   &
   &
   &
   &
   &
   \\
LASSO &
  Estimates &
  $1\rightarrow 2$ &
  $\boldsymbol{\beta}_1$ &
  -0.63 &
  0.77 &
  0.87 &
  0.77 &
  0.00 &
  0.00 &
  0.00 &
  0.00 &
  0.00 &
  0.01 &
  0.00 &
  0.00 &
  0.00 &
  0.01 &
  0.00 &
  0.00 \\
 &
   &
  $1\rightarrow 3$ &
  $\boldsymbol{\beta}_2$ &
  0.86 &
  0.83 &
  0.84 &
  0.76 &
  0.02 &
  0.00 &
  0.00 &
  0.00 &
  0.00 &
  0.00 &
  0.00 &
  0.00 &
  0.00 &
  0.02 &
  0.00 &
  0.00 \\
 &
   &
  $2\rightarrow 3$ &
  $\boldsymbol{\beta}_3$ &
  -0.77 &
  0.84 &
  0.86 &
  0.96 &
  0.00 &
  0.00 &
  0.00 &
  0.00 &
  0.00 &
  0.01 &
  0.00 &
  0.00 &
  0.01 &
  0.01 &
  0.00 &
  0.00 \\
 &
   &
   &
   &
   &
   &
   &
   &
   &
   &
   &
   &
   &
   &
   &
   &
   &
   &
   &
   \\
 & Selection &
  $1\rightarrow 2$ &
  $\boldsymbol{\beta}_1$ &
  1.00 &
  1.00 &
  1.00 &
  1.00 &
  0.49 &
  0.41 &
  0.49 &
  0.30 &
  0.39 &
  0.40 &
  0.38 &
  0.46 &
  0.48 &
  0.37 &
  0.31 &
  0.44 \\
 &
 Frequency  &
  $1\rightarrow 3$ &
  $\boldsymbol{\beta}_2$ &
  1.00 &
  1.00 &
  1.00 &
  1.00 &
  0.44 &
  0.34 &
  0.41 &
  0.31 &
  0.37 &
  0.39 &
  0.27 &
  0.34 &
  0.29 &
  0.31 &
  0.36 &
  0.47 \\
 &
   &
  $2\rightarrow 3$ &
  $\boldsymbol{\beta}_3$ &
  1.00 &
  1.00 &
  1.00 &
  1.00 &
  0.45 &
  0.39 &
  0.29 &
  0.32 &
  0.32 &
  0.21 &
  0.33 &
  0.33 &
  0.24 &
  0.34 &
  0.34 &
  0.39 \\
 &
   &
   &
   &
   &
   &
   &
   &
   &
   &
   &
   &
   &
   &
   &
   &
   &
   &
   &
   \\
ALASSO &
  Estimates &
  $1\rightarrow 2$ &
  $\boldsymbol{\beta}_1$ &
  -0.67 &
  0.82 &
  0.88 &
  0.79 &
  0.00 &
  0.00 &
  0.00 &
  0.00 &
  0.00 &
  0.00 &
  0.00 &
  0.00 &
  0.00 &
  0.00 &
  0.00 &
  0.00 \\
 &
   &
  $1\rightarrow 3$ &
  $\boldsymbol{\beta}_2$ &
  0.88 &
  0.85 &
  0.86 &
  0.78 &
  0.00 &
  0.00 &
  0.00 &
  0.00 &
  0.00 &
  0.00 &
  0.00 &
  0.00 &
  0.00 &
  0.00 &
  0.00 &
  0.00 \\
 &
   &
  $2\rightarrow 3$ &
  $\boldsymbol{\beta}_3$ &
  -0.82 &
  0.90 &
  0.86 &
  0.95 &
  0.01 &
  0.00 &
  0.00 &
  0.00 &
  0.00 &
  0.00 &
  0.00 &
  0.00 &
  0.01 &
  0.00 &
  0.00 &
  0.00 \\
 &
   &
   &
   &
   &
   &
   &
   &
   &
   &
   &
   &
   &
   &
   &
   &
   &
   &
   &
   \\
 & Selection &
  $1\rightarrow 2$ &
  $\boldsymbol{\beta}_1$ &
  1.00 &
  1.00 &
  1.00 &
  1.00 &
  0.03 &
  0.04 &
  0.03 &
  0.05 &
  0.01 &
  0.05 &
  0.04 &
  0.06 &
  0.06 &
  0.02 &
  0.02 &
  0.02 \\
 &
  Frequency &
  $1\rightarrow 3$ &
  $\boldsymbol{\beta}_2$ &
  1.00 &
  1.00 &
  1.00 &
  1.00 &
  0.05 &
  0.03 &
  0.01 &
  0.03 &
  0.03 &
  0.01 &
  0.01 &
  0.02 &
  0.00 &
  0.02 &
  0.02 &
  0.02 \\
 &
   &
  $2\rightarrow 3$ &
  $\boldsymbol{\beta}_3$ &
  1.00 &
  1.00 &
  1.00 &
  1.00 &
  0.05 &
  0.04 &
  0.04 &
  0.07 &
  0.04 &
  0.04 &
  0.04 &
  0.03 &
  0.03 &
  0.05 &
  0.03 &
  0.04 \\ \hline
\end{tabular}%
}
\end{table}
Covariates in groups 1 and 3 are generated from marginal normal distribution as in scenario one where $\text{cov}(Z_{ik},Z_{jk})=\rho$; $(i,j)\in(1,2)\text{or}(3,4)$ while the covariates in groups 2 and 4 follow Bernoulli distribution with $E(Z_{jk})=0.5$ for $j\in\{3,4,8,9,10\}$ and $k\in\{1,2,3\}$. Afterwards, to assess the competence of the proposed method regarding the grouping effect criterion, a grouping effect score (GES) is defined as below:
\begin{eqnarray*}
    \text{GES}=0.2\times \text{$g_1$}+0.2\times \text{$g_2$}+0.3 \times \text{$g_3$}+ 0.3\times \text{$g_4$},
\end{eqnarray*}
where \text{$g_1$} and \text{$g_2$} represent the percentages of the estimated regression coefficients in the first and second group being nonzero. Similarly, \text{$g_3$} and \text{$g_4$} denote the percentages of the regression coefficients in the third and fourth group estimated as zero corresponding to the true values:
\begin{eqnarray*}
\boldsymbol{\beta}_{01}&=&(0.8,0.8,1,1,0,0,0,0,0,0)^\top,\\\boldsymbol{\beta}_{02}&=&(0.8,0.8,1,1,0,0,0,0,0,0)^\top,\\\boldsymbol{\beta}_{03}&=&(0.8,0.8,1,1,0,0,0,0,0,0)^\top.
\end{eqnarray*}
\begin{table}[htb]
\caption{Results on performance of simultaneous covariate selection and estimation of grouping effect (GES)  for data with highly
correlated groups ($\rho=$0.8, 0.9, and 0.95) among covariates and using Bernstein polynomials to approximate the baseline hazard
functions when almost 70\% of the data is right censored.}
\label{groupW}
\resizebox{\textwidth}{!}{%
\begin{tabular}{ccccccccccccccc}
\hline
\multicolumn{15}{c}{$70\%$ censoring rate} 
\\
\hline
       &  &      &  & \multicolumn{5}{c}{$n=300$}                   &  & \multicolumn{5}{c}{$n=500$}                   \\ \cline{5-9} \cline{11-15} 
method &  & $\rho$ &  & GES   & TP    & FP   & MCV  & MMSE (SD)     &  & GES   & TP    & FP   & MCV  & MMSE (SD)     \\ \hline
BAR    &  & 0.8  &  & 0.902 & 11.59 & 0.13 & 0.54 & 0.387 (0.482) &  & 0.961 & 11.94    & 0.09 & 0.15 & 0.282 (0.263) \\
Lasso  &  &      &  & 0.406 & 12.00 & 6.99 & 6.99 & 0.785 (0.489) &  & 0.400 & 12.00    & 9.07 & 9.07 & 0.538 (0.300) \\
ALasso &  &      &  & 0.480 & 11.98 & 3.01 & 3.03 & 0.470 (0.401) &  & 0.544 & 12.00    & 2.86 & 2.86 & 0.371 (0.264) \\
Oracle &  &      &  & 1.00     & 12.00    & 0.00    & 0.00    & 0.320 (0.440) &  & 1.00     & 12.00    & 0.00    & 0.00    & 0.202 (0.303) \\
       &  &      &  &       &       &      &      &               &  &       &       &      &      &               \\
BAR    &  & 0.9  &  & 0.803 & 10.53  & 0.09 & 1.56 & 0.742 (0.464) &  & 0.880 & 11.47 & 0.14 & 0.67 & 0.408 (0.297) \\
Lasso  &  &      &  & 0.403 & 11.97 & 6.15 & 6.18 & 0.891 (0.509) &  & 0.403 & 12.00 & 7.47 & 7.47 & 0.550 (0.307) \\
ALasso &  &      &  & 0.427 & 11.68 & 3.56 & 3.88 & 0.635 (0.413) &  & 0.484 & 11.92 & 3.42 & 3.50 & 0.395 (0.274) \\
Oracle &  &      &  & 1.00     & 12.00    & 0.00    & 0.00    & 0.361 (0.409) &  & 1.00     & 12.00    & 0.00    & 0.00    & 0.335 (0.307) \\
       &  &      &  &       &       &      &      &               &  &       &       &      &      &               \\
BAR    &  & 0.95 &  & 0.738 & 9.16  & 0.14 & 2.98 & 0.739 (0.473) &  & 0.775 & 9.90  & 0.13 & 2.23 & 0.550 (0.297) \\
Lasso  &  &      &  & 0.380 & 11.84  & 5.59 & 5.75 & 0.896 (0.536) &  & 0.391 & 11.94 & 6.61 & 6.67 & 0.606 (0.333) \\
ALasso &  &      &  & 0.288 & 10.99 & 3.85 & 4.86 & 0.688 (0.437) &  & 0.304 & 11.51 & 3.56 & 4.05 & 0.471 (0.301) \\
Oracle &  &      &  & 1.00     & 0.00     & 0.00    & 0.00    & 0.419 (0.356) &  & 1.00     & 12.00    & 0.00    & 0.00    & 0.323 (0.347) \\ \hline
\end{tabular}%
}
\end{table}

\autoref{groupW} and \autoref{groupBH} give results on the GES for the semiparametric and parametric models, respectively. The results of these two tables vote for BAR as the method that gives better results regarding the grouping effect score and classifying the variables into important and non-important classes. Although with the higher correlation within groups, the variable selection performance deteriorates, BAR generally maintains a superior potential over other methods regarding the ability to identify the group structure in clusters of covariates that are non-zero or zero. In other words, BAR can be considered as a more intelligent penalty function in realizing if the variables are clustered. This is particularly important in health data sets where it is common to have some clustering/grouping effects among variables in medical studies.

\begin{figure}[p]
    \begin{subfigure}{.5\textwidth}
    \centering
     \includegraphics[ width=0.6\linewidth]{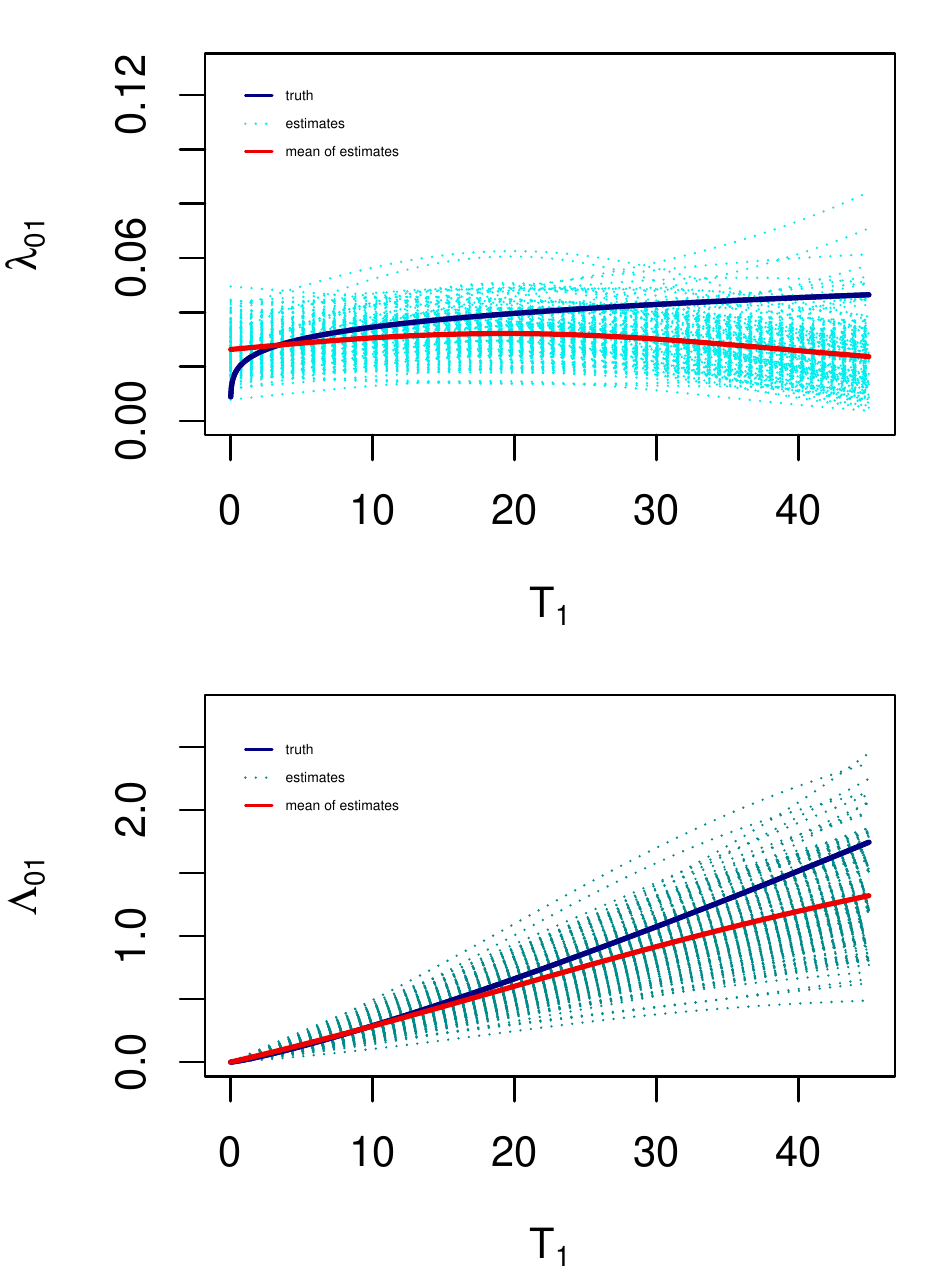} 
    \caption{m=(2,2,3), first transition.}
\end{subfigure}
    \begin{subfigure}{.5\textwidth}
    \centering
      \includegraphics[ width=0.6\linewidth]{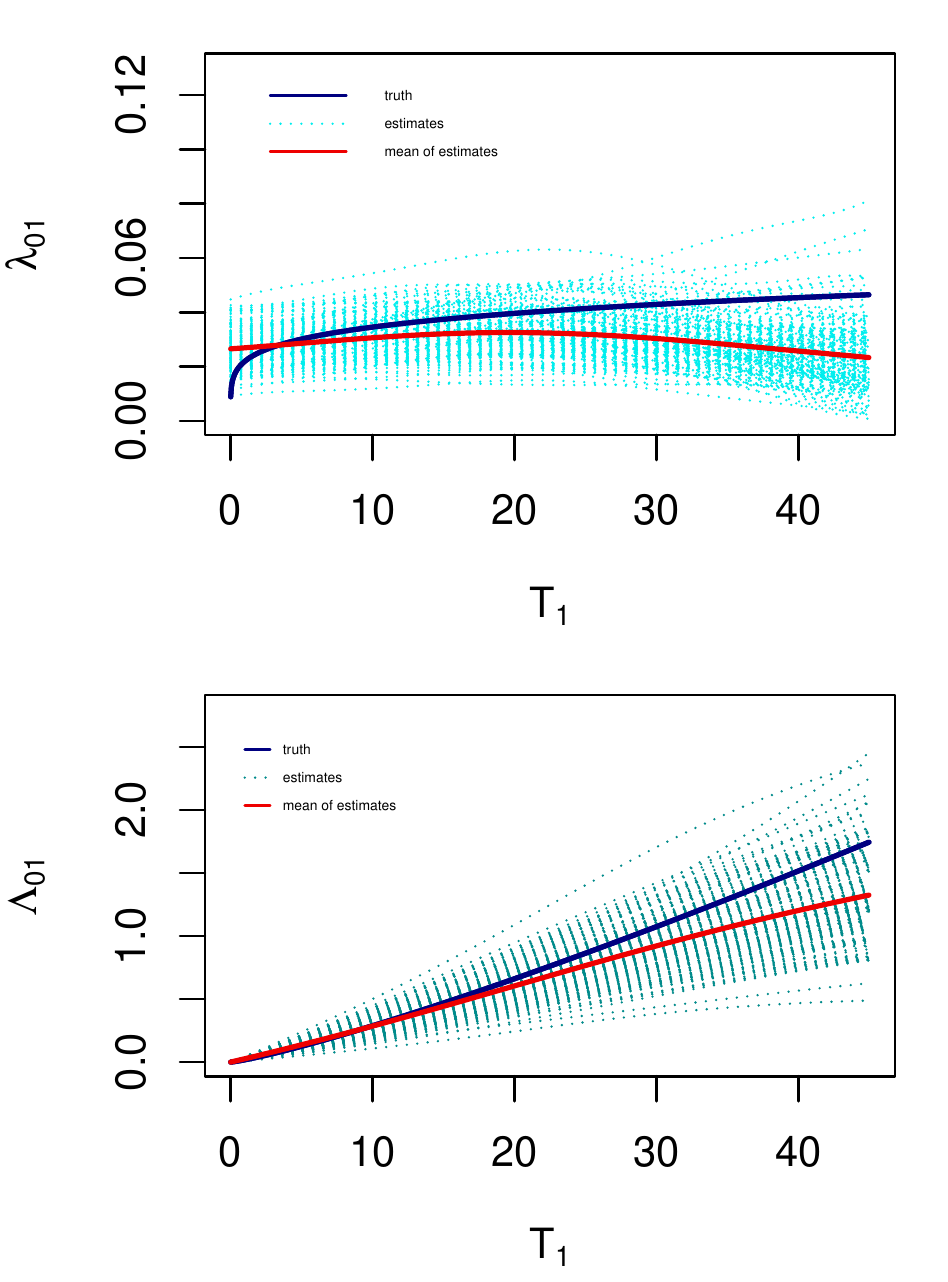} 
    \caption{m=(6,6,6), first transition.}
\end{subfigure}
\newline
\begin{subfigure}{.5\textwidth}
    \centering
      \includegraphics[ width=0.6\linewidth]{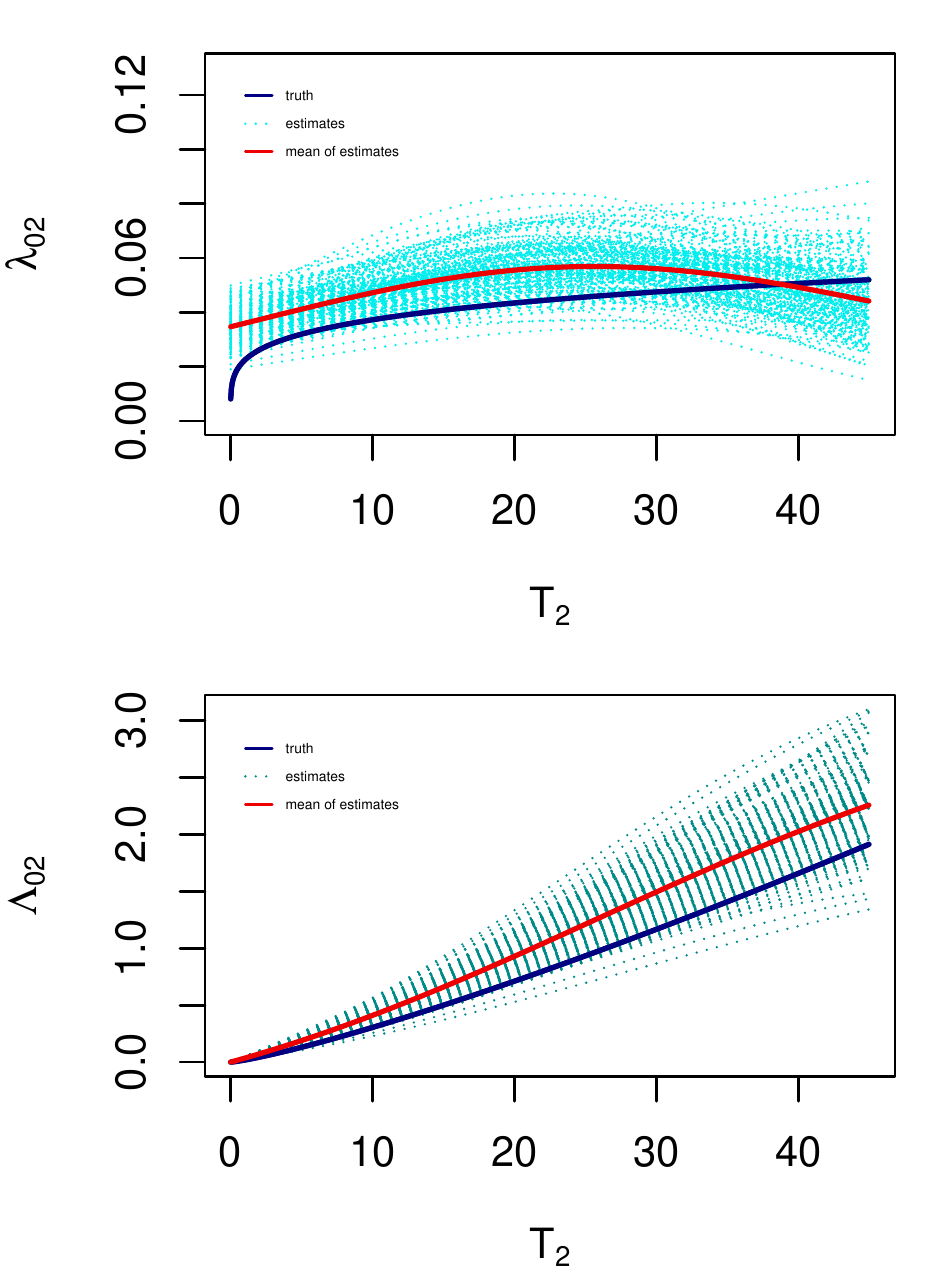}  
    \caption{m=(2,2,3), second transition.}
\end{subfigure}
\begin{subfigure}{.5\textwidth}
    \centering
      \includegraphics[ width=0.6\linewidth]{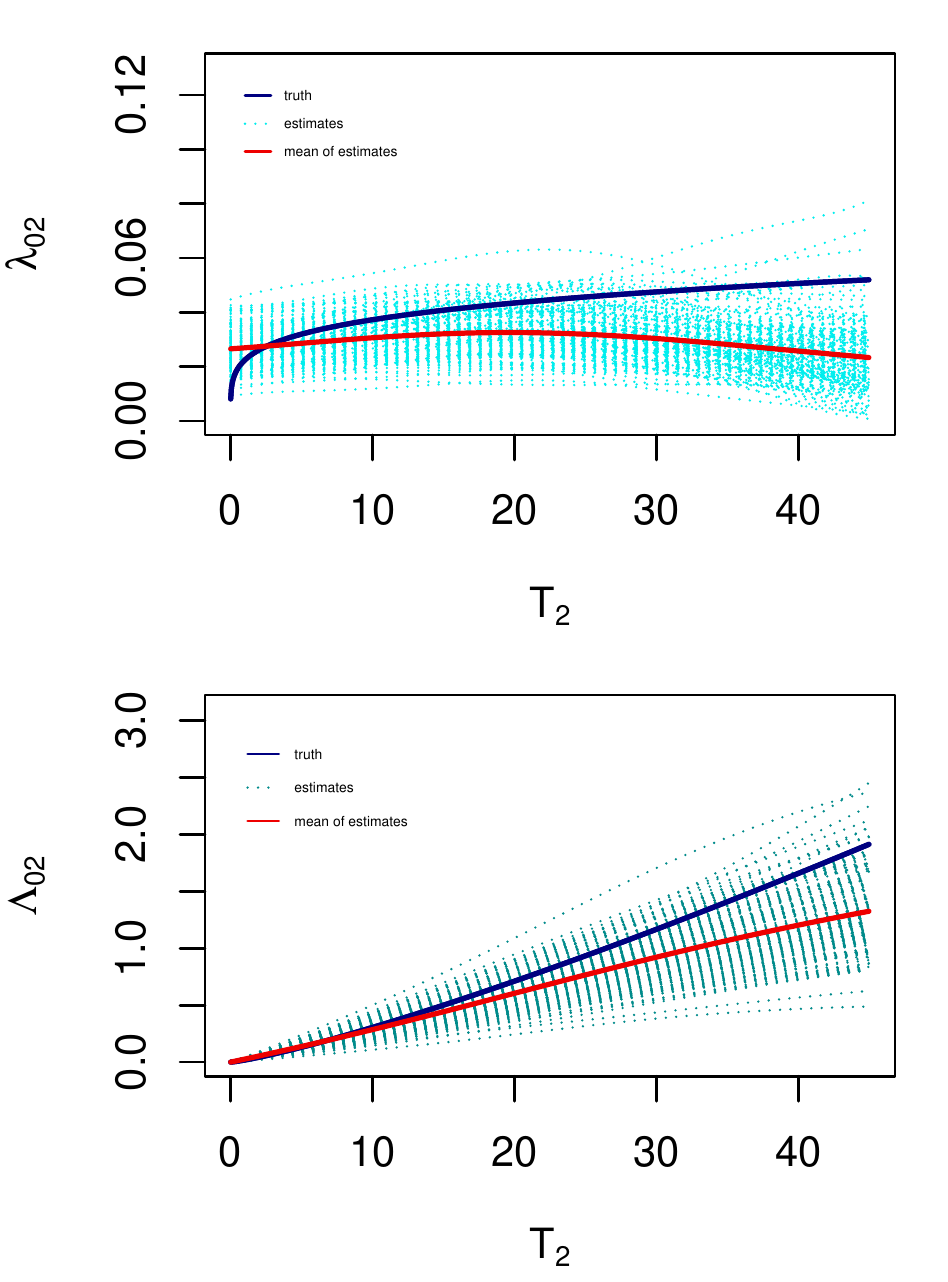}      \caption{m=(6,6,6), second transition.}
\end{subfigure}
\newline
\begin{subfigure}{.5\textwidth}
    \centering
      \includegraphics[ width=0.6\linewidth]{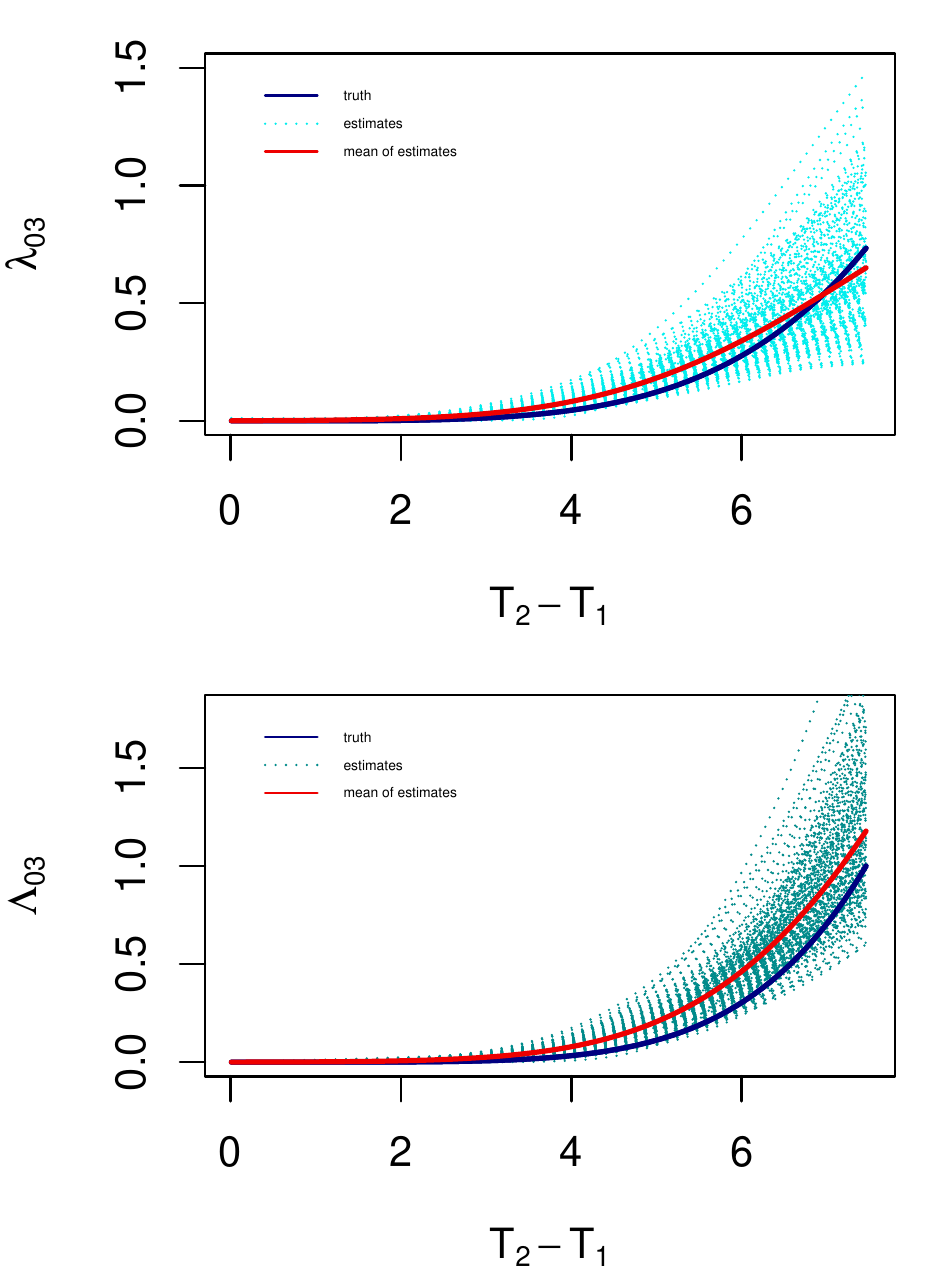}
    \caption{m=(2,2,3), third transition.}
\end{subfigure}
\begin{subfigure}{.5\textwidth}
    \centering
      \includegraphics[ width=0.6\linewidth]{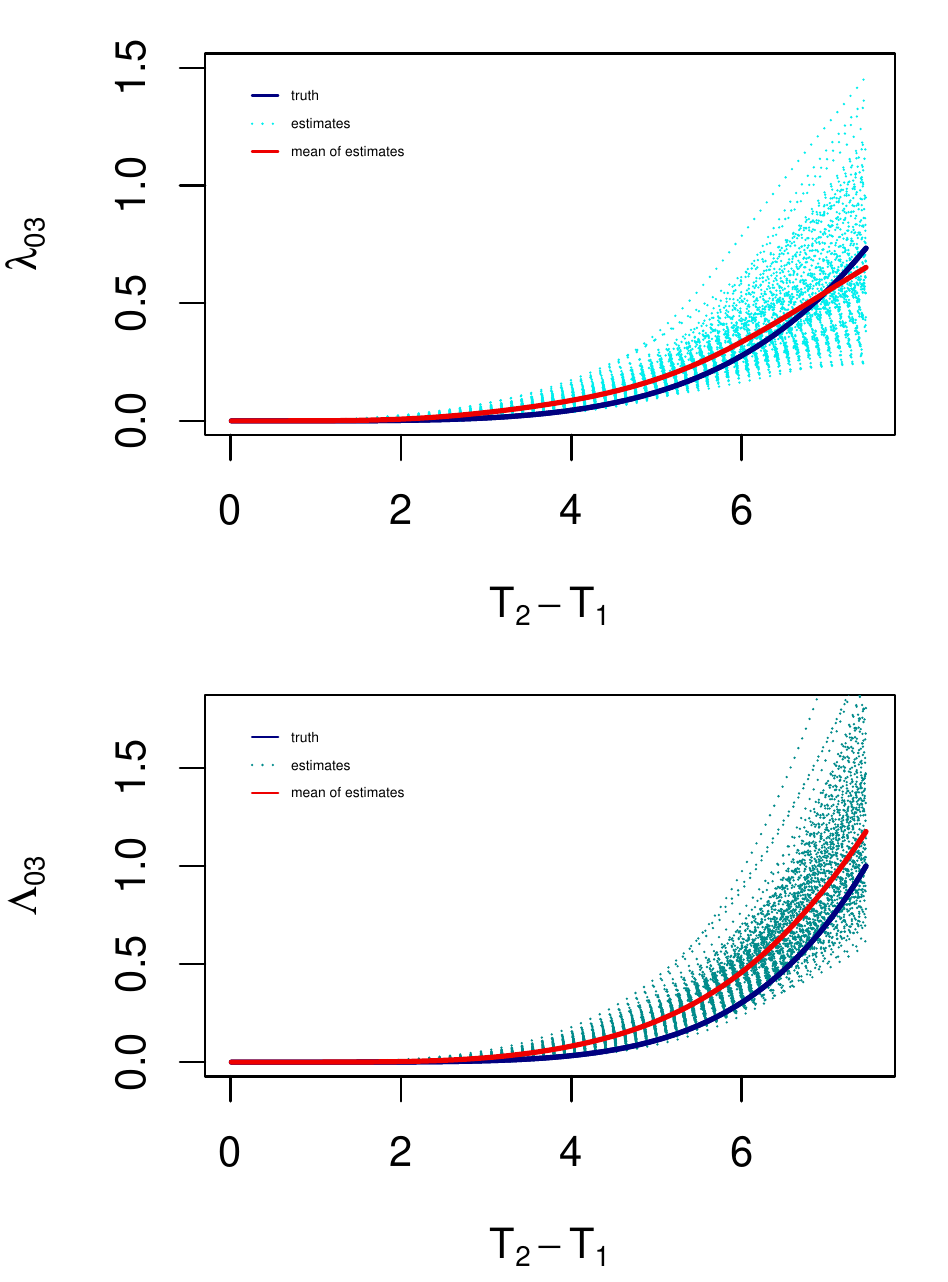}  
    \caption{m=(6,6,6), third transition.}
\end{subfigure}
    \caption{Testing the estimates of baseline hazard functions and cumulative baseline hazard functions under scenario 1 with two different sets of Bernstein polynomial degrees.}
    \label{fig:haz-plot-1}
\end{figure}

\begin{figure}[p]
\label{fig:figureone}
    \begin{subfigure}{.5\textwidth}
    \centering
     \includegraphics[ width=0.6\linewidth]{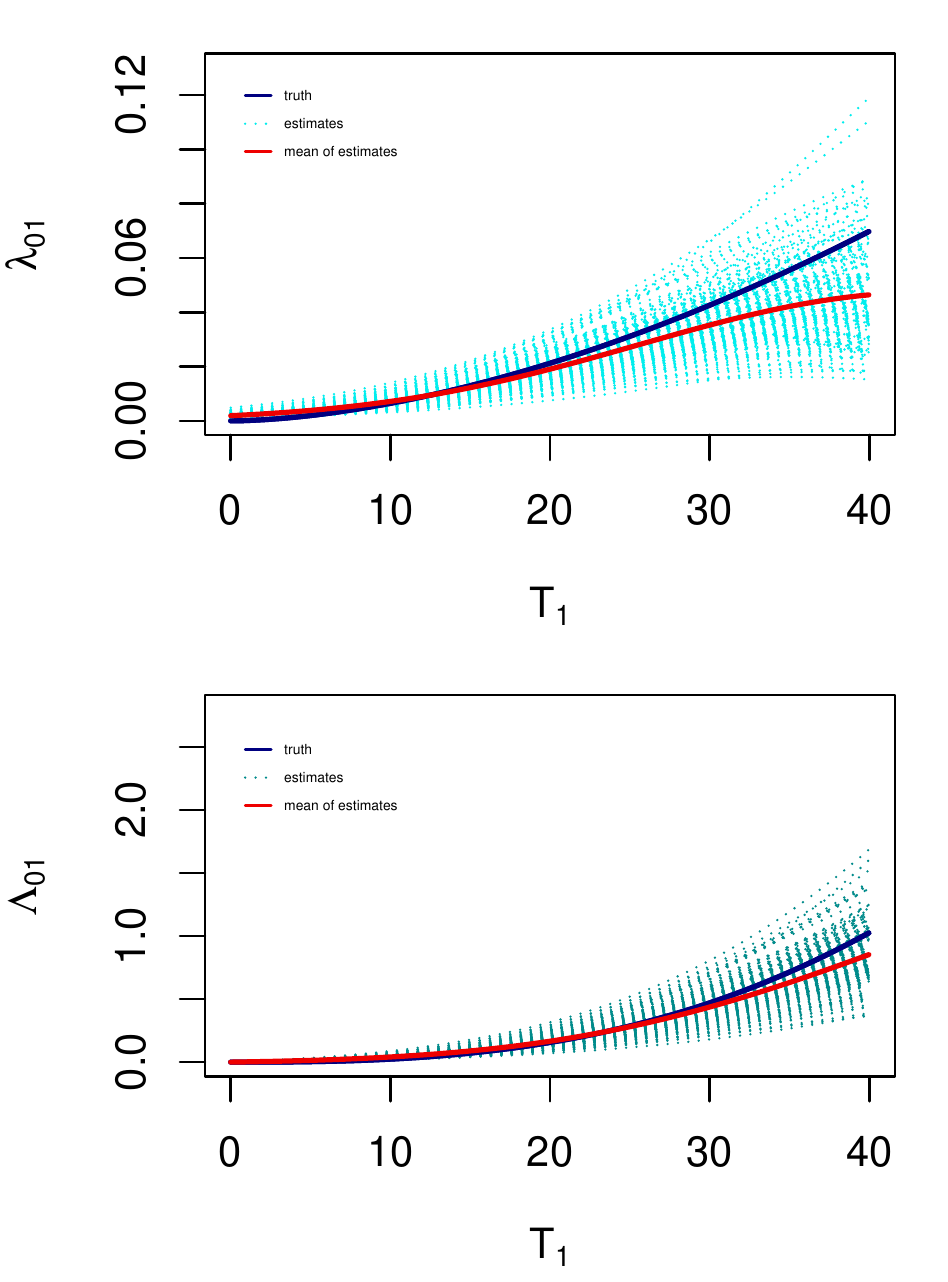} 
    \caption{m=(2,2,3), first transition.}
\end{subfigure}
    \begin{subfigure}{.5\textwidth}
    \centering
      \includegraphics[ width=0.6\linewidth]{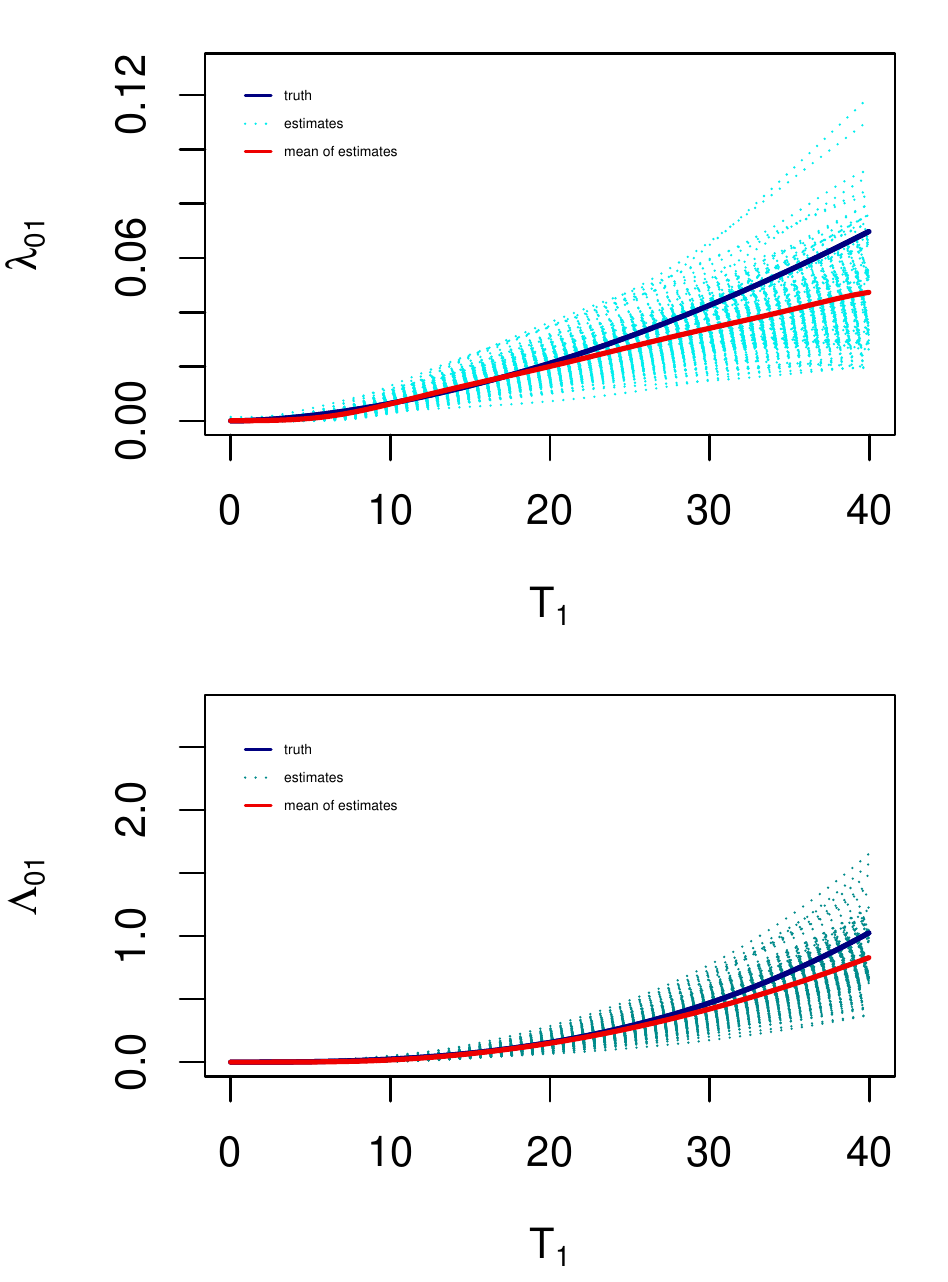} 
    \caption{m=(6,6,6), first transition.}
\end{subfigure}
\newline
\begin{subfigure}{.5\textwidth}
    \centering
      \includegraphics[ width=0.6\linewidth]{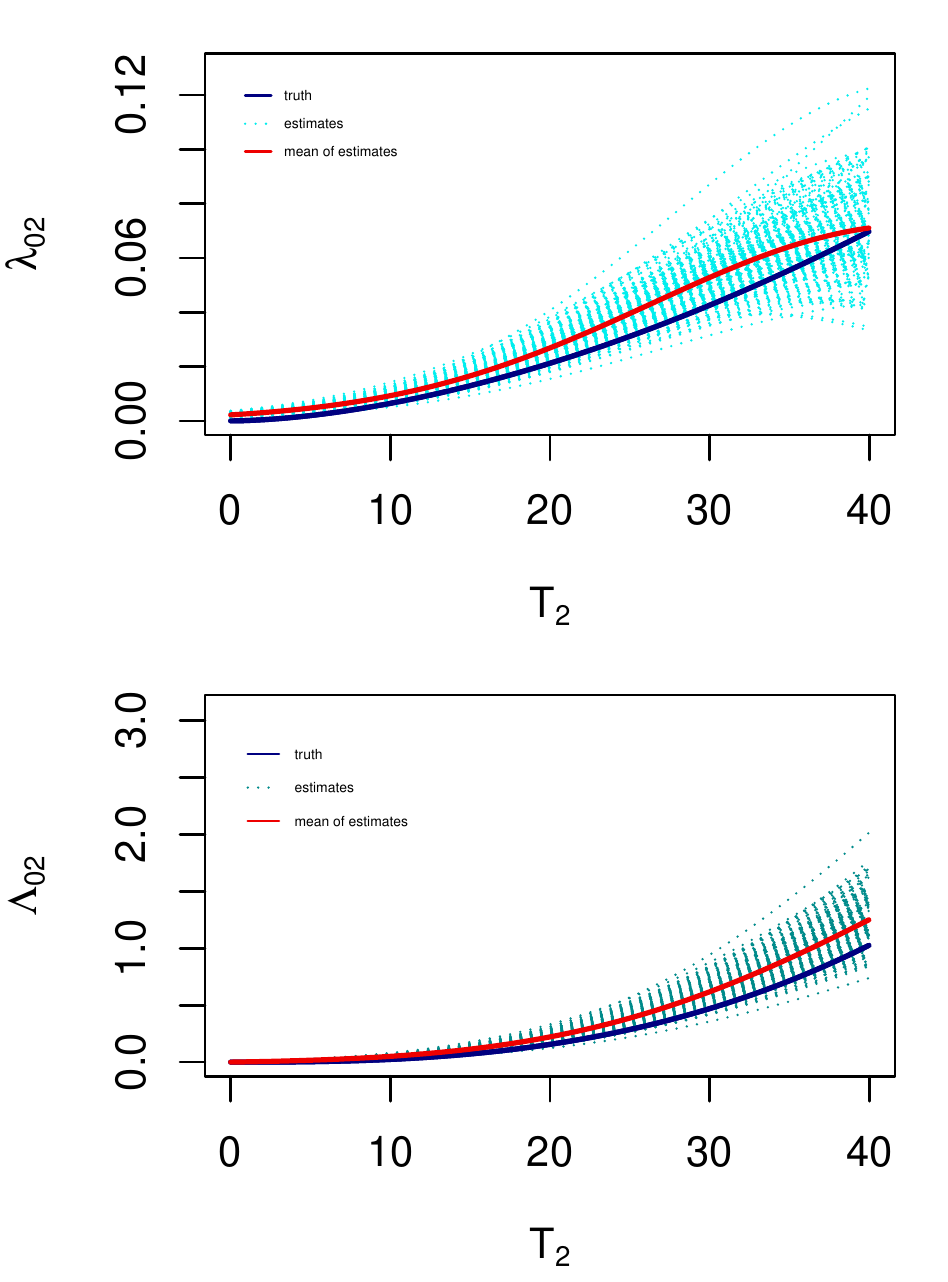}  
    \caption{m=(2,2,3), second transition.}
\end{subfigure}
\begin{subfigure}{.5\textwidth}
    \centering
      \includegraphics[ width=0.6\linewidth]{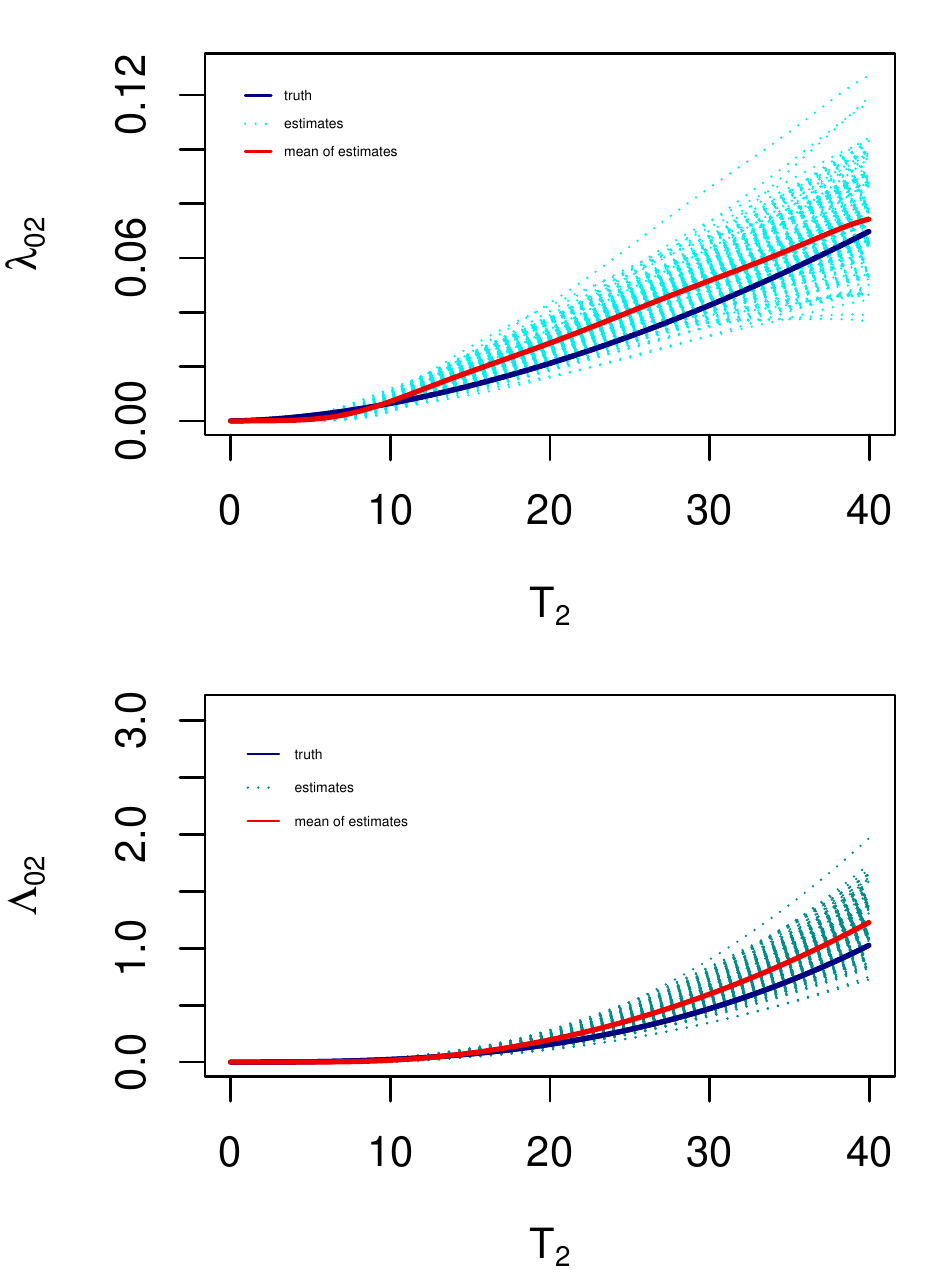}      \caption{m=(6,6,6), second transition.}
\end{subfigure}
    \newline
\begin{subfigure}{.5\textwidth}
    \centering
      \includegraphics[ width=0.6\linewidth]{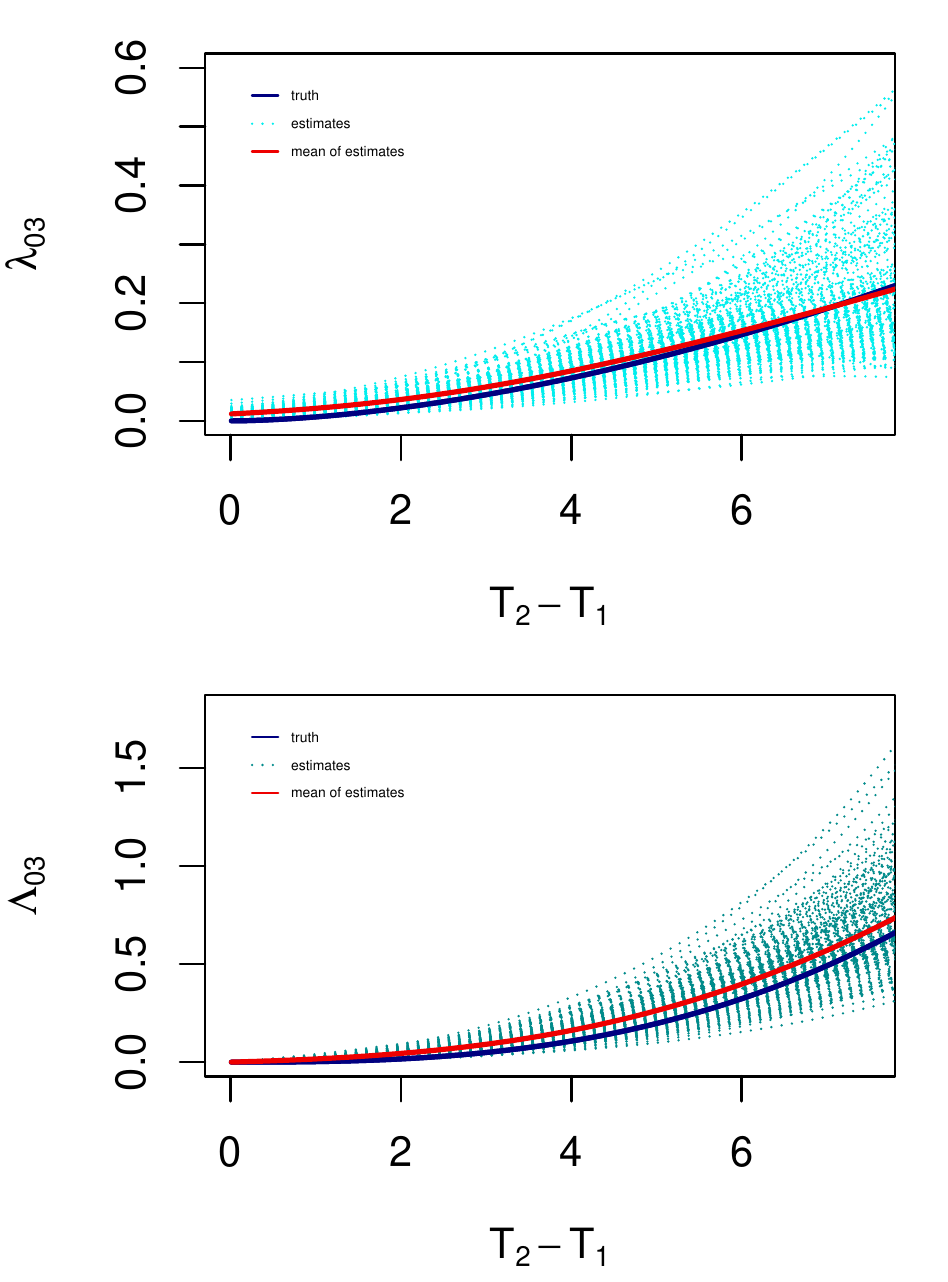}
    \caption{m=(2,2,3), third transition.}
\end{subfigure}
\begin{subfigure}{.5\textwidth}
    \centering
      \includegraphics[ width=0.6\linewidth]{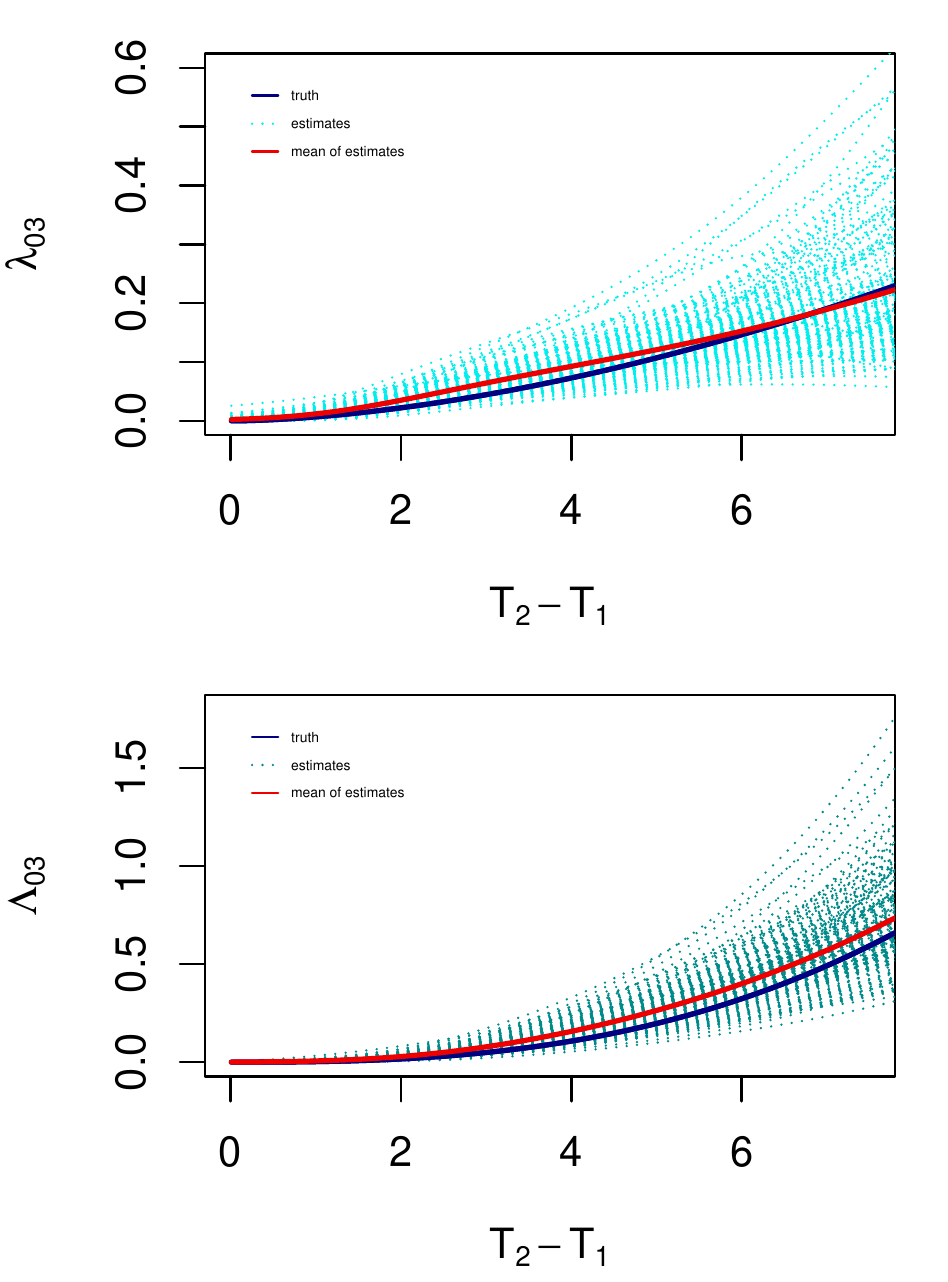}  
    \caption{m=(6,6,6), third transition.}
\end{subfigure}
    \caption{Testing the estimates of baseline hazard functions and cumulative baseline hazard functions under scenario 2  with two different sets of Bernstein polynomial degrees.}
    \label{fig:haz-plot-2}
\end{figure}

Before applying the proposed variable selection procedure to the real-life data, it is noteworthy to investigate the method's robustness to different degrees of Bernstein polynomials. Although in the simulation study, we have set the Bernstein polynomial degree to $m_1=2$, $m_2=2$, and $m_3=3$, we have done some experiments to check the validity of the method under different sets of Bernstein polynomial degrees. For this purpose, we have considered two different scenarios for generating data from Weibull distribution with around $70\%$ right-censored data. Scenario one consists of the parameters that are initially used in the simulation study, and scenario two is generated from the Weibull distribution with parameters $\text{log}(\alpha_1)=1.00$, $\text{log}(\tau_1)=-10.00$, $\text{log}(\alpha_2)=0.80$, $\text{log}(\tau_2)=-9.00$, $\text{log}(\alpha_3)=-6.00$, $\text{log}(\tau_3)=1.00$. These two different parameter settings are interesting to investigate from two points of view. First, simply check the method's robustness to a change in degree. Second, and more importantly, to test the method under different shapes of the baseline hazard functions. We have presented the results of these two scenarios in \autoref{fig:haz-plot-1} and \autoref{fig:haz-plot-2}, respectively. It can be observed that two different sets of degrees tend to give similar results when we vary the degrees of the Bernstein polynomial. Also, another interesting observation from the above-mentioned plots is that,  regardless of the drastically changed shapes of the hazard functions in the first and the second transitions (i.e., with a sharp change at the initial stage and then staying almost plateau with a slight slope), our variable selection method can still perform reasonably well. This highlights the fact that the proposed variable selection method behaves robustly despite difficulties in accurately estimating the baseline hazard functions when their functional forms are complex and hard to catch up by a semiparametric approach during the variable selection step. 

\begin{figure}
    \begin{subfigure}{.5\textwidth}
    \centering
     \includegraphics[ width=0.6\linewidth]{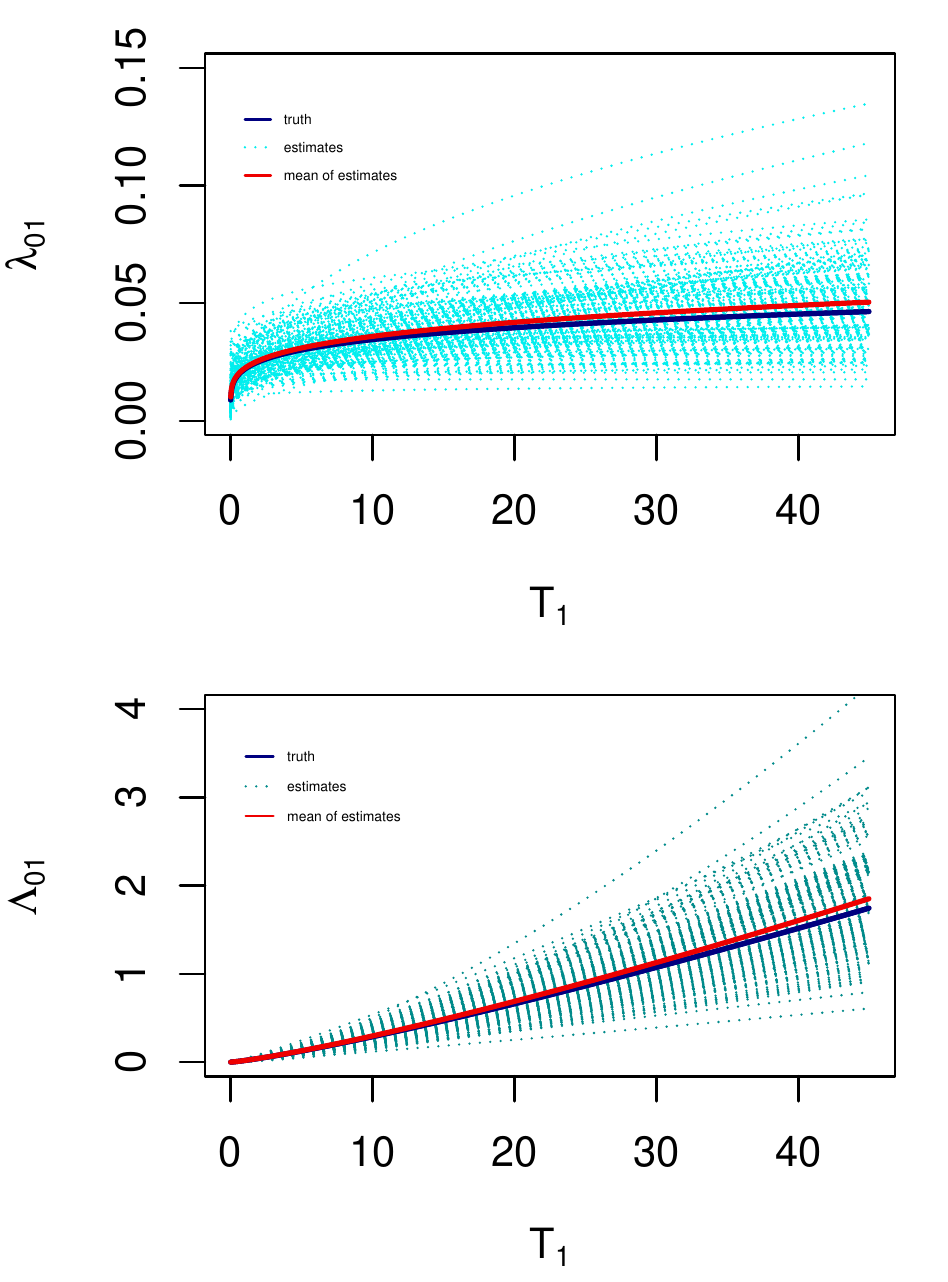} 
    \caption{First transition for the first scenario parameter setting.}
\end{subfigure}
    \begin{subfigure}{.5\textwidth}
    \centering
      \includegraphics[ width=0.6\linewidth]{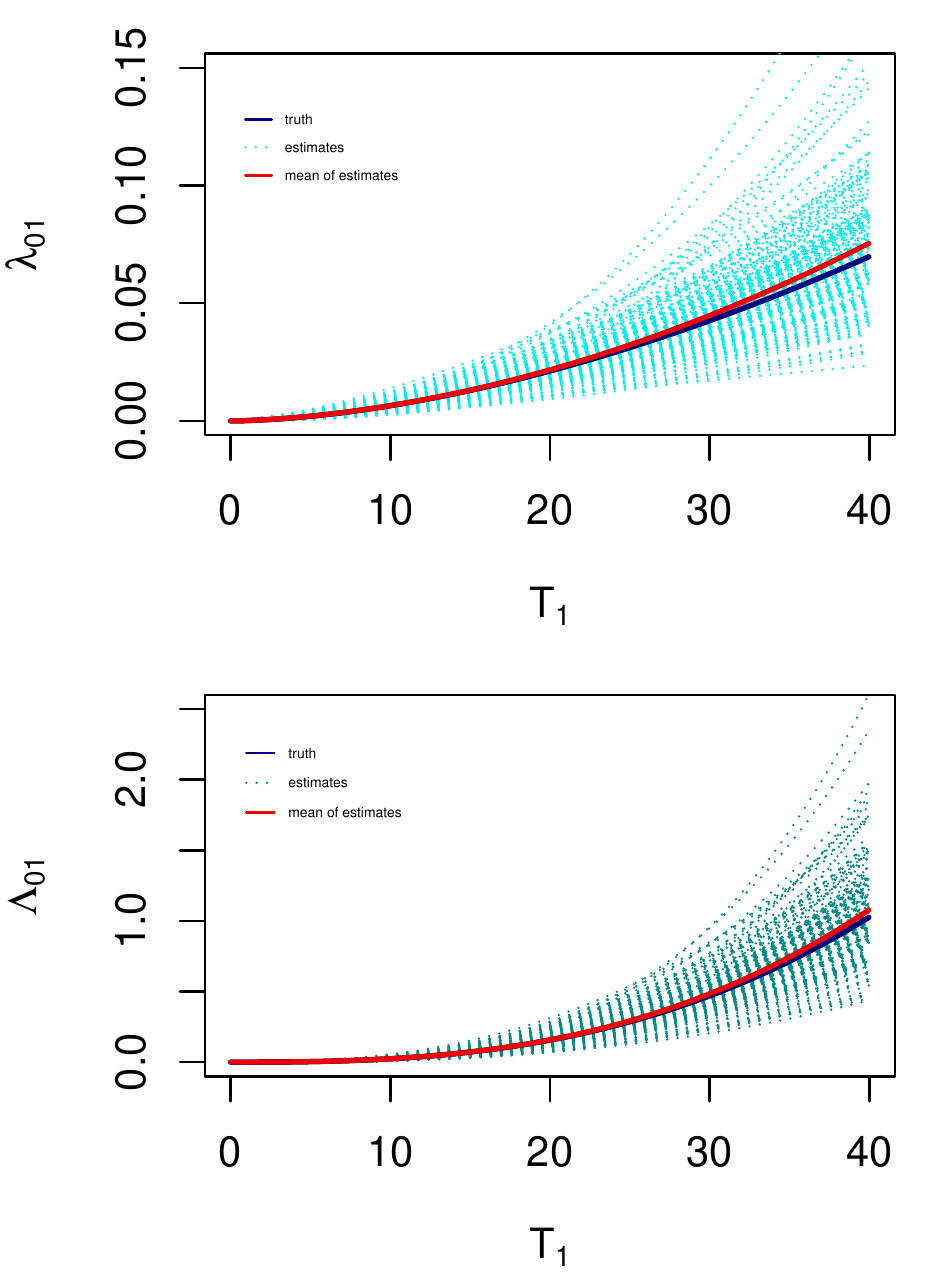} 
    \caption{First transition for the second scenario parameter setting.}
\end{subfigure}
\newline
\begin{subfigure}{.5\textwidth}
    \centering
      \includegraphics[ width=0.6\linewidth]{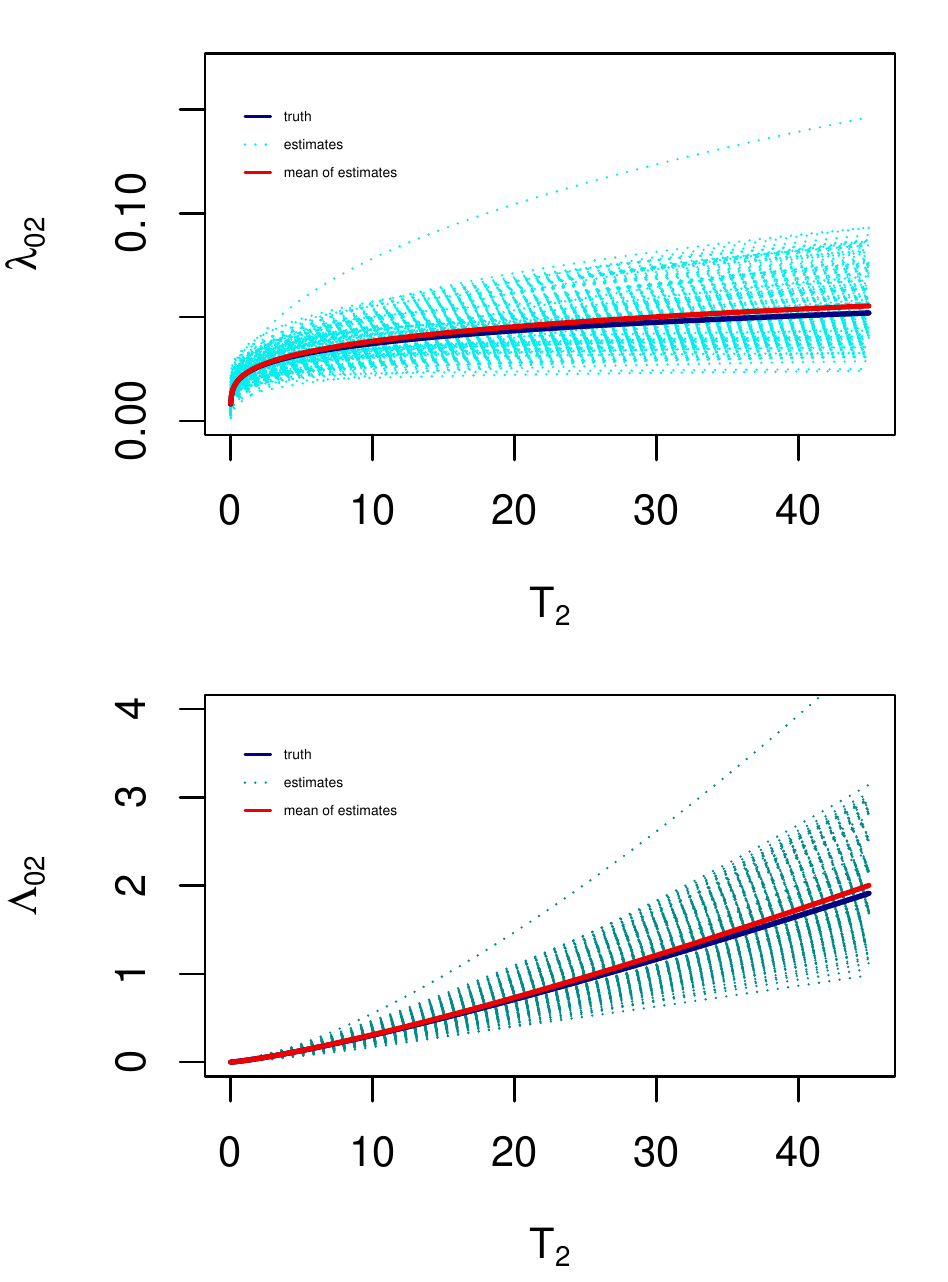}  
    \caption{Second transition for the first scenario parameter setting.}
\end{subfigure}
\begin{subfigure}{.5\textwidth}
    \centering
      \includegraphics[ width=0.6\linewidth]{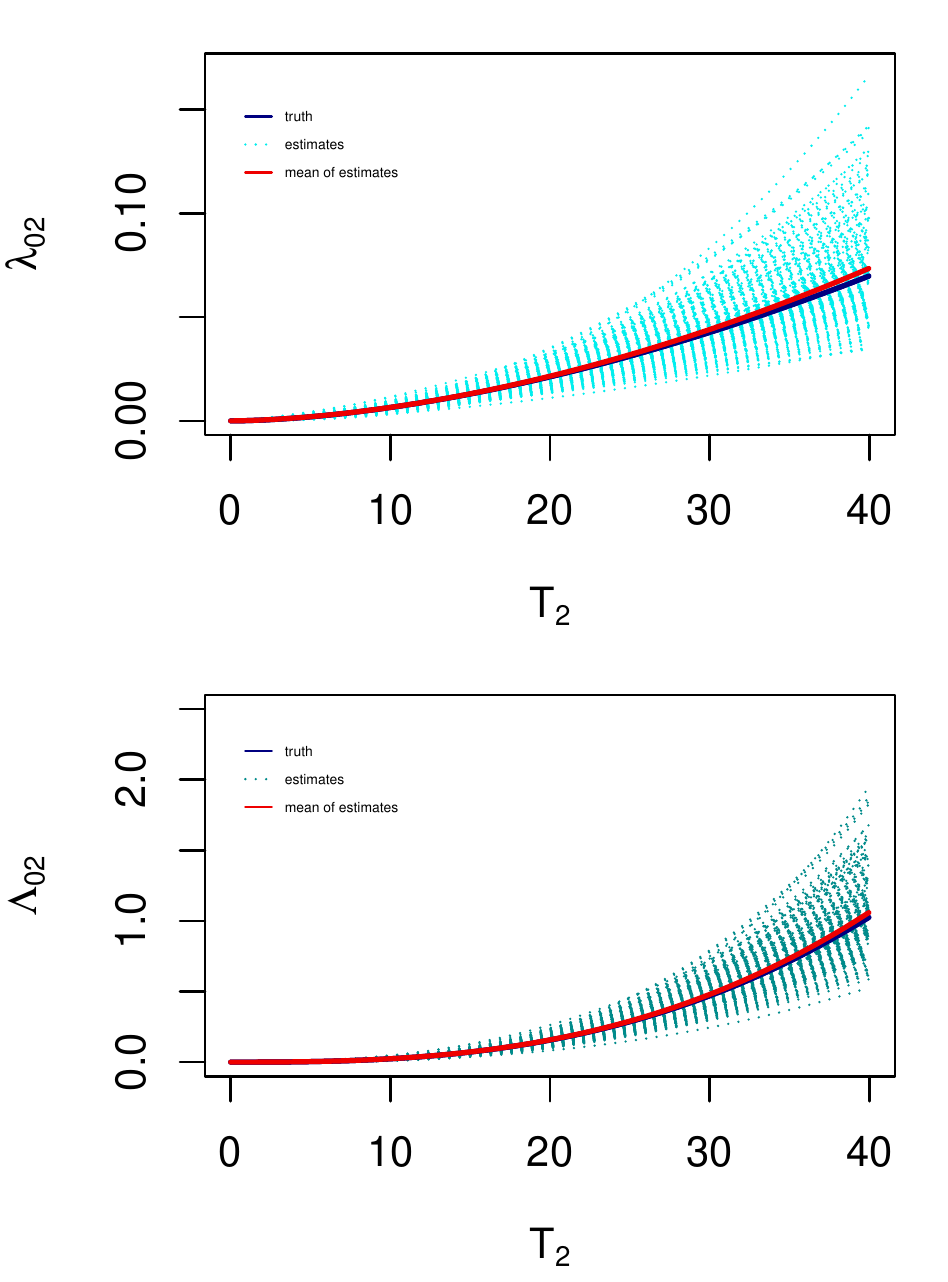}      \caption{Second transition for the second scenario parameter setting.}
\end{subfigure}
    \newline
\begin{subfigure}{.5\textwidth}
    \centering
      \includegraphics[ width=0.6\linewidth]{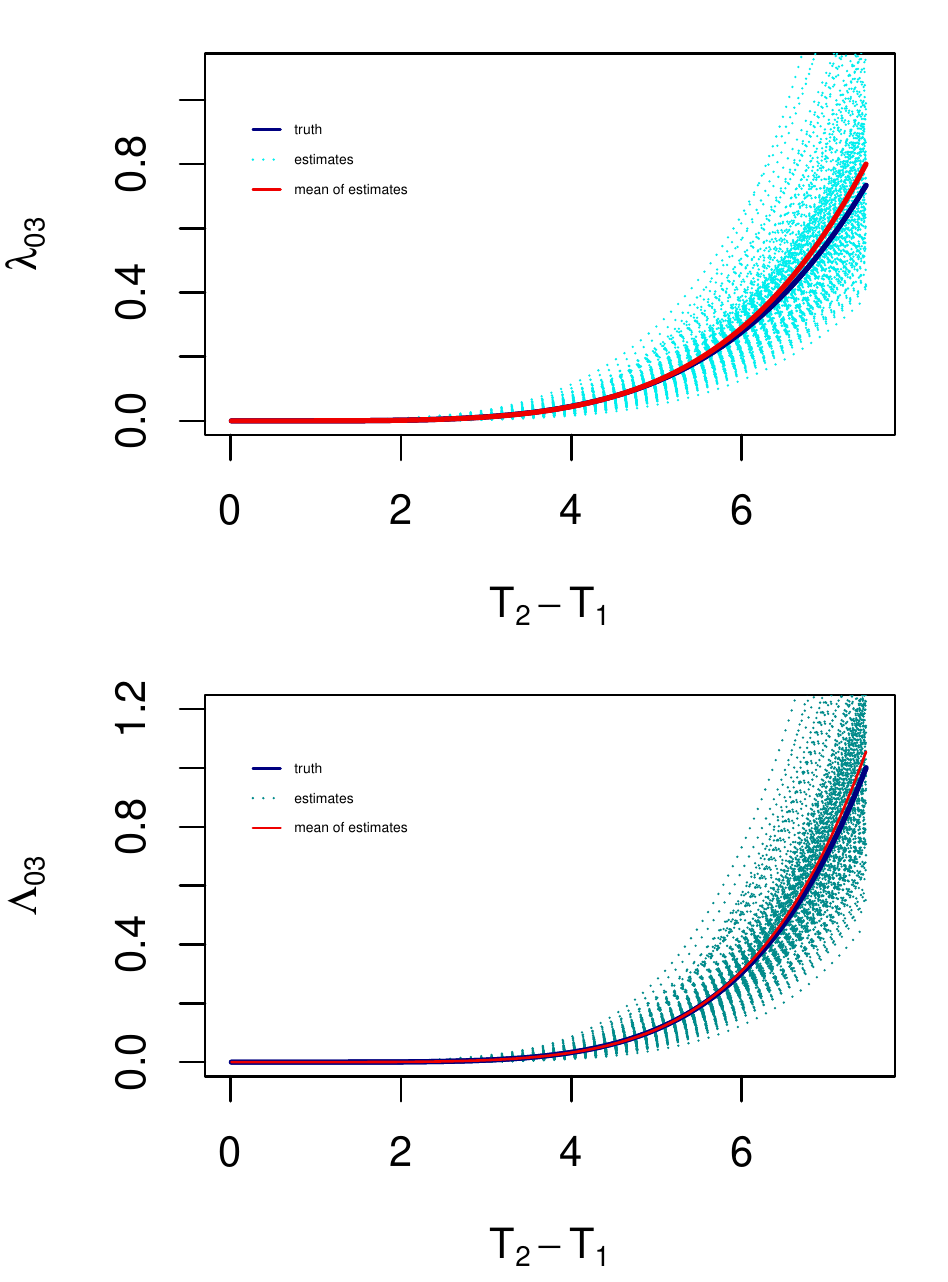}
    \caption{Third transition for the first scenario parameter setting.}
\end{subfigure}
\begin{subfigure}{.5\textwidth}
    \centering
      \includegraphics[ width=0.6\linewidth]{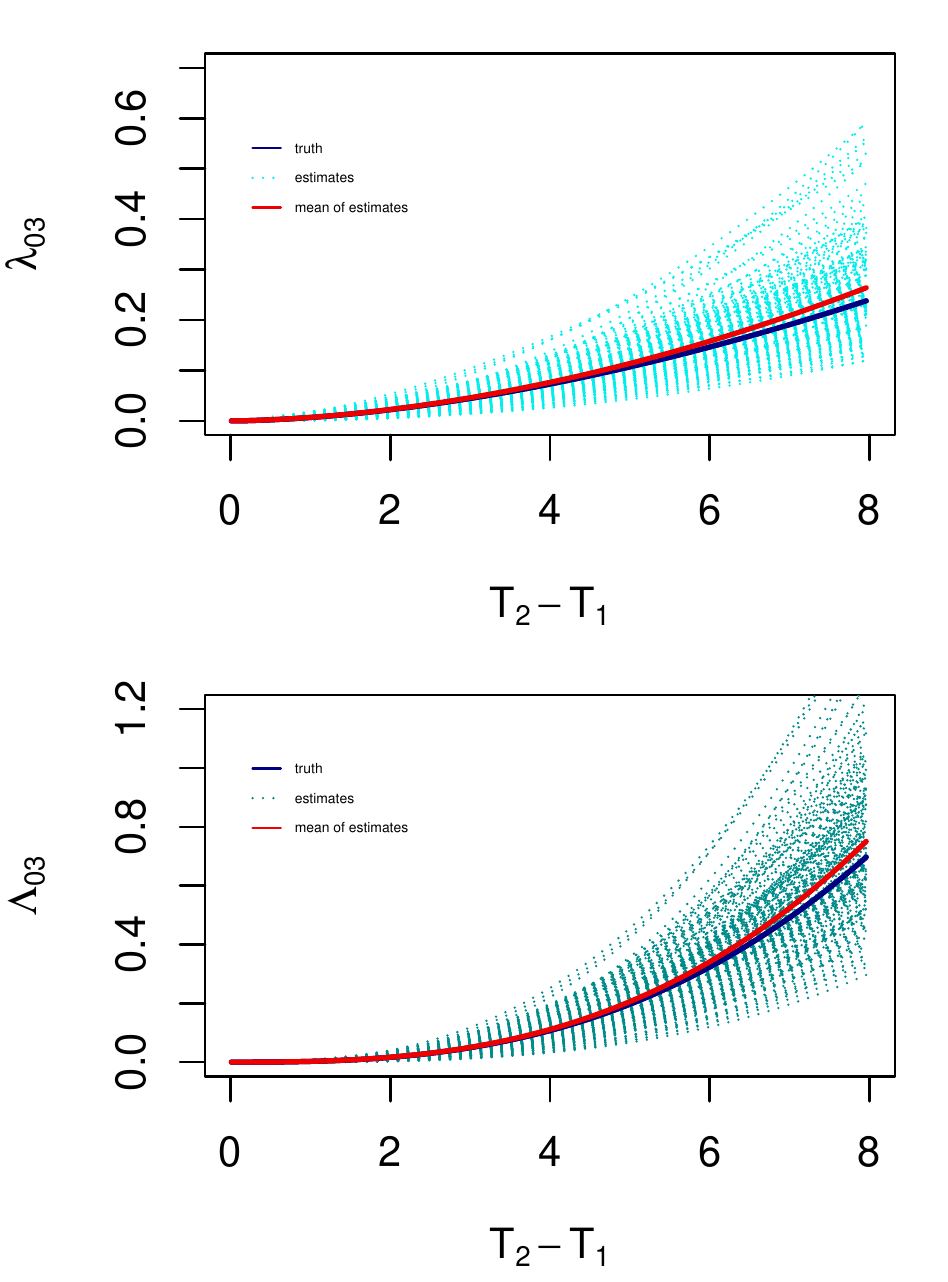}  
    \caption{Third transition for the second scenario parameter setting.}
\end{subfigure}
    \caption{Comparing the true and estimated baseline hazard and cumulative baseline hazard functions under the parametric setting with Weibull distribution.}
    \label{fig:haz-plot-3}
\end{figure}

Based on our experiments, the instability and shape of the hazard functions in different ranges of follow-up times affect the accuracy of the estimates. For the effect of shape, as a piece of evidence, we can inspect the difference in the third transition hazard function's shape and the first two transitions and see how the bias in hazard estimation changes in \autoref{fig:haz-plot-1}. In this simulation study, only around 21\% of the observations fall into the second half of the follow-up time range, while there are around 70\% of right-censored cases in the data. This instability and its extreme shape in the distribution of observations make it challenging for the estimated Bernstein polynomial baseline hazard functions to get closer to the true curves. Parametric estimation under Weibull distribution has also been done under two scenarios, and the results can be found in \autoref{fig:haz-plot-3} in \autoref{sec:additionalresults}. Obviously, they present more accurate results for curve estimation, as the method is parametric. However, as we demonstrated in the examples for variable selection, the proposed method for variable selection is not sensitive to the complexity of the underlying distribution of data and is practically helpful in identifying relevant covariates in modeling semi-competing risks data. 

\section{Real Data Analysis} \label{sec:realdata}
In this section, we illustrate the proposed variable selection method by applying it to a real data set from a colon cancer study. Colorectal cancer (CRC) is the second leading cause of cancer-related death worldwide \citep{dekker2019journal, xie2020comprehensive}. In a research conducted in 1980, 929 patients suffering from stage III colon cancer were randomized to assess the efficacy of a combination of two drugs, levamisole and fluorouracil, as adjuvant therapy after resection of colon carcinoma \citep{moertel1995fluorouracil}. The study involves two events: colon cancer recurrence and death. Although this data set has been analyzed thoroughly in different works under different models, the natural format of the data has not been used to fit a model so far. For instance, under the multivariate failure time model, which does not reflect the natural format of the data, Lin considered the estimation problem \citep{lin1994cox} and \cite{cai2022adaptive} proposed an adaptive bi-level variable selection method. It has also been studied under competing risks setting by \cite{bouvier2015incidence}. However, it is not appropriate to fit these two models mentioned to into this data set. The multivariate failure time model is not a perfect fit because the order of two events, cancer recurrence and death, matters in this study. The competing risks framework is also improper because two risks in the data are not competing events. However, three potentially possible scenarios in this data set make the semi-competing risks setting the perfect natural choice. The three transitions correspond to three paths. From study entry to cancer recurrence or the terminal state (death) after experiencing cancer recurrence and transitioning to death directly without cancer recurrence. The complex data structure in this model presents a challenge in practical data analysis. There are 12 potential risk factors in this study, Lev (treated with only levamisole: yes or no), Lev+FU (treated with a combination of levamisole and fluorouracil: yes or no), sex (male or female), age, obstruct (obstruction of colon by tumor: yes or no), perfor (perforation of colon: yes or no), adhere (adherence of cancer to nearby organs: yes or no), nodes (number of lymph nodes affected by cancer), differ (differentiation of tumor: well, moderate or poor), extent (local extent of tumor: submucosa, muscle, serosa, or contiguous structure), surg (time from surgery to registration: short or long), node4 (more than 4 lymph nodes affected: yes or no). We simultaneously perform 
covariate selection and estimation of covariate effects in each of the three transitions under the semi-competing risks setting.
\begin{table}[H]
\caption{Selected variables and estimated covariate effects for the  colon cancer study.}
\label{realdata}
\resizebox{\textwidth}{!}{%
\begin{tabular}{ccccccccccccccccc}
\hline
\multirow{2}{*}{Variable} &
   &
  \multicolumn{3}{c}{Unpenalized} &
   &
  \multicolumn{3}{c}{Lasso} &
   &
  \multicolumn{3}{c}{ALasso} &
   &
  \multicolumn{3}{c}{BAR} \\ \cline{3-5} \cline{7-9} \cline{11-13} \cline{15-17} 
 &
   &
  CR &
  Death &
  \begin{tabular}[c]{@{}c@{}}Death following \\ CR\end{tabular} &
   &
  CR &
  Death &
  \begin{tabular}[c]{@{}c@{}}Death following \\ CR\end{tabular} &
   &
  CR &
  Death &
  \begin{tabular}[c]{@{}c@{}}Death following\\  CR\end{tabular} &
   &
  CR &
  Death &
  \begin{tabular}[c]{@{}c@{}}Death following\\  CR\end{tabular} \\ \cline{1-1} \cline{3-5} \cline{7-9} \cline{11-13} \cline{15-17} 
         &  &        &        &        &  &        &        &       &  &        &       &       &  &        &       &       \\
Lev      &  & -0.152 & -0.114 & 0.128  &  & -      & -      & 0.055 &  & -      & -     & -     &  & -      & -     & -     \\
Lev+FU   &  & -0.591 & -0.175 & 0.337  &  & -0.416 & -      & 0.241 &  & -0.447 & -     & 0.156 &  & -0.468 & -     & -     \\
Sex      &  & -0.129 & -0.174 & 0.200  &  & -0.109 & -      & 0.154 &  & -      & -     & 0.040 &  & -      & -     & -     \\
Age      &  & -0.004 & 0.004  & 0.012  &  & -0.003 & 0.041  & 0.158 &  & -      & -     & 0.148 &  & -      & 0.026 & 0.015 \\
Obstruct &  & 0.282  & 0.669  & 0.358  &  & 0.140  & -      & 0.142 &  & 0.086  & -     & 0.118 &  & -      & -     & -     \\
Perfor   &  & 0.110  & 0.190  & -0.457 &  & -      & -      & -     &  & -      & -     & -     &  & -      & -     & -     \\
Adhere   &  & 0.340  & 0.503  & 0.266  &  & 0.056  & -      & -     &  & 0.025  & -     & -     &  & -      & -     & -     \\
Nodes    &  & 0.044  & -0.139 & 0.041  &  & 0.052  & 0.071  & 0.042 &  & 0.028  & -     & 0.022 &  & -      & -     & -     \\
Differ   &  & 0.230  & 0.394  & 0.044  &  & 0.168  & -0.263 & 0.070 &  & 0.077  & 0.351 & -     &  & -      & -     & -     \\
Extent   &  & 0.530  & 0.200  & 0.250  &  & 0.540  & 0.210 & 0.242 &  & 0.560  & 0.210 & 0.355 &  & 0.646  & -     & 0.395 \\
Surg     &  & 0.215  & 0.295  & 0.122  &  & 0.169  & -      & -     &  & 0.071  & -     & -     &  & -      & -     & -     \\
Node4    &  & 0.595  & 1.610  & 0.416  &  & 0.499  & -      & 0.345 &  & 0.685  & 0.535 & 0.482 &  & 0.893  & -     & 0.659 \\ \hline
\end{tabular}}
\end{table}
\autoref{realdata} summarizes the results of selected variables and estimated coefficients. It is clear that BAR gives the most sparse result, which is not of a surprise as BAR is built on an $L_0$ penalization approximation using an iteratively reweighted algorithm. Since $L_0$ penalization directly targets the model's cardinality, it should produce the highest sparsity rate. BAR enjoys this feature as an inheritance from $L_0$ penalty, giving a more sparse model compared to other penalty functions \citep{dai2018broken}. It is seen that all the methods identify the combination of levamisole and fluorouracil (Lev+FU) to have a significant effect on reducing the risk of cancer recurrence.
Interestingly, it is also observed that this drug does not have a similarly significant effect on reducing the risk of death without cancer recurrence and/or after experiencing cancer recurrence. This is aligned with the result of the study by \cite{moertel1995fluorouracil}. They mentioned in their paper that levamisole combined with fluorouracil was found to significantly reduce recurrence rates ($p= 0.04$) in patients with surgically treated stage II and stage III colorectal cancer. In contrast, such therapy was not found to be effective in increasing the survival rate (decreasing the odds of death). Our variable selection results are consistent with this finding too.
Based on all three methods, the variable selection results show that patients with more than four positive lymph nodes are at higher risk for returning cancer, followed by death. Additionally, among the three methods, Adaptive Lasso selects this covariate as an important factor in increasing the risk of directly transitioning to death from the initial state. All methods have selected the variable extent. It is evident in \autoref{realdata} that a higher level of colon tumor expansion exposes patients to a higher risk of cancer recurrence or death after it. Furthermore, unlike Lasso and Adaptive Lasso methods, BAR has not identified this variable to affect transitioning to death after study entry. However, it can increase the probability of cancer recurrence and death afterwards.

\begin{figure}[H]
    \centering
    \includegraphics[scale=0.7]{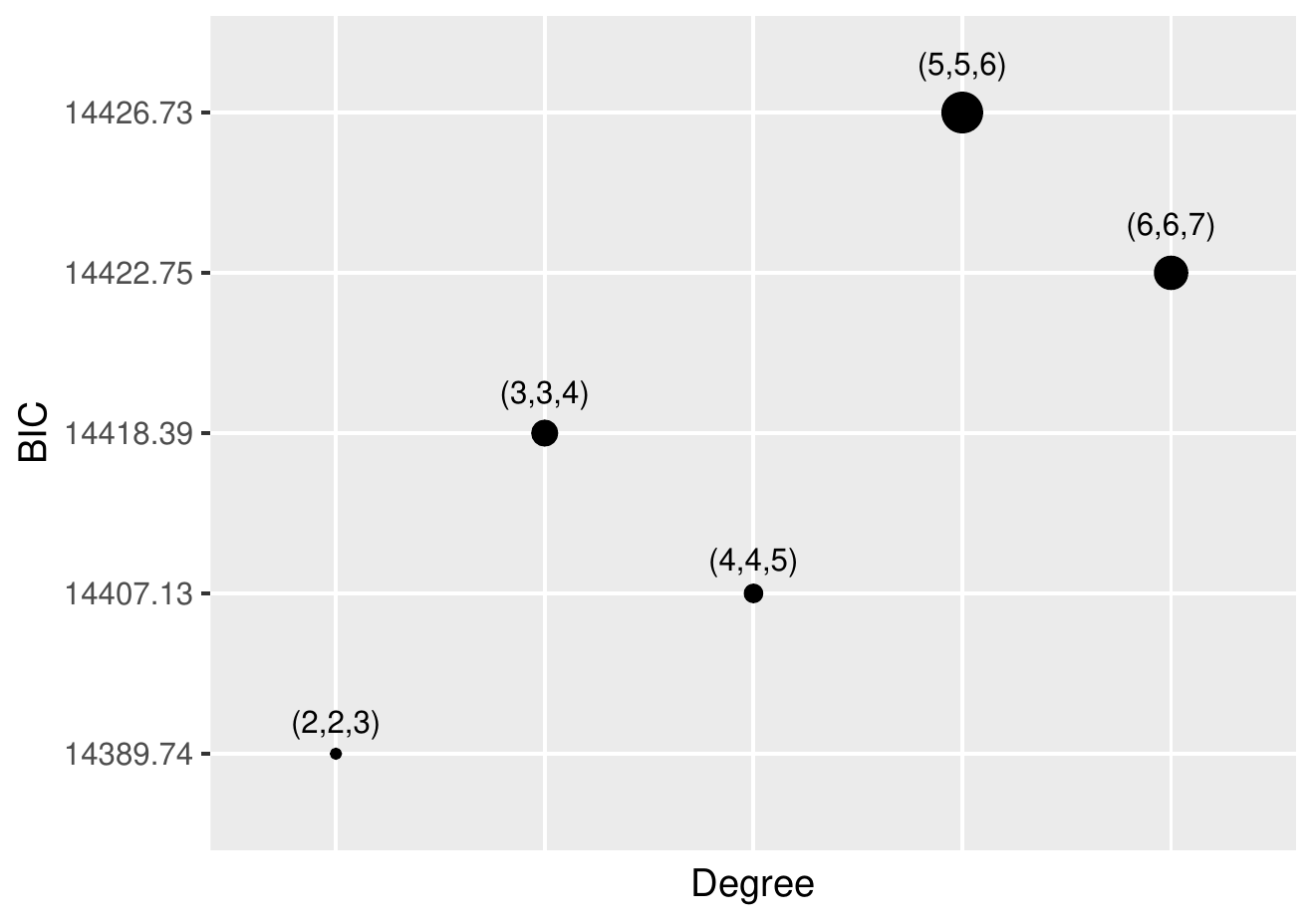}
    \caption{Comparing different sets of degrees for Bernstein polynomials using BIC criterion.}
    \label{fig:BIC}
\end{figure}

Naturally, one needs to set the degree of Bernstein polynomials prior to working with this semiparametric method. In practice, we can use the Bayesian Information Criterion (BIC) to select $\boldsymbol{m}=(m_1,m_2,m_3)$. In this regard, we have tested different combinations of Bernstein polynomial degrees and reported in \autoref{fig:BIC} a sample of the results with the set of degrees that minimizes the Bayesian Information Criterion (BIC) defined as 
\begin{eqnarray*}
\text{BIC}(\boldsymbol{m})=-2\ell_n(\tilde{\boldsymbol{\beta}},\tilde{\boldsymbol{\Phi}}^{BP})+\log(n)\left\{(p_n+1)+\sum_{j=1}^3 (m_j+1)\right\},
\end{eqnarray*}
 where $m_1$, $m_2$, $m_3$, and $n$ are the degrees of Bernstein Polynomials corresponding to the first, second, and third transition and the sample size, respectively. Based on the BIC result on different sets of degrees, we have chosen the set of $(m_1,m_2,m_3)=(5,5,6)$ for the real data analysis. We have also tested some other sets and got the same result. 
\section{Discussion and Concluding Remarks}
\label{sec:discussion}
In this work, we have extended the broken adaptive ridge regression (BAR) to the semiparametric, and parametric illness-death model for the potentially right-censored left truncated data. We employed a shared frailty term to account for the model's dependence between the two events. We have also utilized Bernstein polynomials in our semiparametric approach. Other similar non-parametric methods proposed in the literature can be used for baseline hazards function approximation, such as piecewise constant functions \citep{reeder2022penalized} or B-splines \citep{lee2021fitting}. Although the proposed idea can be implemented using different approaches, the primary motivation for selecting this method is its theoretical and computational advantages over the competing methods. For instance, the differentiability and continuity of Bernstein polynomials approximation are two of its favourable features over piecewise constant functions. This is particularly important when it comes to deriving the log-likelihood function and its first and second derivatives (gradient and Hessian matrix) to facilitate the computation task.
Another advantage of Bernstein polynomials lies in their computational scalability and optimal shape-preserving property among all approximating polynomials
\citep{carnicer1993shape}.
Finally, the fact that they do not require specification of the number of interior knots and their locations makes them superior to other smoothing methods, such as B-splines. In B-splines, one needs to handle the problem of finding these two factors that may control its performance. We have adopted an iteratively reweighted least square algorithm to approximate the non-convex likelihood of this model with a convex function. This is an improvement regarding the computational efficiency of the proposed method compared to the cases where one needs to struggle with optimizing a non-convex function. We have coupled this approximated convex function with a convex penalty function, namely BAR. This recently proposed penalty function possesses many attractive characteristics, such as convexity. The convexity of BAR, along with the likelihood function, implies that one needs to deal with a convex objective function. This is a desired feature in penalized variable selection problems. We have conducted an extensive simulation study to investigate this $L_0$ based penalty function, BAR, compared with two of the most popular $L_1$ based penalty functions. The simulation study indicates the tendency of BAR to produce a lower false positive rate and a more sparse model, which is aligned with the literature findings mentioned above. The other crucial factor that expands the range of BAR favourable features is its closed form. Various complicated algorithms in the literature are proposed to tackle the issue of solving variable selection with the other penalty functions. BAR does not require any of them. BAR penalty also enjoys oracle properties which are investigated in the literature rigorously. It also benefits from the grouping effect property that is shown to beat the other competing penalty functions. We have shown the grouping effect feature of the broken adaptive ridge penalty function under semiparametric and parametric models. This work has the potential to be extended in different directions. For instance, it can be generalized to the case where the shared frailty term is not restricted to following a specific distribution. Another interesting example could be exploring and establishing the proposed method's asymptotic oracle properties and constructing a semiparametric variable selection method for ultra-high dimensional data when $p >> n$.
\bibliographystyle{plainnat}
\bibliography{references}
\clearpage
 \section{Additional Results}
 \label{sec:additionalresults}
In the following, we present some additional simulation study results using a fully parametric method (Weibull distribution)  instead of a semiparametric approach using Bernstein polynomials for approximation. 
 \begin{table}[H]
\caption{Summary of variable selection results in parametric Weibull models with $n=100, 300$, and 500.}
\label{summary-WBH}
\resizebox{\textwidth}{!}{%
\begin{tabular}{ccccccccccc}
\hline
       &  & \multicolumn{4}{c}{$50\%$ censoring rate} &  & \multicolumn{4}{c}{$70\%$ censoring rate} \\ \cline{3-6} \cline{8-11} 
       &  & \multicolumn{9}{c}{$n=100, p=12 \times 3$}                                                          \\ \cline{3-11} 
method &  & TP      & FP     & MCV   & MMSE (SD)      &  & TP      & FP     & MCV   & MMSE (SD)      \\ \cline{1-1} \cline{3-6} \cline{8-11} 
BAR    &  & 11.13   & 0.48   & 1.35  & 1.462 (1.594)  &  & 10.16      & 0.83   & 2.67  & 2.980 (3.626)  \\
Lasso  &  & 11.16   & 3.14   & 3.98  & 5.140 (2.614)  &  &  10.07 & 2.48   & 4.41  & 7.657 (3.460)  \\
ALasso &  & 11.55   & 1.25   & 1.70  & 2.410 (1.876)  &  & 11.08   & 1.82   & 2.73  & 3.105 (3.249)  \\
Oracle &  & 12.00      & 0.00      & 0.00     & 0.743 (0.705)  &  & 12.00      & 0.00      & 0.00     & 1.110 (1.147)  \\ \cline{3-11} 
       &  & \multicolumn{9}{c}{$n=300, p=15 \times 3$}                                                          \\ \cline{3-11} 
BAR    &  & 12.00    & 0.32   & 0.32 & 0.609 (0.454)  &  & 11.98  & 0.43  & 0.45  & 0.465 (0.407)  \\
Lasso  &  & 12.00  & 8.71 & 8.71  & 1.553 (0.607)  &  & 12.00 & 7.12 & 7.14 & 1.427 (0.916)  \\
ALasso &  & 12.00  & 1.20  & 1.20  & 0.991 (0.607)  &  & 12.00   & 1.18   & 1.18  & 0.691 (0.646)  \\
Oracle &  & 12.00      & 0.00      & 0.00     & 0.317 (0.316)  &  & 12.00 & 0.00      & 0.00     & 0.411 (0.295)  \\ \cline{3-11} 
       &  & \multicolumn{9}{c}{$n=500, p=16 \times 3$}                                                          \\ \cline{3-11} 
BAR    &  & 12.00    & 0.23   & 0.23  & 0.595 (0.408)  &  & 12.00 & 0.20   & 0.20 & 0.329 (0.274)  \\
Lasso  &  & 12.00    & 12.55  & 12.55  & 1.147 (0.540)  &  & 12.00   & 10.33  & 10.33  & 0.873 (0.464)  \\
ALasso &  & 12.00   & 0.79   & 0.80  & 0.817 (0.480)  &  & 12.00  & 0.87   & 0.87  & 0.611 (0.390)  \\
Oracle &  & 12.00      & 0.00      & 0.00     & 0.274 (0.226)  &  & 12.00    & 0.00      & 0.00     & 0.262 (0.201)  \\ \hline
\end{tabular}%
}
\end{table}
\begin{table}[htb]
\caption{Results for performance of grouping effect (GES) and variable selection with highly
correlated groups ($\rho=$ 0.8, 0.9, and 0.95) among covariates in Weibull models when 50\% of the data is right censored.}
\label{groupBH}
\resizebox{\textwidth}{!}{%
\begin{tabular}{ccccccccccccccc}
\hline
\multicolumn{15}{c}{$70\%$ censoring rate} 
\\
\hline
       &  &      &  & \multicolumn{5}{c}{$n=300$}                   &  & \multicolumn{5}{c}{$n=500$}                   \\ \cline{5-9} \cline{11-15} 
method &  & $\rho$ &  & GES   & TP    & FP   & MCV  & MMSE (SD)     &  & GES   & TP    & FP   & MCV  & MMSE (SD)     \\ \hline
BAR    &  & 0.8  &  & 0.895 & 11.52 & 0.12 & 0.60 & 0.319 (0.397) &  & 0.970 & 11.92    & 0.05 & 0.13 & 0.164 (0.202) \\
Lasso  &  &      &  & 0.400 & 12.00 & 6.21 & 6.21 & 0.546 (0.432) &  & 0.403 & 12.00    & 8.39 & 8.39 & 0.355 (0.254) \\
ALasso &  &      &  & 0.540 & 11.98 & 2.47 & 2.49 & 0.289 (0.305) &  & 0.561 & 12.00    & 2.42 & 2.42 & 0.213 (0.182) \\
Oracle &  &      &  & 1.00     & 12.00    & 0.00    & 0.00    & 0.192 (0.224) &  & 1.00     & 12.00    & 0.00    & 0.00    & 0.142 (0.112) \\
       &  &      &  &       &       &      &      &               &  &       &       &      &      &               \\
BAR    &  & 0.9  &  & 0.782 & 10.20  & 0.10 & 1.90 & 0.633 (0.452) &  & 0.862 & 11.29 & 0.13 & 0.84 & 0.305 (0.301) \\
Lasso  &  &      &  & 0.407 & 11.96 & 5.53 & 5.57 & 0.580 (0.440) &  & 0.403 & 12.00 & 6.85 & 6.85 & 0.400 (0.261) \\
ALasso &  &      &  & 0.482 & 11.72 & 2.70 & 2.98 & 0.407 (0.294) &  & 0.520 & 11.94 & 2.75 & 2.81 & 0.252 (0.204) \\
Oracle &  &      &  & 1.00     & 12.00    & 0.00    & 0.00    & 0.209 (0.273) &  & 1.00     & 12.00    & 0.00    & 0.00    & 0.154 (0.151) \\
       &  &      &  &       &       &      &      &               &  &       &       &      &      &               \\
BAR    &  & 0.95 &  & 0.745 & 9.17  & 0.12 & 2.95 & 0.599 (0.401) &  & 0.780 & 9.82  & 0.08 & 2.26 & 0.405 (0.256) \\
Lasso  &  &      &  & 0.387 & 11.86  & 4.96 & 5.10 & 0.544 (0.443) &  & 0.397 & 11.94 & 6.04 & 6.10 & 0.367 (0.292) \\
ALasso &  &      &  & 0.355 & 11.12 & 3.10 & 3.98 & 0.402 (0.303) &  & 0.408 & 11.55 & 3.10 & 3.55 & 0.272 (0.220) \\
Oracle &  &      &  & 1.00     & 0.00     & 0.00    & 0.00    & 0.243 (0.317) &  & 1.00     & 12.00    & 0.00    & 0.00    & 0.161 (0.147) \\ \hline
\end{tabular}%
}
\end{table}
\begin{table}[]
\caption{Performance of BAR, LASSO, and ALASSO in terms of estimation accuracy and variable selection frequencies among 100 replications when $n=100$ and $p=36$ under the parametric approach. $\boldsymbol{\beta}_{01}$, $\boldsymbol{\beta}_{02}$, and $\boldsymbol{\beta}_{03}$ denote the true parameter values corresponding to the first, second, and the third transitions.}
\label{selecfreqW1}
\resizebox{\textwidth}{!}{%
\begin{tabular}{cccccccccccccccc}
\hline
\multicolumn{16}{c}{$n=100, p=12 \times 3$} \\ \hline
Method &
  \begin{tabular}[c]{@{}c@{}}Type of \\ Assessment\end{tabular} &
  Transition &
  $\boldsymbol{\beta}$ &
  $X_1$ &
  $X_2$ &
  $X_3$ &
  $X_4$ &
  $X_5$ &
  $X_6$ &
  $X_7$ &
  $X_8$ &
  $X_9$ &
  $X_{10}$ &
  $X_{11}$ &
  $X_{12}$ \\ \hline
- &
True Values &
  $1\rightarrow 2$ &
  $\boldsymbol{\beta}_{01}$ &
  -0.8 &
  1.00 &
  1.00 &
  0.9 &
  0.00 &
  0.00 &
  0.00 &
  0.00 &
  0.00 &
  0.00 &
  0.00 &
  0.00 \\
 &
   &
  $1\rightarrow 3$ &
  $\boldsymbol{\beta}_{02}$ &
  1.00 &
  1.00 &
  1.00 &
  0.9 &
  0.00 &
  0.00 &
  0.00 &
  0.00 &
  0.00 &
  0.00 &
  0.00 &
  0.00 \\
 &
   &
  $2\rightarrow 3$ &
  $\boldsymbol{\beta}_{03}$ &
  -1.00 &
  1.00 &
  1.00 &
  0.9 &
  0.00 &
  0.00 &
  0.00 &
  0.00 &
  0.00 &
  0.00 &
  0.00 &
  0.00 \\
 &
   &
   &
   &
   &
   &
   &
   &
   &
   &
   &
   &
   &
   &
   &
   \\
BAR &
  Estimate &
  $1\rightarrow 2$ &
  $\boldsymbol{\beta}_1$ &
  -0.74 &
  0.84 &
  1.03 &
  0.84 &
  0.03 &
  0.01 &
  -0.01 &
  0.01 &
  0.00 &
  0.00 &
  -0.01 &
  0.01 \\
 &
   &
  $1\rightarrow 3$ &
  $\boldsymbol{\beta}_2$ &
  0.92 &
  0.84 &
  0.94 &
  0.79 &
  0.00 &
  0.01 &
  0.01 &
  0.00 &
  0.00 &
  -0.01 &
  0.00 &
  0.00 \\
 &
   &
  $2\rightarrow 3$ &
  $\boldsymbol{\beta}_3$ &
  -0.90 &
  0.89 &
  0.89 &
  0.98 &
  0.01 &
  0.00 &
  0.00 &
  0.00 &
  0.00 &
  0.00 &
  0.00 &
  0.00 \\
 &
   &
   &
   &
   &
   &
   &
   &
   &
   &
   &
   &
   &
   &
   &
   \\
 & Selection &
  $1\rightarrow 2$ &
  $\boldsymbol{\beta}_1$ &
  0.87 &
  0.91 &
  0.95 &
  0.91 &
  0.06 &
  0.03 &
  0.05 &
  0.03 &
  0.02 &
  0.02 &
  0.03 &
  0.03 \\
 & Frequency
   &
  $1\rightarrow 3$ &
  $\boldsymbol{\beta}_2$ &
  0.99 &
  0.94 &
  0.97 &
  0.98 &
  0.01 &
  0.02 &
  0.03 &
  0.00 &
  0.03 &
  0.02 &
  0.01 &
  0.00 \\
 &
   &
  $2\rightarrow 3$ &
  $\boldsymbol{\beta}_3$ &
  0.86 &
  0.85 &
  0.93 &
  0.97 &
  0.01 &
  0.02 &
  0.02 &
  0.01 &
  0.01 &
  0.00 &
  0.02 &
  0.00 \\
 &
   &
   &
   &
   &
   &
   &
   &
   &
   &
   &
   &
   &
   &
   &
   \\
LASSO &
  Estimate &
  $1\rightarrow 2$ &
  $\boldsymbol{\beta}_1$ &
  -0.22 &
  0.37 &
  0.86 &
  0.59 &
  0.03 &
  0.01 &
  0.00 &
  0.01 &
  0.00 &
  0.00 &
  0.00 &
  0.00 \\
 &
   &
  $1\rightarrow 3$ &
  $\boldsymbol{\beta}_2$ &
  0.84 &
  0.65 &
  0.79 &
  0.56 &
  0.02 &
  0.01 &
  0.01 &
  0.00 &
  0.00 &
  0.01 &
  0.00 &
  0.00 \\
 &
   &
  $2\rightarrow 3$ &
  $\boldsymbol{\beta}_3$ &
  -0.29 &
  0.36 &
  0.70 &
  0.69 &
  0.02 &
  0.00 &
  0.01 &
  0.00 &
  0.00 &
  -0.01 &
  0.00 &
  0.00 \\
 &
   &
   &
   &
   &
   &
   &
   &
   &
   &
   &
   &
   &
   &
   &
   \\
 & Selection &
  $1\rightarrow 2$ &
  $\boldsymbol{\beta}_1$ &
  0.72 &
  0.92 &
  1.00 &
  0.98 &
  0.02 &
  0.17 &
  0.18 &
  0.09 &
  0.15 &
  0.07 &
  0.11 &
  0.02 \\
 & Frequency
   &
  $1\rightarrow 3$ &
  $\boldsymbol{\beta}_2$ &
  1.00 &
  1.00 &
  1.00 &
  1.00 &
  0.22 &
  0.09 &
  0.13 &
  0.09 &
  0.14 &
  0.12 &
  0.11 &
  0.06 \\
 &
   &
  $2\rightarrow 3$ &
  $\boldsymbol{\beta}_3$ &
  0.72 &
  0.84 &
  0.98 &
  1.00 &
  0.13 &
  0.16 &
  0.13 &
  0.09 &
  0.13 &
  0.13 &
  0.12 &
  0.12 \\
 &
   &
   &
   &
   &
   &
   &
   &
   &
   &
   &
   &
   &
   &
   &
   \\
ALASSO &
  Estimate &
  $1\rightarrow 2$ &
  $\boldsymbol{\beta}_1$ &
  -0.51 &
  0.68 &
  0.95 &
  0.74 &
  0.02 &
  0.01 &
  0.00 &
  0.01 &
  0.01 &
  0.00 &
  0.00 &
  0.00 \\
 &
   &
  $1\rightarrow 3$ &
  $\boldsymbol{\beta}_2$ &
  0.80 &
  0.82 &
  0.87 &
  0.68 &
  0.00 &
  0.00 &
  0.01 &
  0.00 &
  0.00 &
  0.00 &
  0.00 &
  0.00 \\
 &
   &
  $2\rightarrow 3$ &
  $\boldsymbol{\beta}_3$ &
  -0.66 &
  0.74 &
  0.77 &
  0.87 &
  0.01 &
  0.00 &
  0.01 &
  0.00 &
  0.00 &
  0.00 &
  0.00 &
  0.00 \\
 &
   &
   &
   &
   &
   &
   &
   &
   &
   &
   &
   &
   &
   &
   &
   \\
 & Selection &
  $1\rightarrow 2$ &
  $\boldsymbol{\beta}_1$ &
  0.90 &
  0.97 &
  0.97 &
  0.94 &
  0.11 &
  0.09 &
  0.13 &
  0.04 &
  0.05 &
  0.05 &
  0.05 &
  0.09 \\
 & Frequency
   &
  $1\rightarrow 3$ &
  $\boldsymbol{\beta}_2$ &
  0.99 &
  1.00 &
  1.00 &
  0.99 &
  0.03 &
  0.02 &
  0.04 &
  0.03 &
  0.02 &
  0.02 &
  0.01 &
  0.01 \\
 &
   &
  $2\rightarrow 3$ &
  $\boldsymbol{\beta}_3$ &
  0.88 &
  0.98 &
  0.95 &
  0.95 &
  0.09 &
  0.06 &
  0.09 &
  0.05 &
  0.05 &
  0.03 &
  0.04 &
  0.05 \\ \hline
\end{tabular}%
}
\end{table}
\begin{table}[]
\caption{Performance of BAR, LASSO, and ALASSO in terms of estimation accuracy and variable selection frequencies among 100 replications when $n=300$ and $p=45$ under the parametric approach. $\boldsymbol{\beta}_{01}$, $\boldsymbol{\beta}_{02}$, and $\boldsymbol{\beta}_{03}$ denote the true parameter values corresponding to the first, second, and the third transitions.}
\label{selecfreqW2}
\resizebox{\textwidth}{!}{%
\begin{tabular}{ccccccccccccccccccc}
\hline
\multicolumn{19}{c}{$n=300, p=15 \times 3$} \\ \hline
Method &
  \begin{tabular}[c]{@{}c@{}}Type of \\ Assessment\end{tabular} &
  Transition &
  $\boldsymbol{\beta}$ &
  $X_1$ &
  $X_2$ &
  $X_3$ &
  $X_4$ &
  $X_5$ &
  $X_6$ &
  $X_7$ &
  $X_8$ &
  $X_9$ &
  $X_{10}$ &
  $X_{11}$ &
  $X_{12}$ &
  $X_{13}$ &
  $X_{14}$ &
  $X_{15}$ \\ \hline
- &
  True Values &
  $1\rightarrow 2$ &
  $\boldsymbol{\beta}_{01}$ &
  -0.80 &
  1.00 &
  1.00 &
  0.90 &
  0.00 &
  0.00 &
  0.00 &
  0.00 &
  0.00 &
  0.00 &
  0.00 &
  0.00 &
  0.00 &
  0.00 &
  0.00 \\
 &
   &
  $1\rightarrow 3$ &
  $\boldsymbol{\beta}_{02}$ &
  1.00 &
  1.00 &
  1.00 &
  0.90 &
  0.00 &
  0.00 &
  0.00 &
  0.00 &
  0.00 &
  0.00 &
  0.00 &
  0.00 &
  0.00 &
  0.00 &
  0.00 \\
 &
   &
  $2\rightarrow 3$ &
  $\boldsymbol{\beta}_{03}$ &
  -1.00 &
  1.00 &
  1.00 &
  0.90 &
  0.00 &
  0.00 &
  0.00 &
  0.00 &
  0.00 &
  0.00 &
  0.00 &
  0.00 &
  0.00 &
  0.00 &
  0.00 \\
 &
   &
   &
   &
   &
   &
   &
   &
   &
   &
   &
   &
   &
   &
   &
   &
   &
   &
   \\
BAR &
  Estimates &
  $1\rightarrow 2$ &
  $\boldsymbol{\beta}_1$ &
  -0.82 &
  0.90 &
  0.93 &
  0.83 &
  0.00 &
  0.00 &
  0.00 &
  0.00 &
  0.00 &
  0.00 &
  0.00 &
  0.00 &
  0.00 &
  0.00 &
  0.00 \\
 &
   &
  $1\rightarrow 3$ &
  $\boldsymbol{\beta}_2$ &
  0.85 &
  0.86 &
  0.90 &
  0.77 &
  0.00 &
  0.00 &
  0.00 &
  0.00 &
  0.00 &
  0.00 &
  0.00 &
  0.00 &
  0.00 &
  0.00 &
  0.00 \\
 &
   &
  $2\rightarrow 3$ &
  $\boldsymbol{\beta}_3$ &
  -0.89 &
  0.88 &
  0.81 &
  0.92 &
  0.00 &
  0.01 &
  0.00 &
  0.00 &
  0.00 &
  0.01 &
  0.00 &
  0.00 &
  0.00 &
  0.00 &
  0.00 \\
 &
   &
   &
   &
   &
   &
   &
   &
   &
   &
   &
   &
   &
   &
   &
   &
   &
   &
   \\
 & Selection &
  $1\rightarrow 2$ &
  $\boldsymbol{\beta}_1$ &
  1.00 &
  1.00 &
  1.00 &
  1.00 &
  0.02 &
  0.03 &
  0.04 &
  0.01 &
  0.01 &
  0.00 &
  0.01 &
  0.01 &
  0.00 &
  0.02 &
  0.02 \\
 & Frequency
   &
  $1\rightarrow 3$ &
  $\boldsymbol{\beta}_2$ &
  1.00 &
  1.00 &
  1.00 &
  1.00 &
  0.00 &
  0.01 &
  0.00 &
  0.01 &
  0.00 &
  0.00 &
  0.01 &
  0.00 &
  0.00 &
  0.00 &
  0.01 \\
 &
   &
  $2\rightarrow 3$ &
  $\boldsymbol{\beta}_3$ &
  1.00 &
  1.00 &
  1.00 &
  1.00 &
  1.00 &
  0.02 &
  0.01 &
  0.01 &
  0.00 &
  0.00 &
  0.00 &
  0.00 &
  0.00 &
  0.00 &
  0.00 \\
 &
   &
   &
   &
   &
   &
   &
   &
   &
   &
   &
   &
   &
   &
   &
   &
   &
   &
   \\
LASSO &
  Estimates &
  $1\rightarrow 2$ &
  $\boldsymbol{\beta}_1$ &
  -0.63 &
  0.73 &
  0.89 &
  0.74 &
  0.02 &
  0.01 &
  0.00 &
  0.00 &
  0.00 &
  0.00 &
  0.00 &
  0.00 &
  0.00 &
  0.00 &
  0.00 \\
 &
   &
  $1\rightarrow 3$ &
  $\boldsymbol{\beta}_2$ &
  0.83 &
  0.80 &
  0.86 &
  0.70 &
  0.01 &
  0.01 &
  0.01 &
  0.00 &
  0.00 &
  0.00 &
  0.00 &
  0.00 &
  0.00 &
  0.00 &
  0.01 \\
 &
   &
  $2\rightarrow 3$ &
  $\boldsymbol{\beta}_3$ &
  -0.68 &
  0.69 &
  0.77 &
  0.82 &
  0.02 &
  0.01 &
  0.00 &
  0.00 &
  0.00 &
  0.01 &
  0.00 &
  0.00 &
  0.01 &
  0.01 &
  0.00 \\
 &
   &
   &
   &
   &
   &
   &
   &
   &
   &
   &
   &
   &
   &
   &
   &
   &
   &
   \\
 & Selection &
  $1\rightarrow 2$ &
  $\boldsymbol{\beta}_1$ &
  1.00 &
  1.00 &
  1.00 &
  1.00 &
  0.33 &
  0.29 &
  0.28 &
  0.28 &
  0.29 &
  0.23 &
  0.28 &
  0.29 &
  0.24 &
  0.23 &
  0.34 \\
 & Frequency
   &
  $1\rightarrow 3$ &
  $\boldsymbol{\beta}_2$ &
  1.00 &
  1.00 &
  1.00 &
  1.00 &
  0.30 &
  0.33 &
  0.21 &
  0.29 &
  0.19 &
  0.27 &
  0.18 &
  0.33 &
  0.28 &
  0.30 &
  0.32 \\
 &
   &
  $2\rightarrow 3$ &
  $\boldsymbol{\beta}_3$ &
  1.00 &
  1.00 &
  1.00 &
  1.00 &
  0.32 &
  0.27 &
  0.15 &
  0.25 &
  0.27 &
  0.17 &
  0.26 &
  0.20 &
  0.22 &
  0.26 &
  0.26 \\
 &
   &
   &
   &
   &
   &
   &
   &
   &
   &
   &
   &
   &
   &
   &
   &
   &
   &
   \\
ALASSO &
  Estimates &
  $1\rightarrow 2$ &
  $\boldsymbol{\beta}_1$ &
  -0.71 &
  0.82 &
  0.91 &
  0.78 &
  0.00 &
  0.00 &
  0.00 &
  0.00 &
  0.00 &
  0.00 &
  0.00 &
  0.00 &
  0.00 &
  0.00 &
  0.00 \\
 &
   &
  $1\rightarrow 3$ &
  $\boldsymbol{\beta}_2$ &
  0.81 &
  0.84 &
  0.88 &
  0.73 &
  0.00 &
  0.00 &
  0.00 &
  0.00 &
  0.00 &
  0.00 &
  0.00 &
  0.00 &
  0.00 &
  0.00 &
  0.00 \\
 &
   &
  $2\rightarrow 3$ &
  $\boldsymbol{\beta}_3$ &
  -0.78 &
  0.80 &
  0.78 &
  0.88 &
  0.00 &
  0.00 &
  0.00 &
  0.00 &
  0.00 &
  0.00 &
  0.00 &
  0.00 &
  0.00 &
  0.00 &
  0.00 \\
 &
   &
   &
   &
   &
   &
   &
   &
   &
   &
   &
   &
   &
   &
   &
   &
   &
   &
   \\
 & Selection &
  $1\rightarrow 2$ &
  $\boldsymbol{\beta}_1$ &
  1.00 &
  1.00 &
  1.00 &
  1.00 &
  0.05 &
  0.03 &
  0.06 &
  0.04 &
  0.08 &
  0.05 &
  0.03 &
  0.03 &
  0.03 &
  0.07 &
  0.06 \\
 & Frequency
   &
  $1\rightarrow 3$ &
  $\boldsymbol{\beta}_2$ &
  1.00 &
  1.00 &
  1.00 &
  1.00 &
  0.01 &
  0.03 &
  0.01 &
  0.03 &
  0.00 &
  0.00 &
  0.00 &
  0.01 &
  0.00 &
  0.04 &
  0.01 \\
 &
   &
  $2\rightarrow 3$ &
  $\boldsymbol{\beta}_3$ &
  1.00 &
  1.00 &
  1.00 &
  1.00 &
  0.05 &
  0.09 &
  0.02 &
  0.07 &
  0.03 &
  0.04 &
  0.04 &
  0.05 &
  0.06 &
  0.04 &
  0.04 \\ \hline
\end{tabular}%
}
\end{table}
\begin{table}[]
\caption{Performance of BAR, LASSO, and ALASSO in terms of estimation accuracy and variable selection frequencies among 100 replications when $n=500$ and $p=48$ under the parametric approach. $\boldsymbol{\beta}_{01}$, $\boldsymbol{\beta}_{02}$, and $\boldsymbol{\beta}_{03}$ denote the true parameter values corresponding to the first, second, and the third transitions.}
\label{selecfreqW3}
\resizebox{\textwidth}{!}{%
\begin{tabular}{cccccccccccccccccccc}
\hline
\multicolumn{20}{c}{$n=500, p=16 \times 3$} \\ \hline
Method &
  \begin{tabular}[c]{@{}c@{}}Type of \\ Assessment\end{tabular} &
  Transition &
  $\boldsymbol{\beta}$ &
  $X_1$ &
  $X_2$ &
  $X_3$ &
  $X_4$ &
  $X_5$ &
  $X_6$ &
  $X_7$ &
  $X_8$ &
  $X_9$ &
  $X_{10}$ &
  $X_{11}$ &
  $X_{12}$ &
  $X_{13}$ &
  $X_{14}$ &
  $X_{15}$ &
  $X_{16}$ \\ \hline
- &
  True Values &
  $1\rightarrow 2$ &
  $\boldsymbol{\beta}_{01}$ &
  -0.8 &
  1.00 &
  1.00 &
  0.9 &
  0.00 &
  0.00 &
  0.00 &
  0.00 &
  0.00 &
  0.00 &
  0.00 &
  0.00 &
  0.00 &
  0.00 &
  0.00 &
  0.00 \\
 &
   &
  $1\rightarrow 3$ &
  $\boldsymbol{\beta}_{02}$ &
  1.00 &
  1.00 &
  1.00 &
  0.9 &
  0.00 &
  0.00 &
  0.00 &
  0.00 &
  0.00 &
  0.00 &
  0.00 &
  0.00 &
  0.00 &
  0.00 &
  0.00 &
  0.00 \\
 &
   &
  $2\rightarrow 3$ &
  $\boldsymbol{\beta}_{03}$ &
  -1.00 &
  1.00 &
  1.00 &
  0.9 &
  0.00 &
  0.00 &
  0.00 &
  0.00 &
  0.00 &
  0.00 &
  0.00 &
  0.00 &
  0.00 &
  0.00 &
  0.00 &
  0.00 \\
 &
   &
   &
   &
   &
   &
   &
   &
   &
   &
   &
   &
   &
   &
   &
   &
   &
   &
   &
   \\
BAR &
  Estimate &
  $1\rightarrow 2$ &
  $\boldsymbol{\beta}_1$ &
  -0.79 &
  0.89 &
  0.93 &
  0.85 &
  0.00 &
  0.00 &
  0.00 &
  0.00 &
  0.00 &
  0.00 &
  0.00 &
  0.00 &
  0.00 &
  0.00 &
  0.00 &
  0.00 \\
 &
   &
  $1\rightarrow 3$ &
  $\boldsymbol{\beta}_2$ &
  0.83 &
  0.86 &
  0.86 &
  0.79 &
  0.00 &
  0.00 &
  0.00 &
  0.00 &
  0.00 &
  0.00 &
  0.00 &
  0.00 &
  0.00 &
  0.00 &
  0.00 &
  0.00 \\
 &
   &
  $2\rightarrow 3$ &
  $\boldsymbol{\beta}_3$ &
  -0.87 &
  0.85 &
  0.80 &
  0.89 &
  0.00 &
  0.00 &
  0.00 &
  0.00 &
  0.00 &
  0.00 &
  0.00 &
  0.00 &
  0.00 &
  0.00 &
  0.00 &
  0.00 \\
 &
   &
   &
   &
   &
   &
   &
   &
   &
   &
   &
   &
   &
   &
   &
   &
   &
   &
   &
   \\
 & Selection &
  $1\rightarrow 2$ &
  $\boldsymbol{\beta}_1$ &
  1.00 &
  1.00 &
  1.00 &
  1.00 &
  0.02 &
  0.00 &
  0.01 &
  0.00 &
  0.02 &
  0.01 &
  0.02 &
  0.00 &
  0.01 &
  0.02 &
  0.00 &
  0.00 \\
 & Frequency
   &
  $1\rightarrow 3$ &
  $\boldsymbol{\beta}_2$ &
  1.00 &
  1.00 &
  1.00 &
  1.00 &
  0.02 &
  0.01 &
  0.00 &
  0.04 &
  0.00 &
  0.00 &
  0.00 &
  0.00 &
  0.00 &
  0.00 &
  0.01 &
  0.00 \\
 &
   &
  $2\rightarrow 3$ &
  $\boldsymbol{\beta}_3$ &
  1.00 &
  1.00 &
  1.00 &
  1.00 &
  0.01 &
  0.02 &
  0.00 &
  0.00 &
  0.00 &
  0.00 &
  0.00 &
  0.00 &
  0.00 &
  0.00 &
  0.00 &
  0.01 \\
 &
   &
   &
   &
   &
   &
   &
   &
   &
   &
   &
   &
   &
   &
   &
   &
   &
   &
   &
   \\
LASSO &
  Estimate &
  $1\rightarrow 2$ &
  $\boldsymbol{\beta}_1$ &
  -0.67 &
  0.79 &
  0.90 &
  0.79 &
  0.02 &
  0.00 &
  0.00 &
  0.00 &
  0.00 &
  0.01 &
  0.00 &
  0.00 &
  0.00 &
  0.00 &
  0.00 &
  0.00 \\
 &
   &
  $1\rightarrow 3$ &
  $\boldsymbol{\beta}_2$ &
  0.82 &
  0.82 &
  0.84 &
  0.74 &
  0.02 &
  0.00 &
  0.00 &
  0.00 &
  0.00 &
  0.00 &
  0.00 &
  0.00 &
  0.00 &
  0.00 &
  0.00 &
  0.00 \\
 &
   &
  $2\rightarrow 3$ &
  $\boldsymbol{\beta}_3$ &
  -0.75 &
  0.74 &
  0.79 &
  0.83 &
  0.01 &
  0.00 &
  0.00 &
  0.00 &
  0.00 &
  0.00 &
  0.00 &
  0.00 &
  0.00 &
  0.00 &
  0.00 &
  0.00 \\
 &
   &
   &
   &
   &
   &
   &
   &
   &
   &
   &
   &
   &
   &
   &
   &
   &
   &
   &
   \\
 & Selection &
  $1\rightarrow 2$ &
  $\boldsymbol{\beta}_1$ &
  1.00 &
  1.00 &
  1.00 &
  1.00 &
  0.45 &
  0.37 &
  0.42 &
  0.37 &
  0.33 &
  0.30 &
  0.37 &
  0.37 &
  0.43 &
  0.38 &
  0.32 &
  0.42 \\
 & Frequency
   &
  $1\rightarrow 3$ &
  $\boldsymbol{\beta}_2$ &
  1.00 &
  1.00 &
  1.00 &
  1.00 &
  0.41 &
  0.35 &
  0.38 &
  0.29 &
  0.35 &
  0.34 &
  0.34 &
  0.35 &
  0.37 &
  0.34 &
  0.36 &
  0.46 \\
 &
   &
  $2\rightarrow 3$ &
  $\boldsymbol{\beta}_3$ &
  1.00 &
  1.00 &
  1.00 &
  1.00 &
  0.39 &
  0.26 &
  0.34 &
  0.28 &
  0.31 &
  0.29 &
  0.29 &
  0.20 &
  0.38 &
  0.37 &
  0.26 &
  0.31 \\
 &
   &
   &
   &
   &
   &
   &
   &
   &
   &
   &
   &
   &
   &
   &
   &
   &
   &
   &
   \\
ALASSO &
  Estimate &
  $1\rightarrow 2$ &
  $\boldsymbol{\beta}_1$ &
  -0.72 &
  0.84 &
  0.92 &
  0.82 &
  0.00 &
  0.00 &
  0.00 &
  0.00 &
  0.00 &
  0.00 &
  0.00 &
  0.00 &
  0.00 &
  0.00 &
  0.00 &
  0.00 \\
 &
   &
  $1\rightarrow 3$ &
  $\boldsymbol{\beta}_2$ &
  0.81 &
  0.85 &
  0.85 &
  0.77 &
  0.00 &
  0.00 &
  0.00 &
  0.00 &
  0.00 &
  0.00 &
  0.00 &
  0.00 &
  0.00 &
  0.00 &
  0.00 &
  0.00 \\
 &
   &
  $2\rightarrow 3$ &
  $\boldsymbol{\beta}_3$ &
  -0.80 &
  0.80 &
  0.79 &
  0.86 &
  0.00 &
  0.00 &
  0.00 &
  0.00 &
  0.00 &
  0.00 &
  0.00 &
  0.00 &
  0.00 &
  0.00 &
  0.00 &
  0.00 \\
 &
   &
   &
   &
   &
   &
   &
   &
   &
   &
   &
   &
   &
   &
   &
   &
   &
   &
   &
   \\
 & Selection &
  $1\rightarrow 2$ &
  $\boldsymbol{\beta}_1$ &
  1.00 &
  1.00 &
  1.00 &
  1.00 &
  0.03 &
  0.02 &
  0.05 &
  0.00 &
  0.05 &
  0.03 &
  0.04 &
  0.03 &
  0.06 &
  0.04 &
  0.03 &
  0.01 \\
 & Frequency
   &
  $1\rightarrow 3$ &
  $\boldsymbol{\beta}_2$ &
  1.00 &
  1.00 &
  1.00 &
  1.00 &
  0.03 &
  0.01 &
  0.00 &
  0.02 &
  0.03 &
  0.01 &
  0.02 &
  0.00 &
  0.01 &
  0.01 &
  0.01 &
  0.01 \\
 &
   &
  $2\rightarrow 3$ &
  $\boldsymbol{\beta}_3$ &
  1.00 &
  1.00 &
  1.00 &
  1.00 &
  0.01 &
  0.03 &
  0.02 &
  0.03 &
  0.02 &
  0.02 &
  0.03 &
  0.02 &
  0.02 &
  0.03 &
  0.01 &
  0.02 \\ \hline
\end{tabular}%
}
\end{table}
\newpage
\end{document}